\documentclass[
  final,
  12pt,
  a4paper,
  english,
]{article}

\usepackage[utf8]{inputenc}
\usepackage[T1]{fontenc}
\usepackage[english]{babel}

\usepackage[
  a4paper,
  top=3cm,
  left=2.25cm,
  right=2.25cm,
  bottom=4cm,
]{geometry}
\usepackage{supertabular}
\usepackage{sidecap}
\sidecaptionvpos{figure}{t}
\usepackage{amsmath}
\usepackage{amsfonts}
\usepackage{amsthm}
\usepackage{amssymb}
\usepackage{comment}
\usepackage{textcomp}
\usepackage{gensymb}
\usepackage{xcolor}
\usepackage{lmodern}
\usepackage[
  tracking=true,
  expansion=true,
  protrusion=true,
  babel,
]{microtype}
\usepackage{fix-cm}
\usepackage[super]{natbib}
\usepackage[list=no]{caption}
\usepackage{graphicx}
\usepackage{booktabs}
\usepackage[mathlines]{lineno}
\usepackage{changepage}
\usepackage{floatrow}
\usepackage{helvet}
\usepackage{mathpazo}

\usepackage{sfmath}
\newcommand{\todo}[1]{{\color{red!80!black} TODO {\itshape\color{orange!70!black} #1}}}

\usepackage{orcidlink}

\usepackage{tikz}
\usetikzlibrary{positioning,arrows,shapes,calc,matrix}
\usetikzlibrary{shapes.misc}
\usepackage{pgf-umlcd}
\usepackage{pgfplots}
\pgfplotsset{compat=1.17} 
\usepackage{adjustbox}

\definecolor{mediumgrey}{rgb}{0.8,0.8,0.8}
\definecolor{lightgrey}{rgb}{0.9,0.9,0.9}
\definecolor{DarkBlue}{rgb}{0.0, 0.0, 0.5}
\definecolor{DarkRed}{rgb}{0.5, 0.0, 0.0}
\definecolor{DarkGreen}{rgb}{0.0, 0.5, 0.0}
\definecolor{Brown}{cmyk}{0.00,0.80,1.00,0.60}
\definecolor{DarkYellow}{rgb}{0.5, 0.5, 0.0}
\definecolor{Grey50}{rgb}{0.5, 0.5, 0.5}
\definecolor{Grey50}{rgb}{0.5, 0.5, 0.5}
\definecolor{keywordcolor}{RGB}{116, 22,83}

\usepackage{hyperref}

\usepackage[nameinlink]{cleveref}
\crefname{section}{Section}{Sections}
\Crefname{section}{Section}{Sections}
\crefname{figure}{Figure}{Figures}
\Crefname{figure}{Figure}{Figures}
\crefname{table}{Table}{Tables}
\Crefname{table}{Table}{Tables}
\crefname{supfigure}{Figure}{Figures}
\Crefname{supfigure}{Figure}{Figures}
\crefname{suptable}{Table}{Tables}
\Crefname{suptable}{Table}{Tables}
\crefname{equation}{Equation}{Equations}
\Crefname{equation}{Equation}{Equations}

\graphicspath{{./graphics/}}
\DeclareRobustCommand{\ie}{i.\,e.~}
\DeclareRobustCommand{\eg}{e.\,g.~}

\newcommand{\articletitle}{Intergenerational Equity in Models of Climate Change Mitigation:\\ Stochastic Interest Rates introduce Adverse Effects, but (Non-linear) Funding Costs can Improve Intergenerational Equity}
\newcommand{\cpfversion}{Version 0.9.6}

\title{\articletitle}
\author{Christian P. Fries \textsuperscript{a,b *} \orcidlink{0000-0003-4767-2034} \and
    Lennart Quante \textsuperscript{c,d} \orcidlink{0000-0003-4942-8254} }
\date{September 27, 2023}

\hypersetup{colorlinks=true}
\hypersetup{linkcolor=red!50!black}
\hypersetup{urlcolor=blue!50!black}
\hypersetup{citecolor=green!50!black}
\hypersetup{pdftitle={\articletitle}}
\hypersetup{pdfauthor={Christian Fries, Lennart Quante}}
\hypersetup{pdfsubject={}}
\hypersetup{pdfkeywords={}{}{}}

\newlength{\widthwidefig}
\newlength{\widthnarrowfig}
\newcommand{\norm}[1]{\left\vert\left\vert #1 \right\vert\right\vert}

\begin{document}

	\maketitle
	\centerline{\small \cpfversion}

%

\vfill

  {
    \footnotesize
    \noindent
    \textsuperscript{a} Department of Mathematics, Ludwig Maximilians University, Munich, Germany\\
    \textsuperscript{b} DZ Bank AG Deutsche Zentral-Genossenschaftsbank, Frankfurt a. M., Germany\\
    \textsuperscript{c} Potsdam Institute for Climate Impact Research (PIK), Member of the Leibniz Association, P.O. Box 6012 03, 14412 Potsdam, Germany\\
    \textsuperscript{d}Institute of Mathematics, University of Potsdam, Potsdam, Germany\\
    \textsuperscript{*} Correspondence to: email@christian-fries.de
}

\vspace{0.5em}

\newpage

\abstract{ 
Assessing the costs of climate change is essential to finding efficient pathways for the transition to a net-zero emissions economy, which is necessary to stabilise global temperatures at any level. In evaluating the benefits and costs of climate change mitigation, the discount rate converting future damages and costs into net-present values influences the timing of mitigation.

Here, we amend the DICE model with a stochastic interest rate model to consider the uncertainty of discount rates in the future. Since abatement reduces future damages, changing interest rates renders abatement investments more or less beneficial. Stochastic interest rates will hence lead to a stochastic abatement strategy.

We introduce a simple stochastic abatement model and show that this can increase intergenerational inequality concerning cost and risk.
    
Analysing the sensitivities of the model calibration analytically and numerically exhibits that intergenerational inequality is a consequence of the DICE model calibration (and maybe that of IAMs in general).

We then show that introducing funding of abatement costs reduces the variation of future cash-flows, which occur at different times but are off-setting in their net-present value. This effect can be interpreted as improving intergenerational effort sharing, which might be neglected in classical optimisation. This mechanism is amplified, including dependence of the interest rate risk on the amount of debt to be financed, \ie considering the limited capacity of funding sources. As an alternative policy optimisation method, we propose limiting the total cost of damages and abatement below a fixed level relative to GDP - this modification induces equality between generations compared to their respective economic welfare, inducing early and fast mitigation of climate change to keep the total cost of climate change below 3\% of global GDP.

Overall, we analyse how different approaches for the valuation of cost under interest rate risk influence optimal mitigation pathways of climate change.
}

\newpage
\tableofcontents

\newpage 
\section{Introduction}
\label{sec:diceExtension:introduction}

Climate change is one of the most significant risks of the next centuries \cite{masson-delmotte_climate_2021}, as demonstrated by its manifold impacts on society \cite{portner_ipcc_2022}. Due to the persistent nature - on human time scales - of the main greenhouse gas carbon dioxide\cite{solomon_persistence_2010}, a transition to (net) zero emissions is necessary to limit global warming to any constant level. To find the optimal pathway towards such an emission-neutral world, a diverse set of approaches have been developed. One traditional method for the assessment of transition pathways are integrated assessment models (IAMs) of the climate and economic systems -- pioneered by Nobel Laureate William Nordhaus -- for an overview of the developments on IAMs see \cite{weyant_contributions_2017,nordhaus_evolution_2018}. 

We focus on the DICE model \cite{Nordhaus2013, Nordhaus2017} as a simple IAM, which is popular to demonstrate the impacts of modifications to the original model due to its simplicity \cite{faulwasser_towards_2018,wirths_permafrost_2018,grubb_modeling_2020,Glanemann2020,Hansel2020}. 
While IAMs were first set up as deterministic, stochastic shocks have been included to consider the risk of tipping points \cite{crost_optimal_2013,lemoine_watch_2014,jensen_optimal_2014,cai_environmental_2015,cai_risk_2016,lemoine_economics_2016,dietz_economic_2021}, natural feedback processes such as permafrost \cite{wirths_permafrost_2018} or abstract catastrophic risks \cite{Ackerman2010a, ikefuji_expected_2020}.
Since most damages of climate change will occur over long time scales, value based decisions on the discount rate and optimisation function have a strong influence on the resulting optimal pathways \cite{Hepburn2007, Gollier2010, van_der_ploeg_simple_2019,VanDerPloeg2020,wagner_carbon_2020}. Stochastic approaches also require more advanced risk evaluations than Monte Carlo averaging to capture the full extent of tail risks \cite{crost_optimal_2013} or inequalities in the distribution of damages \cite{adler_priority_2017}.

All these contributions aim to provide an estimate of the societal costs of emitting one additional ton of carbon dioxide, the so-called social cost of carbon (SCC). Model extensions updating IAMs \cite{Glanemann2020, Hansel2020, dietz_economic_2021,taconet_social_2021} show a wide range of potential SCC, while also econometrics-based SCC estimates for single impacts such as mortality \cite{carleton_valuing_2020} or energy consumption \cite{rode_estimating_2021}, but also general SCC \cite{rennert_comprehensive_2022} emphasise the need to improve the precision of the estimated SCC, especially by including, so far ignored, additional impact channels. 

Discount rates, in particular, are an essential component of IAM modelling. Thus, the discussion between proponents of a descriptive approach based on observation of market returns like in DICE \cite{Nordhaus2017} versus a normative approach to discounting as in the Stern review \cite{Hepburn2007} continues. Recent studies \cite{bauer_rising_2021,newell_discounting_2022} show that dynamic modelling of discount rates might allow a more precise assessment of the costs of damages and abatement.

Here, we do not aim to provide a new, more exact estimate of SCC.

We investigate the DICE model calibration for aspects of inter-generation equity. Measuring the cost per GDP of the optimal (calibrated) climate emission pathway, we see that these costs are unequally distributed among generations.

First, we present an implementation of the DICE model with a flexible time discretisation via an Euler-approximation and couple this model with an extensive modelling toolbox for financial markets \cite{Fries2019} to enable the integration of evaluation methods for complex financial products in IAMs. 

We add stochastic interest rates to the model and introduce a simple stochastic abatement policy. Our numerical experiments show that adapting the abatement policy to the interest rate level may drastically increase the cost and risk of future generations. This is due to a kind of convexity in the dependency structure.

Analysing the sensitivities of the model calibration analytically and numerical exhibits that intergenerational inequality is a consequence of the DICE model calibration (and maybe that of IAMs in general): the calibration balances the \emph{marginal cost changes} of abatement cost and damage cost, weighted by two weights: the sensitivity of the utility function to value changes, which is usually decaying over time, and the discount factor, which is also usually decaying over time. It does not balance the cost or cost per GDP.

As the intergenerational inequality is related to the sensitivity of the utility and the discount factor, we introduce two model extensions related to these two parts that will significantly improve intergenerational equity:
Funding of abatement cost and non-linear discounting of cost, \cite{fries_non-linear_2021}.

We proceed as follows: First, we introduce the detailed specification of our model implementation \cref{sec:diceExtension:diceClassicalRewrite}. Second, we specify our extension with financing dynamics \cref{sec:diceExtension:deferredfinancing}. Third, we introduce metrics to analyse the impacts of the proposed modifications \cref{sec:introMetrics}. We analyse the repercussions of extensions on the valuation dynamics in the model by conducting several numerical experiments \cref{sec:numerical_experiments}. Finally, we provide a summary and conclusion in \cref{sec:discussion_conclusion}.

\clearpage
\section{Time-Continuous DICE Model with Arbitrary Time-Discretisation}
\label{sec:diceExtension:diceClassicalRewrite}

In this section we generalise the classical DICE model to an arbitrary time-discretization. This also constitutes the starting point for our model extensions.

The original model considered a fixed time-step of $\Delta t = 5$. Unfortunately, almost all implementations use parameters that are hard coded using this time step and use an integer $i = 0,1,2,\ldots$ to perform the time step. This makes it sometimes difficult to interpret values and parameters. For example, it may not be obvious if a value is annualised (emission per year) or per time-step (emission per $\Delta t$) and if a transition matrix (or factor) is a 1-year transition matrix or a 5-year transition matrix.

We redefine the model in a mathematical continuous-time form and then apply a Euler discretization. Time will be measured in year-fractions, i.e., $\Delta t = 1$ is one year. The classical model is recovered for a Euler discretization with an equidistant time discretization $t_{i} = i \Delta t$ with $\Delta t =5$.

All state variables $X(t)$ are observable at time $t$ (\eg temperature, carbon concentration). A flux $f(t) := \frac{\partial X}{\partial t}(t)$ is an annualised infinitesimal change (e.g., emission per year). Hence we can describe the change of the state $X$ by the flux $f$ via $\mathrm{d}X(t) = f(t) \mathrm{d}t$ and the Euler discretization of this will be
$X(t_{i+1}) = X(t_{i}) + f(t_{i}) (t_{i+1}-t_{i})$.

Thus, our implementation shows explicitly the time $t_{i}$ and the time-step $\Delta t_{i} = t_{i+1}-t_{i}$ where they apply. A reference implementation of our streamlined model can be found in the open source project \textit{finmath-lib} \cite{Fries2019}.

This model is a basis for further extension, namely to stochastic state variables, which we will discuss in Section~\ref{sec:diceExtension:deferredfinancing}.

\subsection{Temperature}

Temperature is given by a vector in $\mathbb{R}^{2}$
\begin{equation*}
	T(t) \ = \ \left(
	\begin{array}{c}
		T_{\mathrm{AT}} \\
		T_{\mathrm{LO}} \text{,}
	\end{array}
	\right)
\end{equation*}
where $T_{\mathrm{AT}}$ is the temperature in the atmosphere and $T_{\mathrm{LO}}$ is the temperature of land and ocean.

\paragraph{Temperature Forcing}

The temperature forcing $F(t)$ is the (annualised instantaneous) temperature forcing per time ($[F] = \mathrm{K}/\mathrm{year}$). It is modelled as
\begin{equation*}
    F(M;t) \ = \ \mathrm{forcingPerCarbonDoubling} \cdot \log_{2}\left( \frac{M}{M_{0}} \right) \ + \ \mathrm{forcingExternal} \text{,}
\end{equation*}
where $M$ is the carbon in atmosphere ($[M] = \mathrm{GtC}$), $M_{0} = 588 \mathrm{GtC}$, $\mathrm{forcingPerCarbonDoubling} = 3.6813 \cdot \mathrm{K/year}$ and $\mathrm{forcingExternal} = 1 \cdot \mathrm{K/year}$.

\paragraph{Evolution of the Temperature}

The temperature vector $T$ evolves as
\begin{equation*}
    \mathrm{d} T(t) \ = \ \left( \Gamma_{T} \cdot T(t) + F(t) \right) \mathrm{d}t \text{.}
\end{equation*}
The matrix $\Gamma_{T}$ models the transport of heat among the different regimes.

The classical model is recovered when considering an Euler step
\begin{equation*}
    T(t_{i+1}) \ = \ T(t_{i}) + \left( \Gamma_{T} \cdot T(t_{i}) + F(t_{i}) \right) \Delta t_{i} \ = \ \left(1 + \Gamma_{T} \Delta t_{i} \right) \cdot T(t_{i}) + F(t_{i}) \Delta t_{i} \text{.}
\end{equation*}
The matrix $\left( 1 + \Gamma_{T} \cdot 5 \ \mathrm{year} \right)$ would then correspond to the 5-Y transition matrix for the temperature vector of the original model.

\subsection{Carbon Concentration}

Carbon concentration is given by a vector in $\mathbb{R}^{3}$
\begin{equation*}
	M(t) \ = \ \left(
	\begin{array}{c}
		M_{\mathrm{AT}} \\
		M_{\mathrm{UO}} \\
		M_{\mathrm{LO}} \text{,}
	\end{array}
	\right)
\end{equation*}
where $M_{\mathrm{AT}}$ is the concentration in the atmosphere, $M_{\mathrm{UO}}$ is the concentration of the upper ocean, $M_{\mathrm{LO}}$ is the concentration of the lower ocean.

The unit of carbon concentrations is $[M] = \mathrm{GtC}$.

\subsection{Evolution of Carbon in Atmosphere}

The carbon concentration vector $M$ evolves as
\begin{equation*}
    \mathrm{d} M(t) \ = \ \left( \Gamma_{M} \cdot M(t) + c_{\mathrm{C/CO2}} \cdot E(t) \right) \mathrm{d}t \text{.}
\end{equation*}
Here $\Gamma_{M}$ is the infinitesimal generator of the transition matrix and $E$ is the emission per year. The constant $c_{\mathrm{C/CO2}}$ is the unit conversion factor from $[E] = \mathrm{GtCO}_{2}$ to $[M] = \mathrm{GtC}$.

The classical model is recovered when considering an Euler step
\begin{equation*}
	M(t_{i+1}) \ = \ M(t_{i}) + \left( \Gamma_{M} \cdot M(t_{i}) + c_{\mathrm{C/CO2}} \cdot E(t) \right) \Delta t_{i}  \text{.}
\end{equation*}
Considering a 5Y-Euler-step will recover the classical model with the 5Y transition matrix being $\left(1 + \Gamma_{M} \Delta t_{i} \right)$.

\paragraph{Emission}

The emission function $E(t)$ is the (annualised instantaneous) emissions per time ($[E] = \mathrm{GtCO}_{2} / \mathrm{year}$).

Emissions are split into two components: industrial emissions (proportional to the GDP) and external emissions (\eg land use).

The industrial emission is given by
\begin{equation*}
    E_{\mathrm{GDP}}(t) \ = \ \sigma(t) \cdot (1-\mu(t)) \cdot GDP(t) \text{.}
\end{equation*}
Here $GDP(t)$ is the (annualised instantaneous) GDP per time (per year) ($[GDP] = 10^{12} \mathrm{USD} / \mathrm{year}$, $[E_{\mathrm{EX}}(t)] = \mathrm{GtCO}_{2} / \mathrm{year}$).

The parameter $\mu$ describes the fraction of the industrial $GDP$ that has been abated, \ie, that is emission free. Increasing the parameter $\mu$ will create costs (see below).

\subparagraph{Industrial Emission Intensity Function}

The emission intensity function $\sigma(t)$ , \ie $\mathrm{emissionPerEconomicOutput}$, is modelled as
\begin{align*}
    \sigma(t) & \ = \ \sigma_{0} \cdot \exp(- \int_{0}^{t} \delta_{\sigma}(s) \cdot \mathrm{d}s) \\
    \delta_{\sigma}(t) & \ = \ \delta_{\sigma,0} \cdot \exp(- \mathrm{emissionIntensityRateDecay} \cdot t) \text{.}
\end{align*}
The original model uses a time discrete version, where the integral is discretised as
\begin{equation*}
    \sigma(t_{i+1}) \ = \ \sigma(t_{i}) \cot \exp(- \delta_{\sigma}(t_{i}) (t_{i+1}-t_{i}))
\end{equation*}
where
\begin{align*}
    \sigma_{0} = & \mathrm{emissionIntensityInitial} = (38.85 \ \mathrm{GtCO}_{2}) / (105.5 \cdot \mathrm{10^{12} USD}) \text{,} \\
    \delta_{\sigma,0} = & \mathrm{emissionIntensityRateInitial} = 0.0152 / \mathrm{year} \text{,} \\
    & \mathrm{emissionIntensityRateDecay} = -\log(1-0.001) / \mathrm{year/5} \text{.}
\end{align*}
These are the values used in the original model, re-formulated as exponential decay rates.\footnote{In the original model the $\mathrm{emissionIntensityRate}$ is expressed as $\mathrm{emissionIntensityRate} = (1 - 0.001)^{i}$ for a five year time stepping.}

Using the time continuous model makes the emission intensity independent of the time-discretization step-size. The integral can be solved explicitly. In our implementation we used the discretised version to replicate the results of the original model using the original parameters.

For the units we have $[\sigma(t)] = \mathrm{GtCO}_{2} / \mathrm{10^{12} USD}$, and all other parameters being rates with unit $1/\mathrm{year}$.

\subparagraph{External Emissions}

The external emissions $E_{\mathrm{EX}}(t)$ is an (annualised instantaneous) emissions per time ($[E_{\mathrm{EX}}] = \mathrm{GtCO}_{2} / \mathrm{year}$). 

\subparagraph{Total Emissions}

The total emissions is just the sum of $\mathrm{emissionIndustrial}$ and $\mathrm{emissionExternal}$
\begin{equation*}
    E(t) \ = \ E_{\mathrm{GDP}}(t) + E_{\mathrm{EX}}(t)
\end{equation*}

\subsection{Cost}

The cost functions are given by abatement and (climate related) damage. Both are unit-less fractions of the GDP.

\paragraph{Abatement Function}

The abatement function is a unit less function $t \mapsto \mu(t)$ that describes a fraction of GDP related emission that are abated.
That is, the fraction $\mu(t)$ of the GDP is emission free, while the faction $1-\mu(t)$ is causing emissions.

The abatement function $\mu$ is a free parameter of the model, which we may calibrate.

\paragraph{Abatement Cost}

The abatement cost function is a \textit{cost per (abated) emission} and is given by
\begin{equation*}
    \Lambda(t) = \mathrm{backstopPriceInitial} \cdot \exp(- \mathrm{backstopPriceDecayRate} \cdot t) \cdot \frac{1}{\alpha} \mu(t)^{\alpha} \text{,}
\end{equation*}
where $\mathrm{backstopPriceInitial} = 550.0/1000.0 \cdot 10^{12} \ \mathrm{USD}/\mathrm{GtCO}_{2}$ and $\mathrm{backstopPriceDecayRate} = -\log(1-0.025)/5$ and the abatement exponent $\alpha = 2.6$.

The unit of $\Lambda$ is $[\Lambda] = 10^{12} \mathrm{USD}/\mathrm{GtCO}_{2} = 1000 \mathrm{USD}/\mathrm{tCO}_{2}$.

This choice of the $\mathrm{backstopPriceDecayRate}$ will recover the value of the original model, which is
\begin{equation*}
    \exp(- \mathrm{backstopPriceDecayRate} \cdot t) \ = \ (1-0.025)^{t/5} \text{.}
\end{equation*}

The function $\mu \mapsto \mu^{2.6}/2.6$ is zero for $\mu = 0$ and small for $\mu = 0.03$ (the initial value). For $\mu = 1$ the slope of that function is $1$, such that the $\mathrm{backstopPriceInitial}$ can be interpreted as the limit price to achieve full (100\%) abatement and that 100\% of abatement has an average price of $550/2.6 \ \mathrm{USD}/\mathrm{tCO}_{2}$.

The abatement cost apply to the industrial emissions only. The abatement costs thus are
\begin{equation*}
    \Lambda(t) \cdot E_{\mathrm{GDP}}(t) \text{.}
\end{equation*}
The unit is $10^{12} \ USD$ \text{.}

\paragraph{Damage}

The model assumes that an increase in temperature leads to damages, \ie reduction of the GDP. The damage function is a unit-less function $t \mapsto \Omega(t)$ that represents a fraction of the GDP to be damaged. It is given by
\begin{equation*}
    \Omega(t) = \frac{a_{2} T(t)^{2}}{1 + a_{2} T(t)^{2}} \text{,}
\end{equation*}
where $a_{2} = 0.00236 \cdot 1/K^{2}$.

The fraction $\Omega(t)$ is removed from the GDP and is not available for investment or consumption.

\paragraph{Total Cost}

The total (annualised) costs are
\begin{equation*}
    C(t) \ = \ \Omega(t) \cdot GDP(t) + \Lambda(t) \cdot E_{\mathrm{GDP}}(t) \text{.}
\end{equation*}

\subsection{Economics}

The GDP is reduced by the $C(t)$ and the remaining amount is split between investment and consumption, where the distribution if determined by the savings rate $s(t) \in [0,1]$, \ie investment in $t$ is given by
\begin{equation*}
    I(t) := s(t) (GDP(t)-C(t))
\end{equation*}
and thus consumption in $t$ is
\begin{equation*}
    C_C(t)\footnote{In the original model, consumption is denoted as $C(t)$. Since we will refer to total costs as $C(t)$, we use $C_C(t)$ for consumption.} := (1-s(t)) (GDP(t)-C(t))
\end{equation*}

\subsubsection{Welfare}

Given the consumption $C_C(t)$, the time-$t$ utility $U(t)$ is defined via
\begin{equation*}
    V(t) \ = L(t) \left(\left(\frac{C_C(t)}{\frac{L(t)}{1000}}\right)^{1-\alpha}-1\right)/(1-\alpha),
\end{equation*}
where $L(t)$ is the population at time $t$ and $\alpha$ the elasticity of marginal utility. The original model used $\alpha = 1.45$.

The integrated discounted utility is the total welfare
\begin{equation*}
    W \ := \ \int_{0}^{T_{\max}} V(t) \frac{N(0)}{N(t)} \mathrm{d}t \text{,}
\end{equation*}
where we allow for a time-dependent discount factor\footnote{The original model used a constant discount rate $r(t) = r = \mathrm{const.}$. We will late allow for stochastic interest rates $r$.},
\begin{equation*}
    N(t) \ := \ \exp\left( -\int_{0}^{t} r(s) \mathrm{d}s \right) \text{.}
\end{equation*}

\subsubsection{GDP}

Capital $K(t)$ and $GDP(t)$ evolve as in the classical DICE model, \ie
\begin{equation*}
    K(t_{i+1}) = (1-\delta) K(t_{i}) + I(t_i),
\end{equation*}
where $\delta$ is a capital depreciation factor per time step, in the classical model 10\% per 5 years.

Thus, with $A(t)$ being the productivity and $\gamma$ the elasticity of substitution between capital and labor, 
\begin{equation*}
    GDP(t_{i+1}) = A(t_{i+1}) K(t_{i+1})^{\gamma} \frac{L(t_{i+1})}{1000}^{1-\gamma}.
\end{equation*}

Here, productivity $A(t)$ is modelled as
\begin{equation*}
    A(t_{i+1}) = \frac{A(t_{i})} {(1 - ga * \exp(- deltaA * t))^{\frac{\delta t}{5}}},
\end{equation*}
   where initial productivity growth rate $ga= 0.076$ and productivity growth decreases by $deltaA=0.005$.

Population $L(t)$ is assumed to follow a deterministic path described by
\begin{equation*}
    L(t_{i+1}) = L(t_{i}) \frac{L(\infty)}{L(t_{i})}^{g t_{i+1}},
\end{equation*}
   using $L(\infty)= 11500$ and $g=0.134/5$, such that the unit of population is [mil persons].

\clearpage

\section{IAM with Deferred Non-Linear Stochastic Financing Cost}
\label{sec:diceExtension:deferredfinancing}

We propose modifications to integrated assessment models to investigate the effect of stochastic interest rates, funding, and non-linear financing cost \cite{fries_non-linear_2021}. These may be also incorporated into more complex IAMs. We use the simple DICE model to illustrate potential effects.

Our choice to make interest rates stochastic is only exemplary. The interest rate is an important factor in linking present abatement cost to the avoided future damage cost. While one may introduce stochasticity in many other state variables, the use of an abatement policy adapted to the economic factors (interest rates) will already lead to all other state variables becoming stochastic.

\subsection{Model Extensions}

\subsubsection{Stochastic Interest Rates and using a Risk Measure}
\label{sec:introducingStochInterestRates}

The classical way in which interest rates are modelled in an IAM is via a discount factor. A time-$t$ value $V(t)$ is derived in the model, then discounted and aggregated to a final value forming the objective function
\begin{equation*}
    \int_{0}^{T} V(t) \exp\left( - \int_{0}^{t} r(s) \mathrm{d}s \right) \mathrm{d}t \text{.}
\end{equation*}

We model stochastic interest rates $r$. Our implementation allows to use a general discrete forward rate model (LIBOR Market Model). The experiments in Section~\ref{sec:numerical_experiments} were conducted with a classical Hull-White model. The model provides the stochastic numeráire $N$ (and thus discount factor) as well as the stochastic forward rate $FR$.

If interest rates are stochastic but the function $V(t)$ remains deterministic, adding stochastic interest rates does not change the interaction with the IAM since
\begin{align*}
    \mathrm{E} \left( \int_{0}^{T} V(t) \exp\left( - \int_{0}^{t} r(s) \mathrm{d}s \right) \mathrm{d}t \right) & \ = \ \int_{0}^{T} V(t) \mathrm{E} \left( \exp\left( - \int_{0}^{t} r(s) \mathrm{d}s \right) \right) \mathrm{d}t \\
	& \ = \ \int_{0}^{T} V(t) \exp\left( - \int_{0}^{t} \bar{r}(s) \mathrm{d}s \right) \mathrm{d}t \text{,}
\end{align*}
with $\bar{r}(t) \ = \ - \frac{\partial}{\partial t} \log\left( \mathrm{E} \left( \exp\left( - \int_{0}^{t} r(s) \mathrm{d}s \right) \right) \right)$. Thus, in a classical IAM the discount factor can be interpreted as an expectation.
Due to the lack of a feedback that allows to adjust the abatement path and resulting damages, adding stochastic interest rates does not introduce changes in the model dynamics.

\subsubsection{Stochastic Abatement Policy}

As interest rate levels and term-structure have a strong impact on the optimal abatement policy, it is natural to introduce a stochastic abatement policy, i.e., $\mu$ will become stochastic. This reflects the possibility that the abatement policy can be adjusted to the interest rate scenarios.
Thus abatement policy $t \mapsto \mu(t)$ becomes a stochastic process that adapts to the changes in interest rates.
A simple example for a stochastic abatement model is a parametric one, where the abatement speed is a (linear) function of the interest rate level $r$, e.g.,
\begin{equation}
    \mu(t,\omega) = \min\left( \mu_{0} + \left( a_{0} + a_{1} \cdot r(t,\omega) \right) \cdot t , 1.0 \right) \text{.}
\end{equation}

Moving to a stochastic abatement model, \emph{all} model quantities become stochastic. This then introduces stochastic cost, and hence a notion of risk.

\subsubsection{Funding of Abatement Costs}
\label{sec:diceExtension:fundingOfAbatement}

The geophysical part of the model produces damages. Future damages may be reduced by performing abatement. Both, abatement and damage are associated with costs. The abatement costs and the damage costs are compensated instantaneously. This is modelled by reducing the time $t_{i}$ GDP by the  time $t_{i}$ costs. The remainder is then available for consumption or investment. Investment adds to the capital which determines the GDP of the next time $t_{i+1}$.

We introduce the ability of funding of cost, especially for abatement cost. As abatement is a planned process of societal relevance, it is reasonable to assume that abatement cost are covered by a loan for which interest rate corresponds to the current discount rate.\footnote{Our model allows to apply a funding spread, \ie higher interest rates for loans than for discounting, but this is not considered in this general introduction of the phenomenon.}

We define $C_{\mu}(t)$ as the instantaneous abatement costs in time $t$, which might be funded for a period $\Delta T_{\mathrm{A}}$.

Thus abatement costs of $C_{\mu}(t)$ are accrued with the forward rate $FR(t,t+\Delta T_{\mathrm{A}};t)$ observed in $t$, such that the realized abatement cost $C_{\mathrm{A}}$ at time $t+ \Delta T_{\mathrm{A}}$ are
\begin{equation}
    \label{eq:dice:fundingOfAbatementCost}
    C_{\mathrm{A}}(t + \Delta T_{A}) \ := \ C_{\mu}(t) \ \left( 1 + FR(t,t+\Delta T_{A};t) \Delta T_{\mathrm{A}} \right).
\end{equation}
For the case of no funding, i.e. $\Delta T_{\mathrm{A}} = 0$, $C_{\mu} = C_{\mathrm{A}}$.

In a standard model for risk-neutral valuation, this change would have no effect, as the accruing is compensated by the discounting. However, the change might introduce an effect in the DICE model, due to the way how abatement and damage cost are associated and due to the time-preference included in the utility function.

\medskip

Since damages occur instantaneously, we do not consider funding of these and thus total cost is given by
\begin{equation*}
    C(t) \ := \ C_{\mathrm{A}}(t) + C_{\mathrm{D}}(t) \text{.}
\end{equation*}

\subsubsection{Non-Linear Financing Costs}
\label{sec:nonlinear_discounting}

For an unsecured financial cash-flow its present value is defined by a discount factor times the cash-flow. If the future cash-flow is subject to default, the discount factor is lower, reflecting the additional value reduction due to the risk of (partial) default. As default is not an option for future damage cost, it appears as if the risk-free discount factor should apply. However, since no hedging strategy exists, a risk free funding is not possible. Thus additional cost may occur to secure the unsecured funding \cite{fries_non-linear_2021}. Since damage cost may become very large - much larger than funds provided by standard financial markets - it is reasonable to assume that these additional funding cost become (over-proportionally) large for larger cost. We model this by optionally adding non-linear financing cost using a non-linear funding model \cite{fries_non-linear_2021}. 

This model\cite{fries_non-linear_2021} allows, that the discount factor may depend on the magnitude of the cash-flow. Thus the funding of larger cash-flows requires a premium to compensate for a larger default risk or other frictions. For our application, this means that larger cost get a larger weight.

Our model modification is now a modification of the damage cost. Let $C^{\circ}_{\mathrm{D}}(t)$ denote the damage cost of the classical model, i.e.

We then define the effective damage costs as
\begin{equation*}
    C_{\mathrm{D}}(t) \ = \  C^{\circ}_{\mathrm{D}}(t) \cdot DC(C^{\circ}_{\mathrm{D}}(t); t) \text{,}
\end{equation*}
where $DC(C^{\circ}_{\mathrm{D}}(t); t)$ is the \emph{default compensation factor}. It is somewhat similar to the inverse of a discount factor, describing the over-proportional cost to fund large projects.

As $C^{\circ}_{\mathrm{D}}(t)$ represents a time-value, it is natural that the default compensation factor depends on a renormalised value only. We allow two different renormalisations: either with the numéraire $N(t)$ or with the GDP $Y(t)$. Hence, our model for the default compensator is
\begin{equation}
    \label{eq:nonlineardiscounting:relativetoNumeraire}
    DC(C^{\circ}_{\mathrm{D}}(t); t) \ = \ DC^{N}(C^{\circ}_{\mathrm{D}}(t); t) \ := \ DC^{*}( \frac{C^{\circ}_{\mathrm{D}}(t)}{N(t)} ) \text{,}
\end{equation}
or, alternatively,
\begin{equation}
    \label{eq:nonlineardiscounting:relativetoGDP}
    DC(C^{\circ}_{\mathrm{D}}(t); t) \ = \ DC^{Y}(C^{\circ}_{\mathrm{D}}(t); t) \ := \ DC^{*}( \frac{C^{\circ}_{\mathrm{D}}(t)}{Y(t)} ) \text{.}
\end{equation}

The factor $DC^{*}(x)$ depends on the size of $x$. For small $x$ we have $DC^{*}(x) = 1$, but for large $x$ we may have factors $> 1$. Obviously, this will penalise large spikes in the costs.

\smallskip

As the abatement cost are usually smaller and part of a more planned process, we do not consider a default compensation factor for the abatement cost. Our implementation would allow this, however.
The total cost is given by
\begin{equation*}
    C(t) \ := \ C_{\mathrm{A}}(t) + C_{\mathrm{D}}(t) \text{.}
\end{equation*}

\subsubsection{Closing the Model: modelling a Sustainable Equilibrium for the Far Future}

If we simulate welfare $V(t)$ over a finite time horizon $[0,T]$ and calculate the discounted total welfare $\int_{0}^{T} \frac{V(t)}{N(t)} \mathrm{d}t$ (where $\frac{1}{N(t)}$ denotes the discount factor), we have to ensure that the remainder $\int_{T}^{\infty} \frac{V(t)}{N(t)} \mathrm{d}t$ is negligible. In other words, we have to ensure that the model does not ``cheat'' by moving the cost beyond the time horizon. This is maybe even more important in our modification since it allows to defer the costs, but the issue could also exist in the classical model.

\subsection{Alternative Objective Functions and Constraints}

Given that state variables are stochastic, the question arises of what are suitable objective functions and constraints.

Now that the time $t$ welfare (utility) $V(t)$ is a random variable, the total welfare\footnote{Using a time-discretization with constant time steps $\Delta t_{i} = \Delta t$, the definition \eqref{eq:aggregatedDiscountedWelfare} differs by a factor of $\Delta t$ from the definition in the classical model. For the maximisation this makes no difference.}
\begin{equation}
    \label{eq:aggregatedDiscountedWelfare}
    V^{*} \ := \ \int_{0}^{T} V(t) \frac{N(0)}{N(t)} \mathrm{d} t \ \approx \ \sum_{i=0}^{n} V(t_{i}) \frac{N(t_{0})}{N(t_{i})} \Delta t_{i}
\end{equation}
is a random variable too.

As we show in Section~\ref{sec:distributionOfCost}, maximising the expected total welfare - as the classical model does - will introduce some intergenerational inequality as one may exchange future burden for present benefits. This effect is already present in the classical model (see  \cref{fig:costOverTimeDiscounted-reduced}).

With stochastic state variables, we could perform a (risk-neutral) valuation and consider the expectation of the discounted welfare as our objective function. But risk-neutral valuation is only admissible if there are means of hedging, \ie neutralisation of risks through financial instruments. 

As this is not a-priori clear for all damages and costs in an IAM, we additionally consider a risk measure applied to the, e.g., discounted welfare. In this case, the policy optimisation becomes sensitive to the stochastic properties of the interest rates, as different stochastic realisations introduce different time-dependent weights that take the place of the discount factor. A natural choice would be the expected shortfall of discounted welfare, since it is a coherent risk measure \cite{tasche_expected_2002,acerbi_coherence_2002} and thus suitable to asses tail-risks of distributions, which would be neglected if one would consider \eg Value-at-Risk, \ie the $\alpha$ level quantile of the observed random variable.

\subsubsection{Expected Shortfall}
The expected shortfall measures the expected return of value random variable $X$ in the $\alpha$ worst cases, \ie
\begin{equation}
\label{eq:expectedShortfall}
    ES_{\alpha}\left(X\right) := \frac{1}{\alpha} \int_0^{\alpha} q_u(X) du,
\end{equation}
where $q_u(X)$ is the quantile function of the distribution of $X$.

\subsubsection{Altering the Time-Preference}
\label{sec:objectiveFunction:pnorm}

Using the aggregated discounted welfare \eqref{eq:aggregatedDiscountedWelfare} calibration will balance short term welfare gains with long-term welfare losses.

While this may seem reasonable for ones individual utility, it may induce intergenerational inequality, as we show in \cref{sec:distributionOfCost}.
A possible modification of the objective function that addresses this problem is to allow the aggregation of time-preferences welfare only over a limited time interval, then apply some other norm to the resulting value vector, e.g., a norm that is creating a preference for equi-distribution of the vector elements.

An objective function that achieves this is a $p$-norm of the discounted welfare (with $0 < p < 1$, since we like to maximise the value but penalise inequality).

Optionally one may allow some inter-temporal aggregation and discounting, for example within the lifespan of a generation $\Delta T^{\mathrm{g}}$ (e.g., for $\Delta T^{\mathrm{g}} = 100$). Introduce the time-$t$ average discounted welfare as
\begin{equation}
    \bar{V}^{\Delta T^{\mathrm{g}}}(t) \ := \ \frac{1}{\Delta T^{\mathrm{g}}} \int_{t-\Delta T^{\mathrm{g}}}^{t} V(s) \frac{N(t)}{N(s)} \mathrm{d}s
\end{equation}
we may then apply as some norm to $V(t)/N(t)$ and define the models objective function as
\begin{equation}
    W \ := \ \norm{ \frac{\bar{V}^{\Delta T^{\mathrm{g}}}(t)}{N} } \text{.}
\end{equation}

The classical model is recovered if we choose $\Delta T^{\mathrm{g}} = 0$ such that $\bar{V}^{\Delta T^{\mathrm{g}}}(t) = V(t)$, then choose the $L_{1}$-norm, i.e.,
\begin{equation}
    W \ := \ \norm{ \frac{\bar{V}^{\Delta T^{\mathrm{g}}}(t)}{N} }_{L{1}} \ = \ \int_0^{T} \frac{V(t)}{N(t)} \ \mathrm{d}t \text{.}
\end{equation}

A norm that would favour more evenly distributed discounted welfare is a p-norm with $0 < p < 1$. We show the model calibration results with an $\frac{1}{4}$-norm in Section~\ref{sec:result:modelWithPNorm}.
We observe that using this approach improves the intergenerational equity, however, a similar (or even better) result can also be achieved using the non-linear discounting, as it is penalising large local peaks in the cost function.

\clearpage
\section{Derived Quantities providing Model Insights}
\label{sec:introMetrics}
The classical objective function of the model is the integrated discounted welfare. Apart from this there are other quantities that provide insights or may serve as a condition or objective function. The \textit{social cost of carbon} is the most common metric to summarise results.

\subsection{Social Cost of Carbon}

The social cost of carbon is defined as the random variable
\begin{equation}
    SCC(t, \omega) \ := \ -1000 \cdot \frac{\partial V(t,\omega)}{\partial E(t,\omega)} \Big/ \frac{\partial V(t,\omega)}{\partial C(t,\omega)} \text{.}
\end{equation}
The $SCC$ is the time $t$ marginal price of emmiting one additional unit of carbon. It's unit is $[SCC] = \frac{\mathrm{USD}}{\mathrm{tCO}_{2}}$.

In what follows we will rarely consider the $SCC$ and rather focus on other quantities. However, we like to note that depicting the $SCC$ in this form may be misleading. As it is price, one should rather consider the discounted, or, numéraire relative $SCC$, i.e.,
\begin{equation}
    SCC^{N}(t, \omega) \ := \ -\frac{1000}{N(t,\omega)} \cdot \frac{\partial V(t,\omega)}{\partial E(t,\omega)} \Big/ \frac{\partial V(t,\omega)}{\partial C(t,\omega)} \text{.}
\end{equation}

\subsection{Distribution of Cost}

Adding to the aggregated discounted welfare, it provides some insights to analyse the distribution of the cost $C(t)$, which is the sum of the abatement cost and the damage cost $C(t) = C_{\mathrm{A}}(t) + C_{\mathrm{D}}(t)$.

This allows us to separate expected gains from the economy and the modelled economic degradation from climate damages.

Already in the classical model the temporal distribution of cost is not even among different generations, which may raise some questions with respect to intergenerational equity.
One might argue that this effect is reflected by discounting.
However, this argument may be incomplete, since the aggregated and discounted utility is the utility of different generations.

As cost enters into the final welfare though application of the utility and discounting, we consider the \textit{cost-to-value-weight}
\begin{equation*}
    \frac{\partial V(t)}{\partial C(t)} \text{.}
\end{equation*}

\bigskip

In a stochastic model, additional differences may arise regarding risk and financing cost related to these cost.

\subsection{Sensitivity of Damage Cost and Abatement Cost to Policy Changes}

In order to understand how the presented model extensions affect modelling results, we analyse the sensitivity of damage cost and abatement cost to policy changes.

The effect of a local change of the abatement policy $\mu(t)$ to the abatement cost is immediate. Its effect to the damage cost is  a reduction of future damages. This establishes a dependency on the model inherent inter-temporal weights attributed to costs.

\subsubsection{Value Change per Cost Change}

To understand how abatement and damage are connected we analyse first the sensitivity of the welfare function to both parts of the costs, $\frac{\partial V}{\partial C(t)}$. 

The objective function of the model is the value or welfare $V$. As $V$ depends on the cost, two transformations take place:
First, the value is defined as utility. Since the utility function is convex, there is some saturation, resulting in a over time strongly decaying weight applied to the cost. Second, the value is discounted. For the model of a constant positive discount rate, the discount factor constitutes an exponentially decaying weight applied to the cost.

This quantity is just the product of the discount factor and the slope of the utility function. Note that not only the discount factor is decreasing by time, but also the slope of the utility function, as long as overall utility keeps increasing.

\subsubsection{Sensitivity of Cost to Policy Changes}

A change in the abatement policy in time $t$ affects abatement cost $C_{\mathrm{A}}(t)$ only at time $t$, whereas it affects damage cost $C_{\mathrm{D}}(s)$ for possibly all $s > t$. The inter-temporal cost structure of a change in the abatement policy $\mu(t)$ is given by
\begin{equation}
    \label{eq:dice:sensiOfCostToAbetement}
    \frac{\frac{\partial C_{\mathrm{D}}(s)}{\partial \mu(t)}}{\frac{\partial C_{\mathrm{A}}(t)}{\partial \mu(t)}} \text{.}
\end{equation}
\Cref{eq:dice:sensiOfCostToAbetement} gives the damage cost per abatement cost due to a policy change in $t$. As the objective function of the model is $V$, the model is weighting the cost by discounting and utility calculation. For the equilibrium state the relevant sensitivity is
\begin{equation}
    \label{eq:dice:sensiOfCostToAbetementValue}
    \frac{\partial V}{\partial C}  \frac{\frac{\partial C_{\mathrm{D}}(s)}{\partial \mu(t)}}{\frac{\partial C_{\mathrm{A}}(t)}{\partial \mu(t)}} \text{.}
\end{equation}

\subsubsection{Example: One Parameter Abatement Model}

Consider the one-parameter deterministic abatement model
\begin{equation*}
    \label{eq:abatementModel:determ:oneparam}
    \mu(t) \ = \ \min \left( \mu(0) + \frac{1 - \mu(0)}{T^{\mu=1}} \ t , 1.0 \right) \text{,}
\end{equation*}
where the parameter $T^{\mu=1}$ represents the time for reaching 100\% abatement. The optimal parameter, representing the equilibrium state of the model, fulfills
\begin{equation}
    \label{eq:dice:costAsCashflowStructure}
    \frac{\mathrm{d}}{\mathrm{d} T^{\mu=1}} \ \int_0^{T} V(t) \frac{N(0)}{N(t)} \mathrm{d}t \ = \ 
    \int_{0}^{T} \frac{\partial C_{\mathrm{A}}(t) + C_{\mathrm{D}}(t)}{\partial T^{\mu=1}} \ \frac{\partial V}{\partial C(t)} \ \frac{N(0)}{N(t)} \ \mathrm{d}t \ = \ 0 \text{.}
\end{equation}

\cref{eq:dice:costAsCashflowStructure} allows us to interpret a change in a parameter of the abatement policy as a change in the cash-flow structure of a financial product. Here $C_{\mathrm{A}}(t) + C_{\mathrm{D}}(t)$ corresponds to the time-$t$ cash-flow and $\frac{\partial V}{\partial C(t)} \mathrm{d}t$ to the discount factor.

From \cref{eq:dice:costAsCashflowStructure} we can also see why the model calibration may result in some intergenerational inequality: The calibration does not minimise some temporal norm of the cost (say, an $L_{2}$-norm, which would lead to a more even distribution). The calibration is equating the marginal cost changes. Maximising the total discounted value, we find at the maximum
\begin{align}
    \nonumber \frac{\mathrm{d}}{\mathrm{d} T^{\mu=1}} \ \int_0^{T} V(t) \frac{N(0)}{N(t)} \mathrm{d}t
    & \ = \ \int_{0}^{T} \frac{\mathrm{d}V(t)}{\mathrm{d}C(t)} \frac{\mathrm{d}C(t)}{\mathrm{d} T^{\mu=1}} \ \frac{N(0)}{N(t)} \mathrm{d}t \\
    & \ = \ \int_{0}^{T} \frac{\mathrm{d}V(t)}{\mathrm{d}C(t)} \left( \frac{\mathrm{d}C_{\mathrm{D}}(t)}{\mathrm{d} T^{\mu=1}} + \frac{\mathrm{d}C_{\mathrm{A}}(t)}{\mathrm{d} T^{\mu=1}} \right) \ \frac{N(0)}{N(t)} \mathrm{d}t \ \stackrel{!}{=} \ 0
    \label{eq:cost_objective}
\end{align}
Since $\mathrm{d}C_{\mathrm{D}} / \mathrm{d} T^{\mu=1} > 0$ and $\mathrm{d}C_{\mathrm{A}} / \mathrm{d} T^{\mu=1} < 0$ we see from \cref{eq:cost_objective} that the model equates the \emph{increment} of the cost of damage to the \emph{increment} of the cost of abatement, both weighted by the value-to-cost-sensitivity $\mathrm{d}V / \mathrm{d}C$.
\begin{equation*}
    \int_{0}^{T} \frac{\mathrm{d}V(t)}{\mathrm{d}C(t)} \frac{\mathrm{d}C_{\mathrm{D}}(t)}{\mathrm{d} T^{\mu=1}} \ \frac{N(0)}{N(t)} \mathrm{d}t
    \ + \
	\int_{0}^{T} \frac{\mathrm{d}V(t)}{\mathrm{d}C(t)} \frac{\mathrm{d}C_{\mathrm{A}}(t)}{\mathrm{d} T^{\mu=1}} \ \frac{N(0)}{N(t)} \mathrm{d}t
	\ \stackrel{!}{=} \ 0
\end{equation*}
This criterion is plausible if a single agent optimises its cost, but may be problematic if the costs are distributed between different generations.

The two cost components are weighted by $\frac{\mathrm{d}V(t)}{\mathrm{d}C(t)}$ and $\frac{N(0)}{N(t)}$. Both weights decaying over time. The first weight is the sensitivity of the utility function to changes in cost. It decays as utility becomes more saturated. The second wights is the discount factor at the interest rate $r$.

As abatement cost occur early, but damage cost occur late, introducing a time-shift in the abatement cost via the funding \cref{eq:dice:fundingOfAbatementCost} will reduce the variance in the change of the cash-flow structure when changing the abatement policy. Funding of abatement cost will improve the alignment of the utility sensitivity weight $\frac{\mathrm{d}V(t)}{\mathrm{d}C(t)}$  of damage cost and abatement cost.

For the discount factor $\frac{N(0)}{N(t)}$ we introduce a non-linear discounting that models a non-linear increasing in compensating very large cost.
 
Our numerical experiment show that both modifications improve the inter-generational equity in the sense of the temporal distribution of cost.

\clearpage
\section{Stylised Effects in Numerical Experiments}
\label{sec:numerical_experiments}

In this section we discuss some qualitative effects of the model extensions and present associated numerical experiments conducted with the model framework.

The first experiments are based on the re-implementation of the classical DICE model - without adding stochastic interest rates. We then add our model modifications (funding and non-linear discounting) to the model with deterministic rates and analyse the factors that impact the abatement policies and cost structure. We repeat this analysis for the model with stochastic interest rates and stochastic abatement policy, where now risk is becoming an additional aspect.

%
%

\subsection{Model Calibration}

The classical model is calibrated by choosing the parameters $\mu(t_{i})$ and $s(t_{i})$, $i=1,\ldots,n$ such that they maximise $W$,
\begin{equation*}
    W \ := \ \int_{0}^{T} V(t) \frac{N(0)}{N(t)} \mathrm{d}t \text{.}
\end{equation*}

\subsubsection{Abatement and Savings Rate Model: Parameter Reduction}

To analyse the impact of changing model parameters, we sometimes restrict the model to two very simple one-parametric sub-models for the abatement model and the savings rate model.
We assume a deterministic model
\begin{equation}
    \label{eq:abatementModelDeterministic}
    \begin{split}
    \mu(t) & \ = \ \min\left( \mu_{0} + a_{0} \cdot t , 1 \right) \text{,} \\
    s(t) & \ = \ s_{0} \text{.}
    \end{split}
\end{equation}
Within the reduced model \cref{eq:abatementModelDeterministic}, we determine the pair $(a_{0},s_{0})$ that maximises the welfare.
This abatement speed parameter $a_{0}$ can be translated to
\begin{equation*}
    T^{\mu=1} \ := \ \left( 1.0 - \mu_{0} \right) / a_{0} \text{,}
\end{equation*}
the time where the model reaches 100\% abatement.

\begin{figure}[hp]
    \centering
    \includegraphics[width=0.45\textwidth]{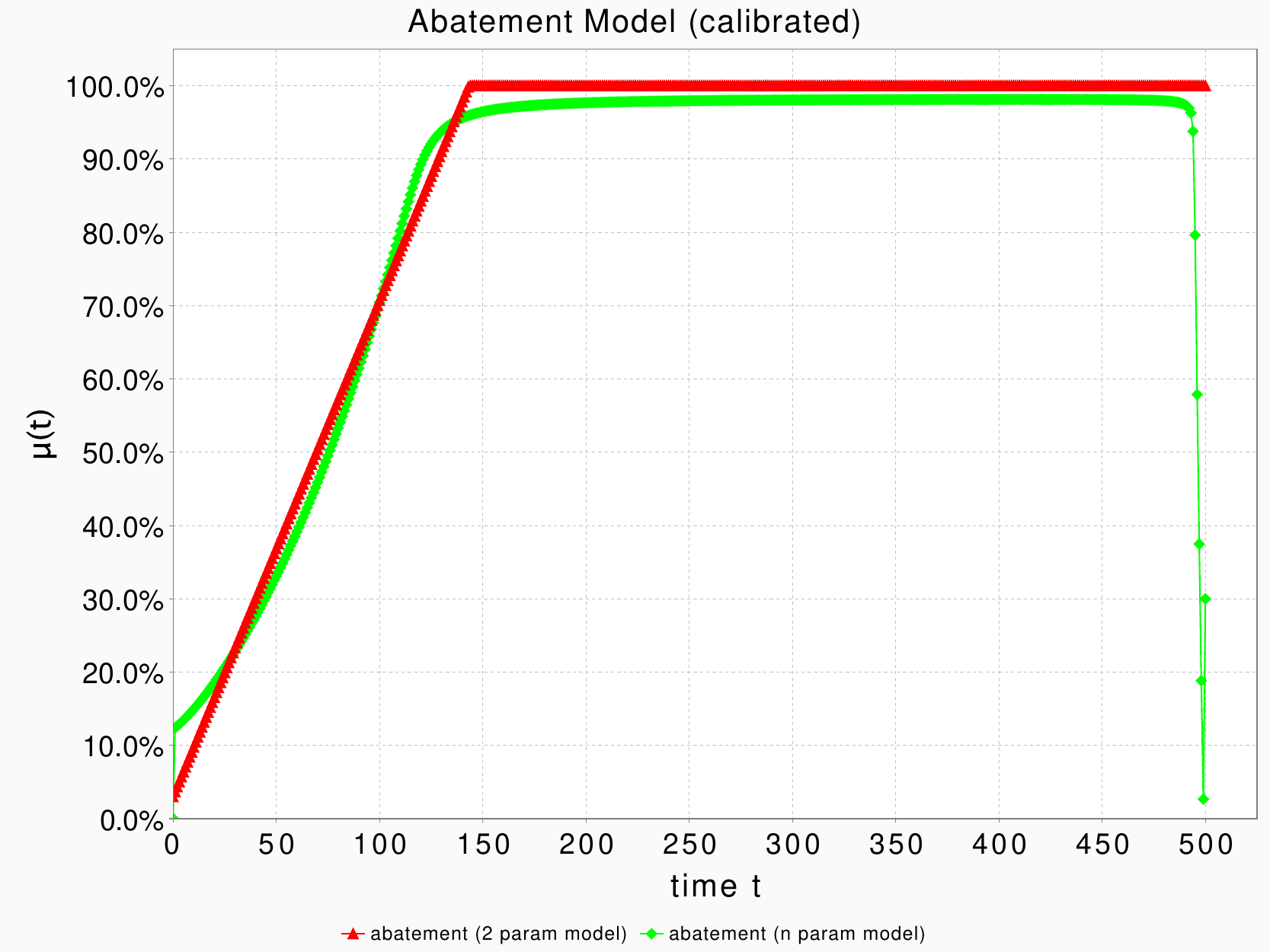}
    \includegraphics[width=0.45\textwidth]{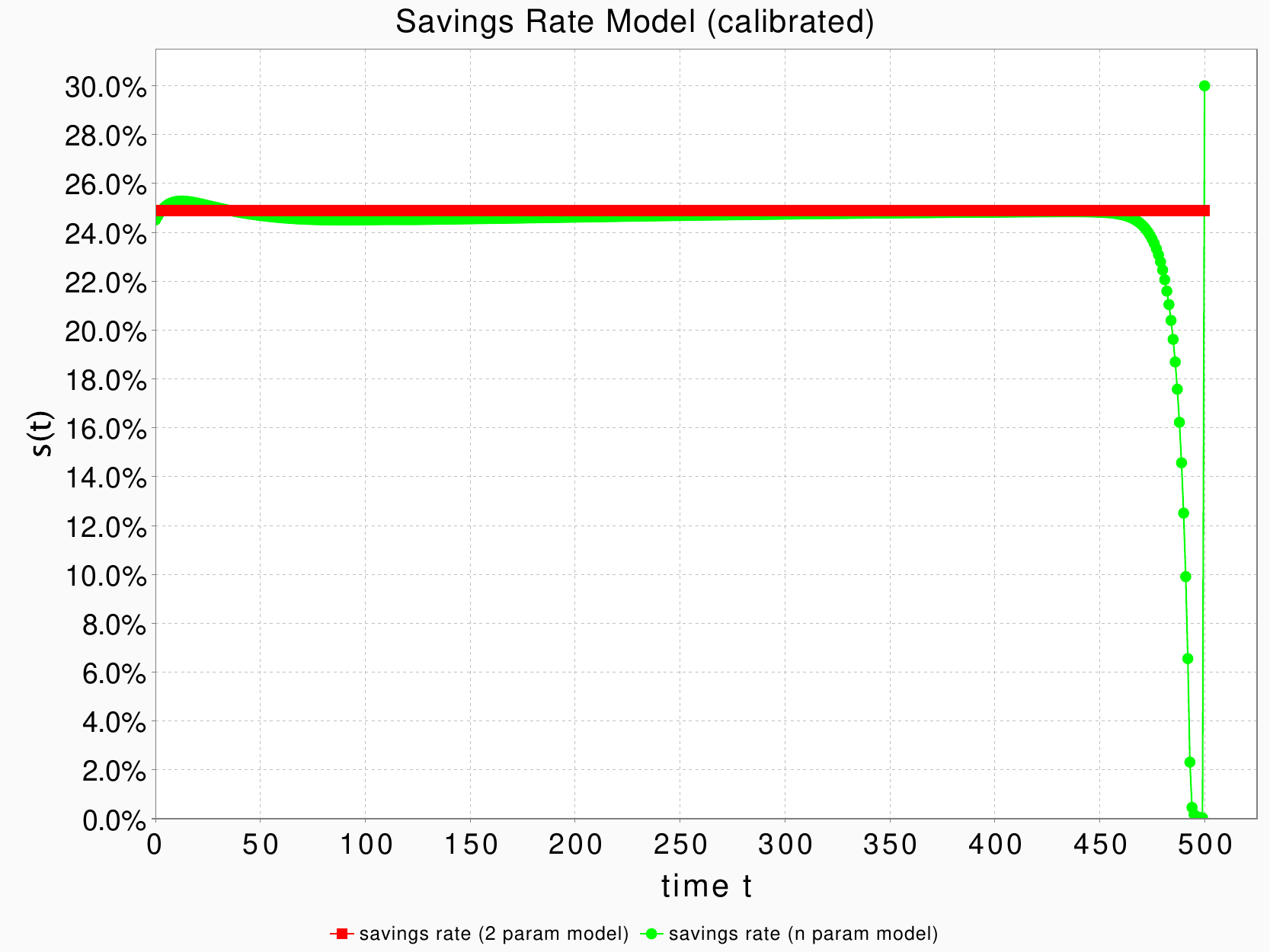}
    \caption{The calibrated model (maximising welfare) using two parameters (red) versus 1000 parameters (green).\footnote{The model with a free functional form shows a strong drop in the model parameters shortly before the time horizon. The finite time-horizon of the model implies that all is consumed and nothing is saved before the time-horizon. See Section~\ref{sec:closingthemodel}.}}
    \label{fig:calibration}
\end{figure}

It turns out the reduced model is fairly close to the full model, where every $\mu(t_{i})$ and every $s(t_{i})$ is a free parameter. The following \cref{fig:calibration} compares the optimal abatement and savings rate paths of the two parameter model to the full model having 1000 free parameters (500 time-steps).

%
%

\clearpage
\subsection{Distribution of Costs - intergenerational Equity in the Calibrated Model}
\label{sec:distributionOfCost}

We analyse the cost function $C(t)$ in the calibrated model.\footnote{The figures of this subsection can be reproduced by class \texttt{ClimateModelExperimentCostDistribution}.} As discounting and utility still play a role within a generation, but maybe less across generations, we additionally depict a running average of the cost, the generational average discounted cost
\begin{equation}
    C^{\mathrm{g}}(t) := \int_{t-\Delta T^{\mathrm{g}}}^{t} C(s) \frac{N(t)}{N(s)} \mathrm{d}s \text{,}
\end{equation}
where for example $\Delta T^{\mathrm{g}} = 100$ is chosen as the lifetime of a generation.

\smallskip

Since the cost structure of the full model (where $\mu(t_{i})$ and $s(t_{i})$ are calibrated) is similar to the cost structure of the reduced two parameter model, we consider the calibrated reduced model in the following.

\begin{figure}[hp]
    \centering
    \includegraphics[width=0.9\textwidth]{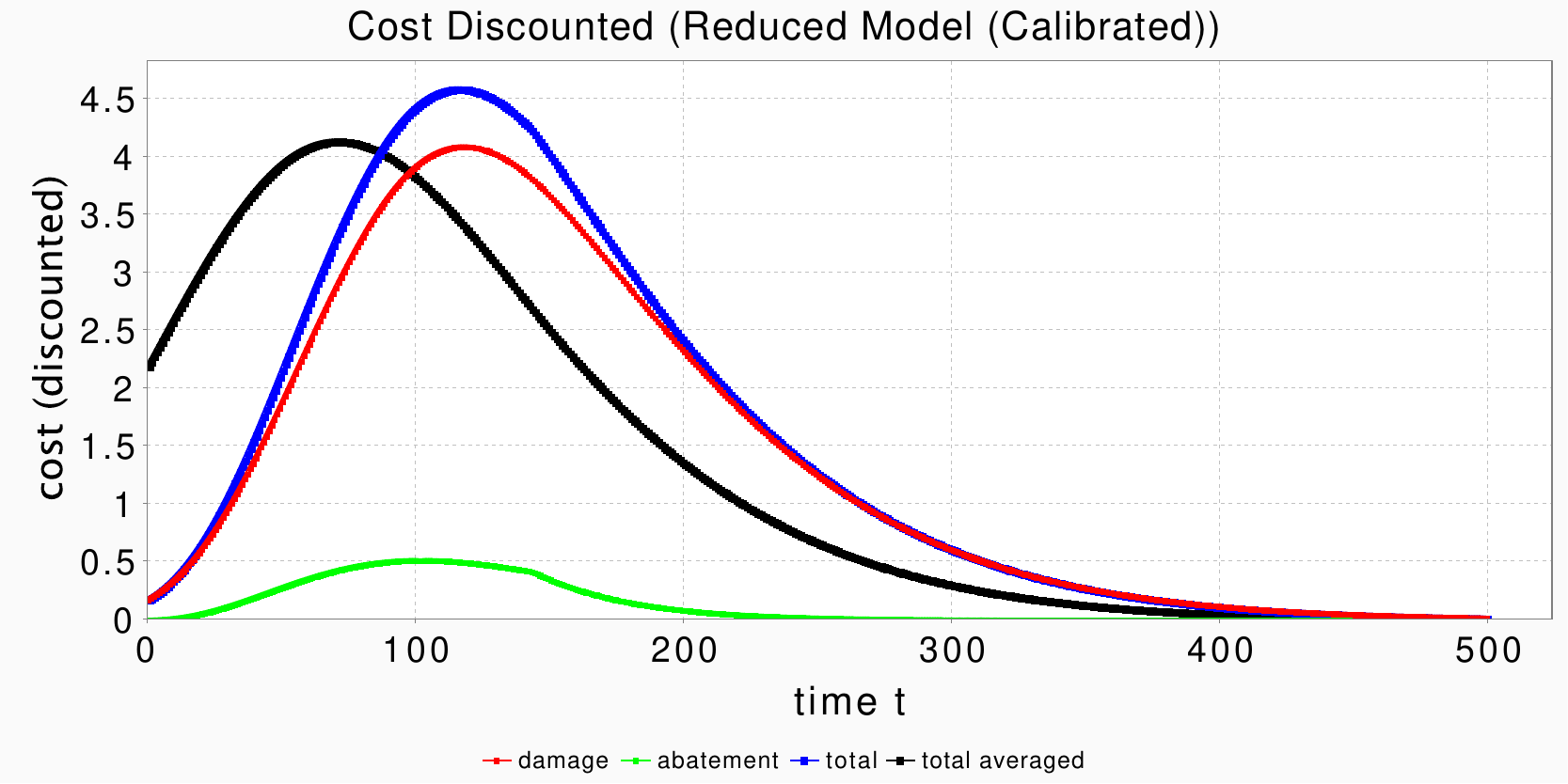}
    \caption{The cost in the calibrated model. Damage cost (red), abatement cost (green) and the sum of the two (blue). Black shows a forward running average over a time window of 100 years.}
    \label{fig:costOverTimeDiscounted-reduced}
\end{figure}

For the calibrated two-parameter model, \cref{fig:costOverTimeDiscounted-reduced} depicts the cost due to damage $C_{\mathrm{D}}(t)$ (red) and due to abatement $C_{\mathrm{A}}(t)$ (green). It turns out, that the costs and the total cost (blue) are not distributed evenly over time, even including discounting.

Abatement cost will decline once 100\% abatement has been reached due to technical progress. Damage cost will level once 100\% abatement has been reached. Finally, damage declines due to the diffusion of atmospheric carbon into the ocean and the resulting decreasing temperature.

In addition, we plot a 100-year running (discounted) average that approximately illustrates the generational cost (assuming a generation to live 100 years).

\begin{figure}[hp!]
    \centering
    \includegraphics[width=0.9\textwidth]{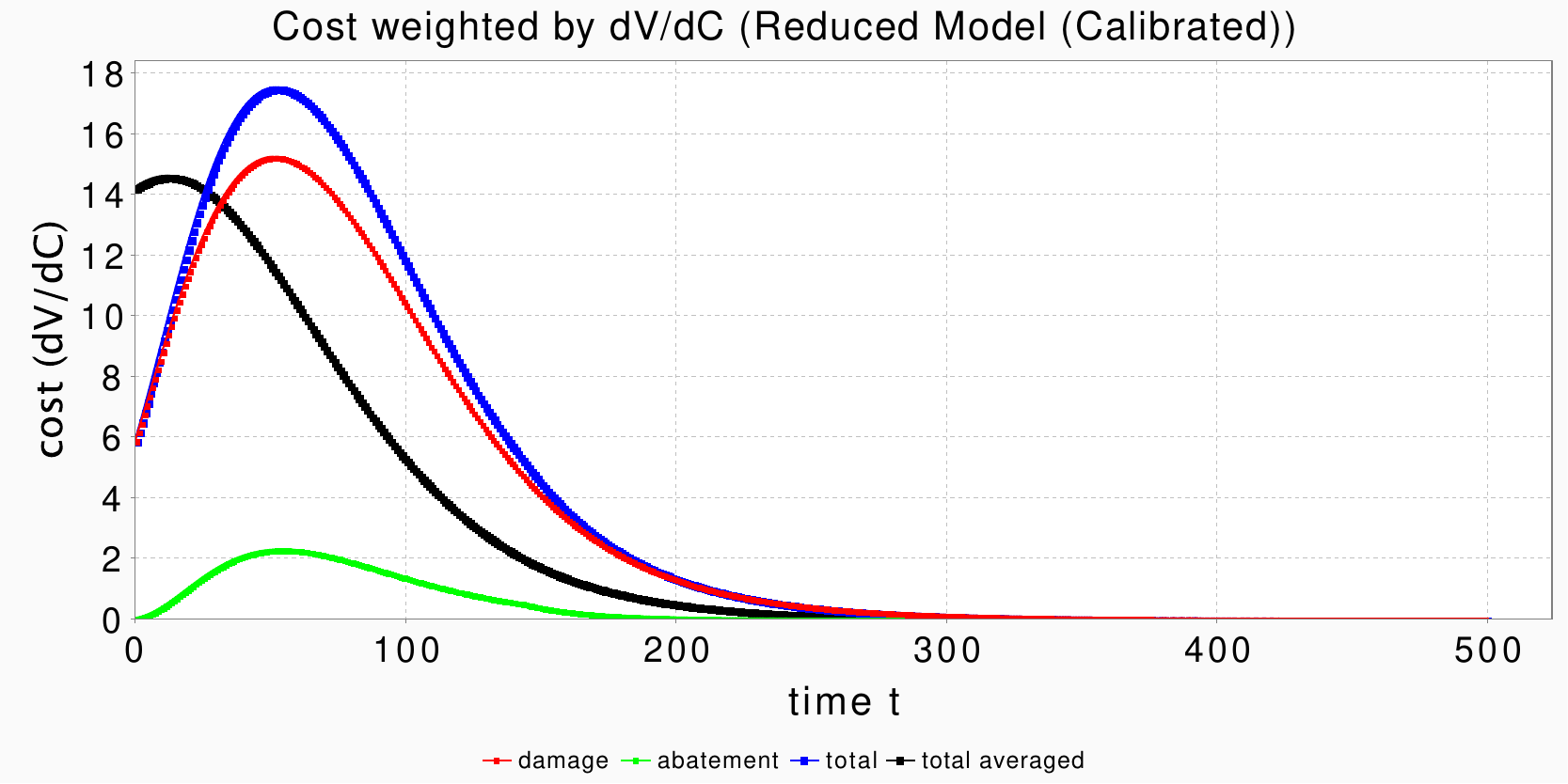}
    \caption{The cost weighted by $\frac{\mathrm{d}V}{\mathrm{d}C}$ in the calibrated model. Damage cost (red), abatement cost (green) and the sum of the two (blue). Black shows a forward running average over a time window of 100 years.}
    \label{fig:costOverTimeWeightedVC-reduced}
\end{figure}

\begin{figure}[hp]
    \centering
    \includegraphics[width=0.9\textwidth]{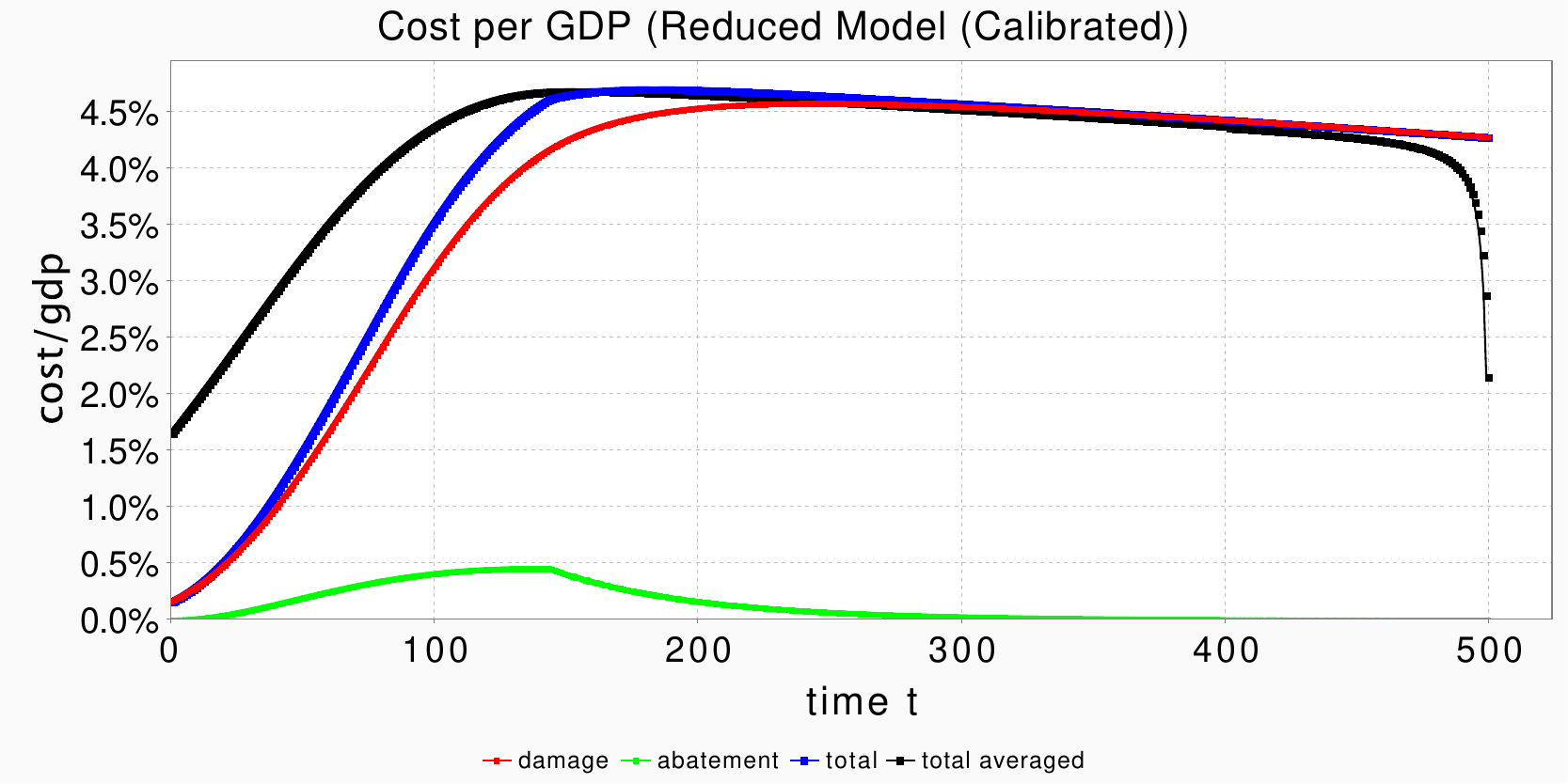}
    \caption{The cost as percentage of the GDP in the calibrated model. Damage cost (red), abatement cost (green) and the sum of the two (blue). Black shows a forward running average over a time window of 100 years.}
    \label{fig:costOverTimePerGDP-reduced}
\end{figure}

If we apply the cost-to-value weight $\frac{\partial V(t)}{\partial C(t)}$, we obtain the distribution shown in \cref{fig:costOverTimeWeightedVC-reduced}.
Alternatively, \cref{fig:costOverTimePerGDP-reduced} shows the cost as a percentage of the GDP, which is not discounted, as one could argue that each generation should measure the burden in relation to its GDP.


Regardless of which of the relations one considers, it is obvious that the abatement cost are comparably low and that large damages are accepted instead of further increasing the abatement cost.

\begin{figure}[hp]
    \centering
    \includegraphics[width=0.9\textwidth]{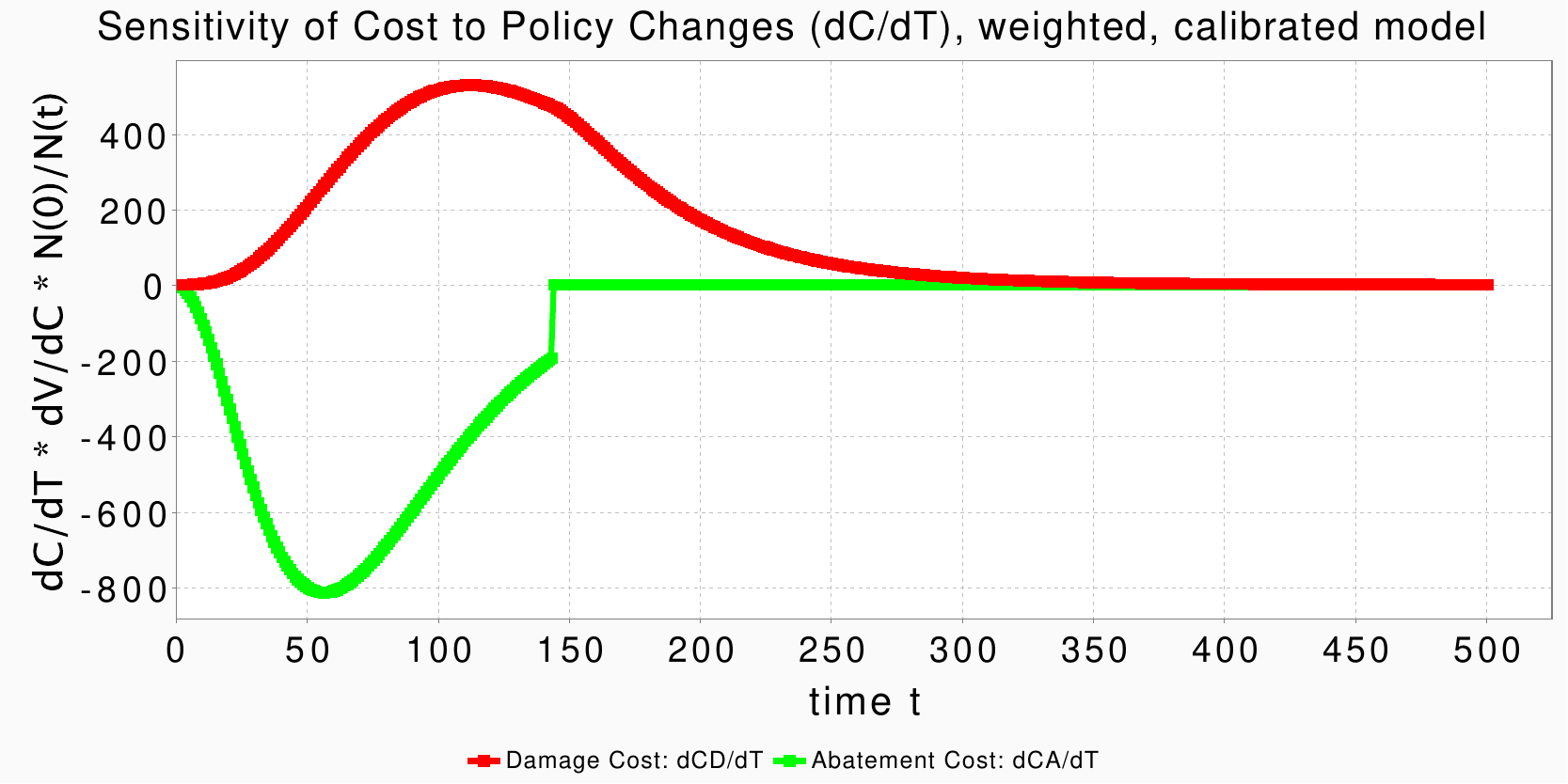}
    \caption{The sensitivity of cost to policy changes, i.e., the two integrands from expression~\eqref{eq:cost_objective}. The calibration of the model consist of matching the area under the red curve to the area over the green curve.}
    \label{fig:costSensitivity-reduced}
\end{figure}

It should be noted that this is a feature of the model calibration. The effect that we see is explained by Equation~\eqref{eq:dice:costAsCashflowStructure} and~\eqref{eq:cost_objective}. The model calibration does not level the temporal cost distribution, it levels the temporal cost changes, see Figure~\ref{fig:costSensitivity-reduced}. This motivates model extensions that improve the temporal equity of costs.

\smallskip

\begin{figure}[hp]
    \centering
    \includegraphics[width=0.9\textwidth]{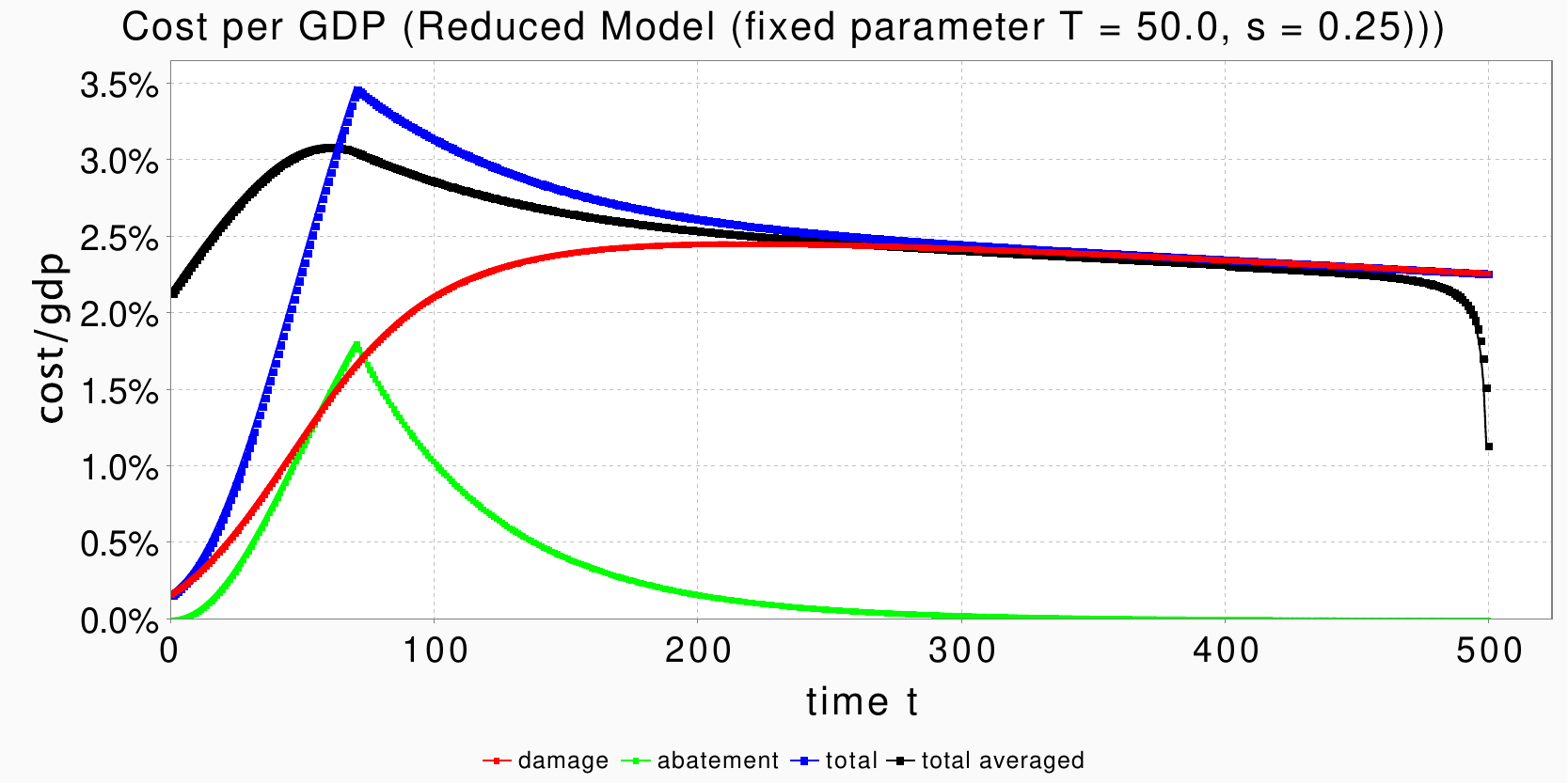}
    \caption{The cost relative to the GDP in a model with fixed parameters (non-calibrated). Damage cost (red), abatement cost (green) and the sum of the two (blue). Black shows a forward running average over a time window of 100 years.}
    \label{fig:costOverTimePerGDP-fixed}
\end{figure}

Moving to a not-calibrated abatement policy changes this result.
\cref{fig:costOverTimePerGDP-fixed} shows the costs in a model with the parameters set to $T^{\mu=1}=50.0$ $s=0.25$ (earlier abatement), whereas the calibrated model had $T^{\mu=1}=103.4$ $s=0.248$.

In the model with earlier abatement the cost structure appears to be improved with respect to intergenerational equity. However the discounted aggregated utility is not optimal. Considering the burden relative to the GDP, the model with earlier abatement appears to be a significant improvement.
\begin{figure}[hp]
    \centering
    \includegraphics[width=0.9\textwidth]{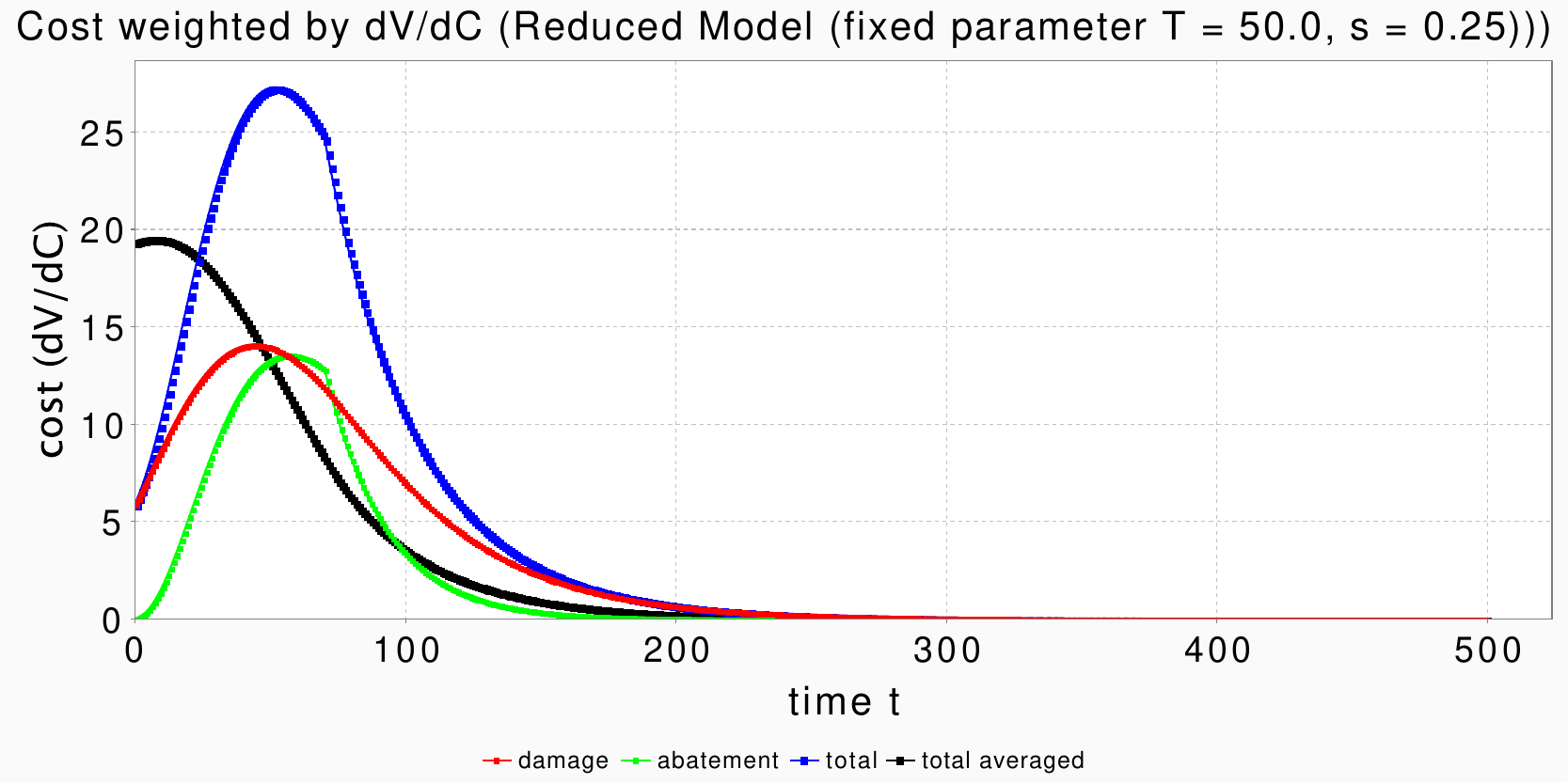}
    \caption{The cost weighted with $\frac{\partial V}{\partial C}$ in model with fixed parameters. Damage cost (red), abatement cost (green) and the sum of the two (blue). Black shows a forward running average over a time window of 100 years.}
    \label{fig:costOverTimeWeightedVC-fixed}
\end{figure}

\clearpage
\subsubsection{Modification of the Objective Function towards intergenerational Equity}
\label{sec:result:modelWithPNorm}

Comparing the generational average discounted cost per GDP (black) in \cref{fig:costOverTimePerGDP-fixed} to \cref{fig:costOverTimePerGDP-reduced} motivates a modification of the objective towards intergenerational equity. Taking the $p$-norm (see Section~\ref{sec:objectiveFunction:pnorm}) as objective function improves the situation.

\begin{figure}[hp]
    \centering
    \includegraphics[width=0.9\textwidth]{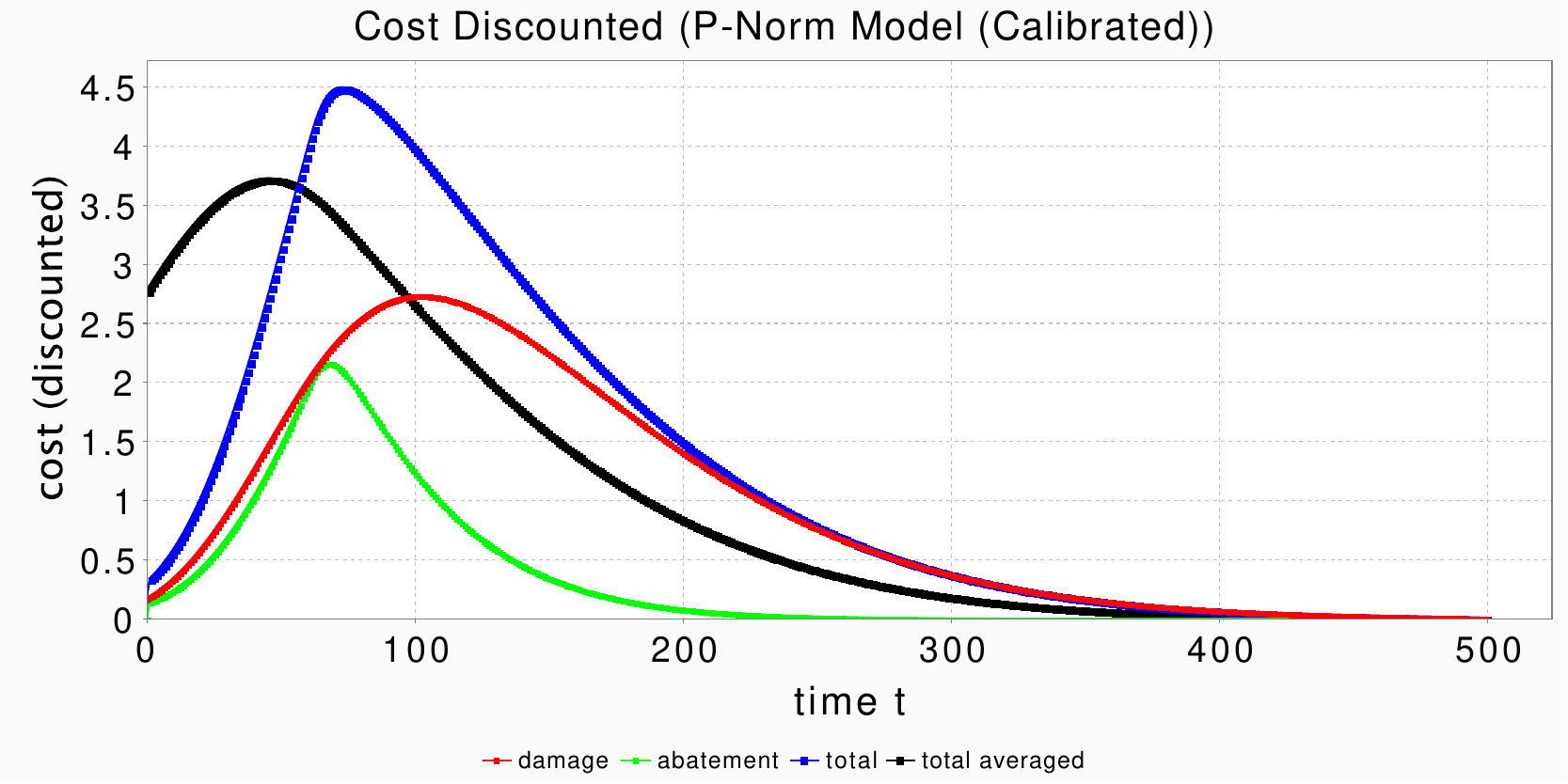}
    \caption{The cost in the calibrated model using the expectation of the $p=\frac{1}{4}$-norm with $p=1/4$ as objective function. Damage cost (red), abatement cost (green) and the sum of the two (blue). Black shows a forward running average over a time window of 100 years.}
    \label{fig:costOverTimeDiscounted-pnorm4}
\end{figure}

\begin{figure}[hp]
    \centering
    \includegraphics[width=0.9\textwidth]{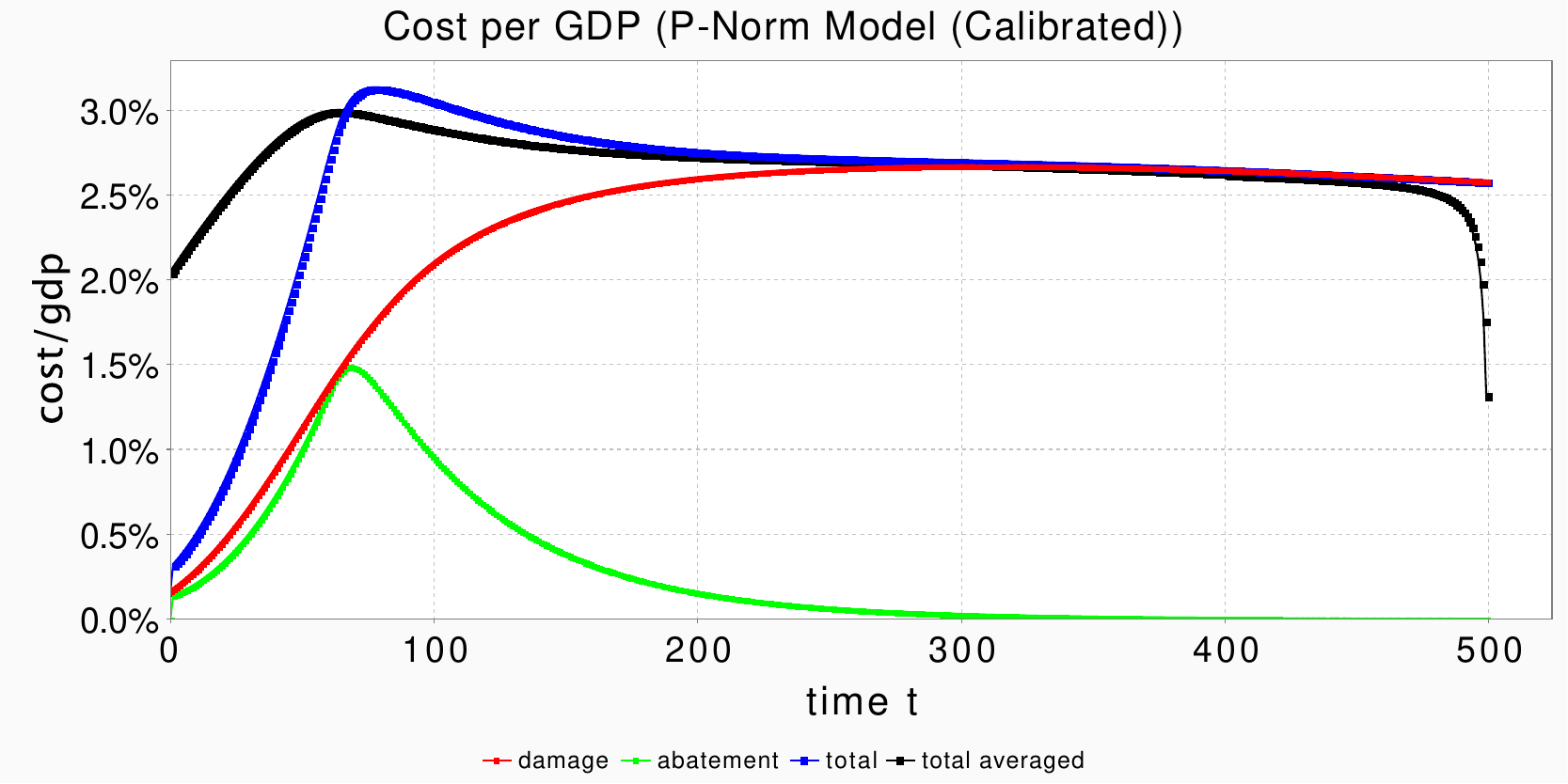}
    \caption{The cost relative to the GDP in calibrated model using the expectation of the $p=\frac{1}{4}$-norm with $p=1/4$ as objective function. Damage cost (red), abatement cost (green) and the sum of the two (blue). Black shows a forward running average over a time window of 100 years.}
    \label{fig:costOverTimePerGDP-pnorm4}
\end{figure}

\cref{fig:costOverTimeDiscounted-pnorm4} and~\ref{fig:costOverTimePerGDP-pnorm4} show the results for $p = 1/4$.
We see in \cref{fig:costOverTimePerGDP-pnorm4} that the $p$-norm indeed modifies the calibration result towards more intergenerational equity.

While the $p$-norm levels the distribution of cost, it is a maybe someone unmotivated modification of the model. We will see in Sections~\ref{sec:dice:results:fundingofabatementcost} and~\ref{sec:dice:results:nonlineardiscounting} that funding of abatement cost and non-linear discounting of damage cost will also improve the cost structure.

%
%

\clearpage
\subsection{Sensitivity of Damage Cost and Abatement Cost to Policy Changes}
\label{sensitivity_cost_policy}

In order to understand how our model extensions affect the model calibration, it is helpful to gain a deeper understanding of the sensitivity of damage cost and abatement cost to changes of the abatement policy.

The effect of a local change in the abatement policy $\mu(t)$ on the abatement cost is immediate in time $t$. Its effect to the damage cost comes with a delay. The question is, how does the change to damage cost distributed over time $s > t$, given that there is a change of the abatement policy in time $t$.

The distribution of costs over time establishes a dependency on the model inherent intertemporal weights attributed to costs.

To understand how abatement and damage are connected, we first analyse the sensitivity of the welfare function $V(t)$ with respect to cost $C(t)$.

\subsubsection{Value Change per Cost Change}

The objective function of the model is the value or welfare. As cost enter into value, two transformations take place:
\begin{itemize}
    \item First, the value is defined in terms of the utility function. Since the utility function is convex, there is some saturation, resulting in a over time strongly decaying weight applied to the cost.

    \item Second, the value is discounted. For the model of a constant positive discount rate, the discount factor constitutes an exponentially decaying weight applied to the cost.
\end{itemize}
The combined effect is contained in the sensitivity
\begin{equation*}
    \frac{\partial V(t)}{\partial C(t)}
\end{equation*}

\subsubsection[Sensitivity of Cost to Policy Changes]{Sensitivity of Cost $C(s)$ to Policy Changes $\mu(t)$}

We consider the dependency of the damage cost $C_{\mathrm{D}}(s)$ occurring at a later time $s$, with respect to a change in the abatement policy $\mu(t)$. We express this relative to the required abatement cost, i.e., we consider the function
\begin{equation}
    \label{eq:weigetedAbatementEffect}
    s \mapsto \frac{\frac{\partial V}{\partial C_{D}(s)} \frac{\partial C_{D}(s)}{\partial \mu(t)}}{\frac{\partial V}{\partial C_{A}(t)} \frac{\partial C_{A}(t)}{\partial \mu(t)}}
\end{equation}

The quantity $\frac{\partial C_{D}(s)}{\partial \mu(t)}$ is the change of the damage cost in time $s$ created by a change in the abatement policy in time $t$ ($t<s$). Costs are discounted and also weighted by marginal utility, and the total weight is given by $\frac{\partial V}{\partial C_{D}(s)}$ such that the product of the two is the impact on the welfare resulting from the change in damage by a change in the abatement policy.

The quantity $\frac{\partial C_{A}(t)}{\partial \mu(t)}$ is the change of the abatement cost in time $t$ created by a change in the abatement policy in time $t$. Likewise, this is weighted by $\frac{\partial V}{\partial C_{A}(t)}$ to calculate the impact on the welfare that results from the change in the abatement cost resulting from a change in the abatement policy.

\begin{figure}[hp]
    \centering
    \includegraphics[width=0.9\textwidth]{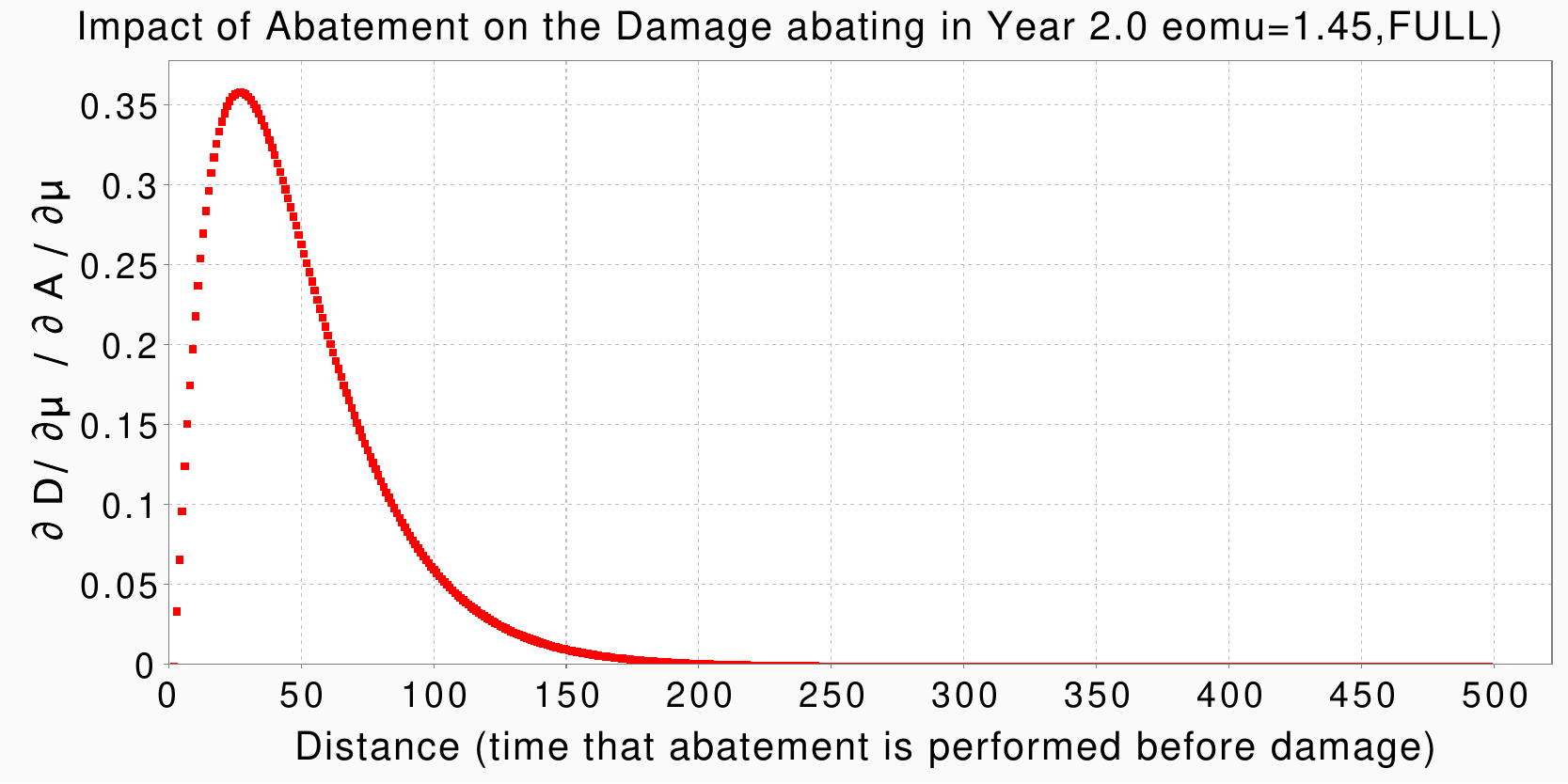}
    \caption{The value of $\frac{\partial C_{D}(s)}{\partial \mu(t)} / \frac{\partial C_{A}(t)}{\partial \mu(t)} \cdot \frac{\frac{\partial V}{\partial C(s)}}{\frac{\partial V}{\partial C(t)}}$ for $t = 2.0$: how much damage cost per abatement cost are seen in time $s \geq t$. Translated to the value function (including the effects of utility saturation and discounting.}
    \label{fig:sensitivityDamageToAbatement:full}
\end{figure}

The function~\eqref{eq:weigetedAbatementEffect} describes how much damage in time $s > t$ results from a change in the abatement policy relative to the corresponding change in the abatement cost in time $t$. Figures \ref{fig:sensitivityDamageToAbatement:full} show the function~\eqref{eq:weigetedAbatementEffect} at time $t = 2.0$. From this we may see that the model has some inherent latency: a policy change in $t=2$ (generating the abatement cost in $t=2$) will create the largest change to damage cost in at a much later time, given that we consider discounting and utility via the factor $\frac{\partial V}{\partial C_{D}(s)}$. This function is the one that is relevant to the model calibration, as the model optimises not the cost, but the discounted value (welfare). 

We see that a change in the abatement policy creates damages during the next 100 years, peaking in 20 years at approx. 18\% of the abatement cost.
The average of the integral under the curve, \ie the weighted average over all abatement times, should be 1, as the model is in its equilibrium state, where abatement cost and damage costs are balanced.

\smallskip

\begin{figure}[hp]
    \centering
    \includegraphics[width=0.9\textwidth]{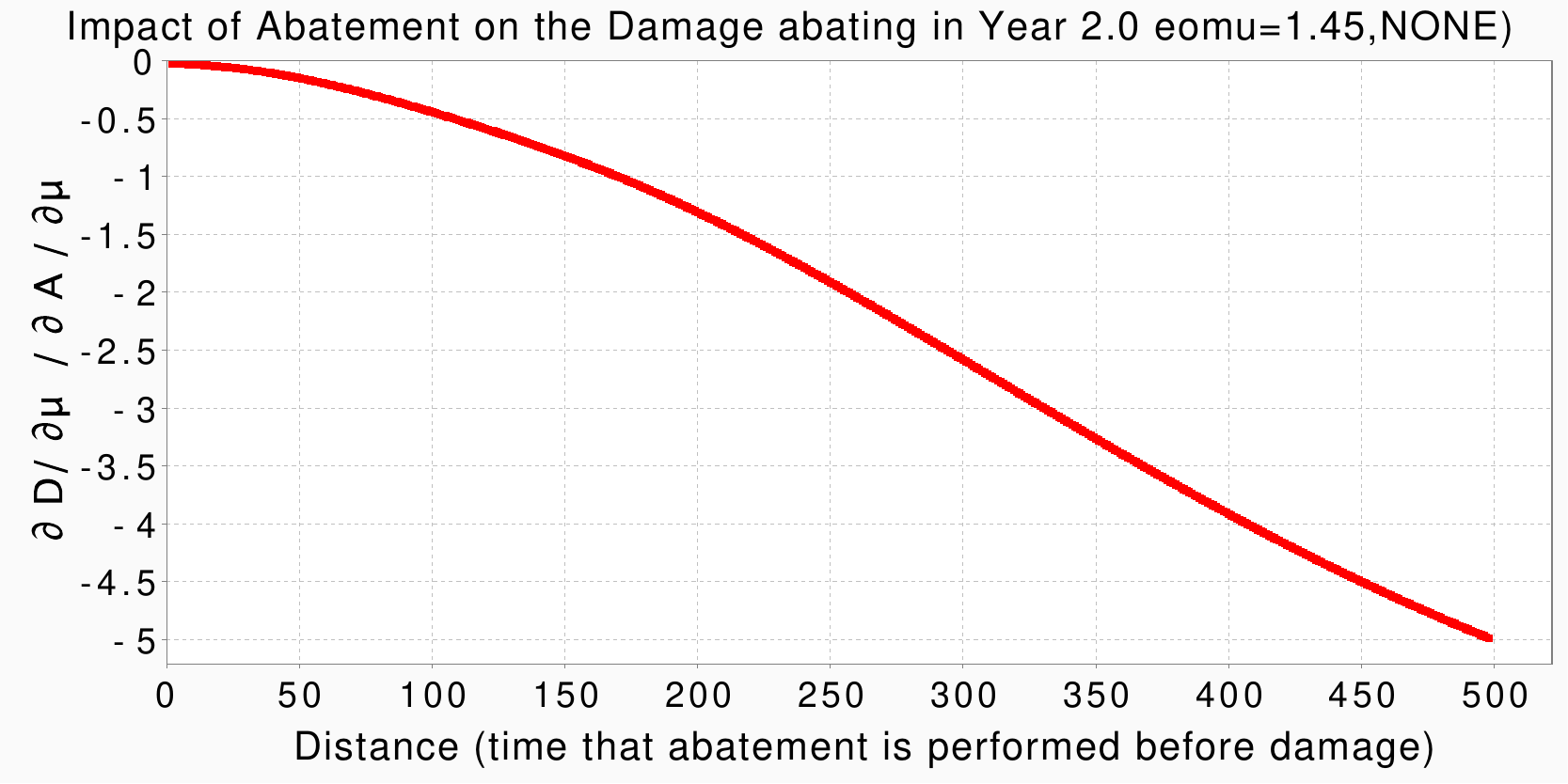}
    \caption{The value of $\frac{\partial C_{D}(s)}{\partial \mu(t)} / \frac{\partial C_{A}(t)}{\partial \mu(t)}$ for $t = 2.0$: how much damage cost per abatement cost are seen in time $s \geq t$.}
    \label{fig:sensitivityDamageToAbatement:none}
\end{figure}

We can now separate the role of the discount curve, by considering the pure relation of the cost, without any weight, that is
\begin{equation}
    \label{eq:dice:costfunction-noweight}
    s \mapsto \frac{\frac{\partial C_{D}(s)}{\partial \mu(t)}}{\frac{\partial C_{A}(t)}{\partial \mu(t)}}
\end{equation}
\cref{fig:sensitivityDamageToAbatement:none} shows the Function~\eqref{eq:dice:costfunction-noweight}. A change in the abatement policy creates a slowly growing (accumulating) damage. In our model calibration, this growth rate of damage increases for some 200 years starting from the time where the abatement policy was changed.

\cref{fig:sensitivityDamageToAbatement:full} results from \cref{fig:sensitivityDamageToAbatement:none} by applying the weight
\begin{equation*}
    \frac{\frac{\partial V}{\partial C(s)}}{\frac{\partial V}{\partial C(t)}} \text{.}
\end{equation*}
This weight comprises the discounting and the marginal utility, where the discounting is the more stronger effect.

\smallskip

\begin{figure}[hp]
    \centering
    \includegraphics[width=0.9\textwidth]{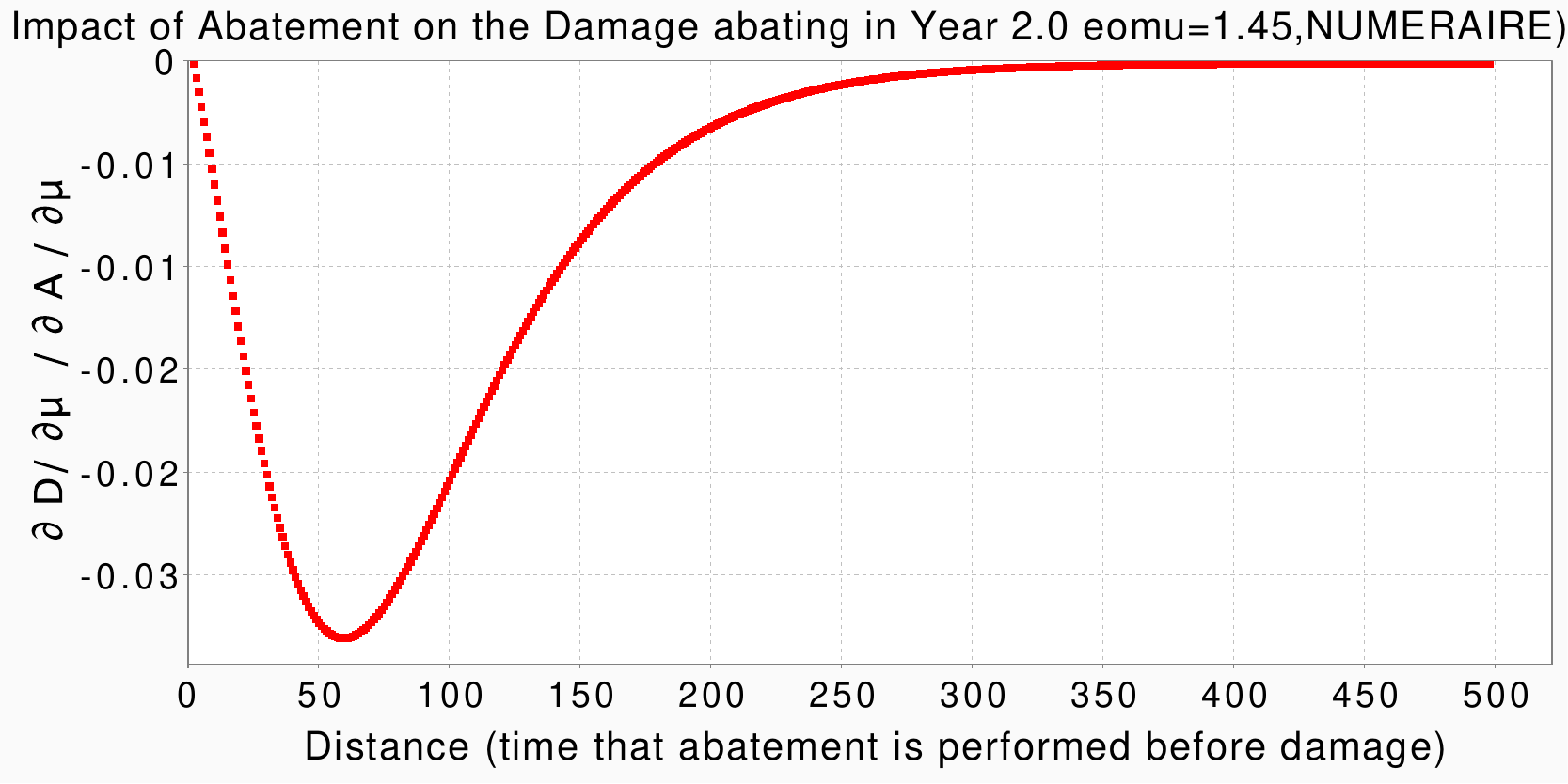}
    \caption{The value of $\frac{\partial C_{D}(s)}{\partial \mu(t)} / \frac{\partial C_{A}(t)}{\partial \mu(t)} \cdot \frac{N(t)}{N(s)}$ for $t = 2.0$: how much damage cost per abatement cost are seen in time $s \geq t$. Weighted with the discounting that is part of the value function.}
    \label{fig:sensitivityDamageToAbatement:NUMERAIRE}
\end{figure}

\begin{figure}[hp]
    \centering
    \includegraphics[width=0.9\textwidth]{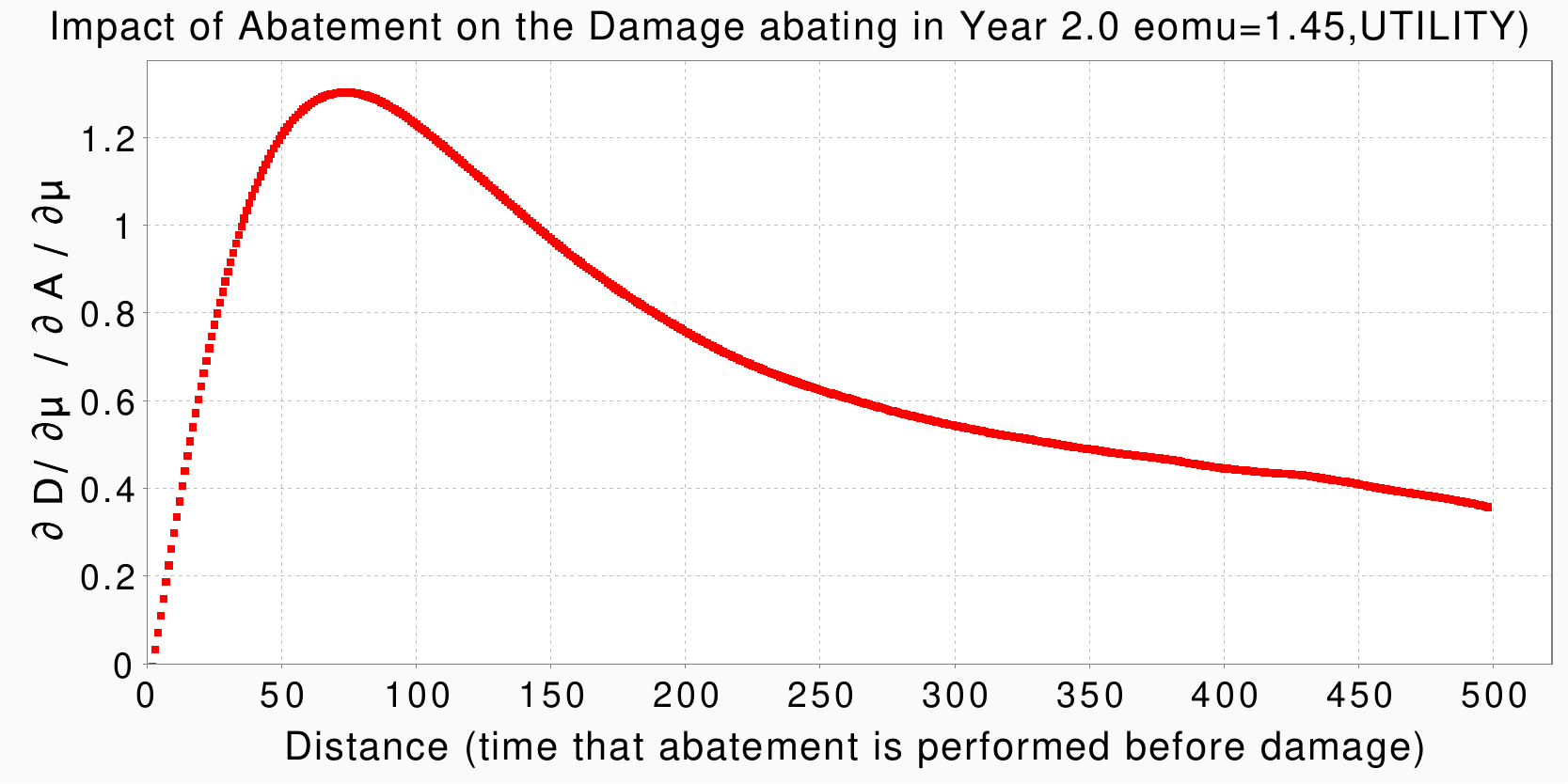}
    \caption{The value of $\frac{\partial C_{D}(s)}{\partial \mu(t)} / \frac{\partial C_{A}(t)}{\partial \mu(t)} \cdot \frac{\frac{\partial V}{\partial C(s)}}{\frac{\partial V}{\partial C(t)}} / \frac{N(t)}{N(s)}$ for $t = 2.0$: how much damage cost per abatement cost are seen in time $s \geq t$. Translated to the value function via the utility, but removing the discounting.}
    \label{fig:sensitivityDamageToAbatement:UTILITY}
\end{figure}

\cref{fig:sensitivityDamageToAbatement:NUMERAIRE} shows the same analysis, if the effect of discounting is added, but the effect of utility is not present, whereas \cref{fig:sensitivityDamageToAbatement:UTILITY} show the same analysis, if the effect of the saturation of the utility function is added, but the effect of the discounting is removed.

\begin{figure}
    \centering
    \includegraphics[width=0.9\textwidth]{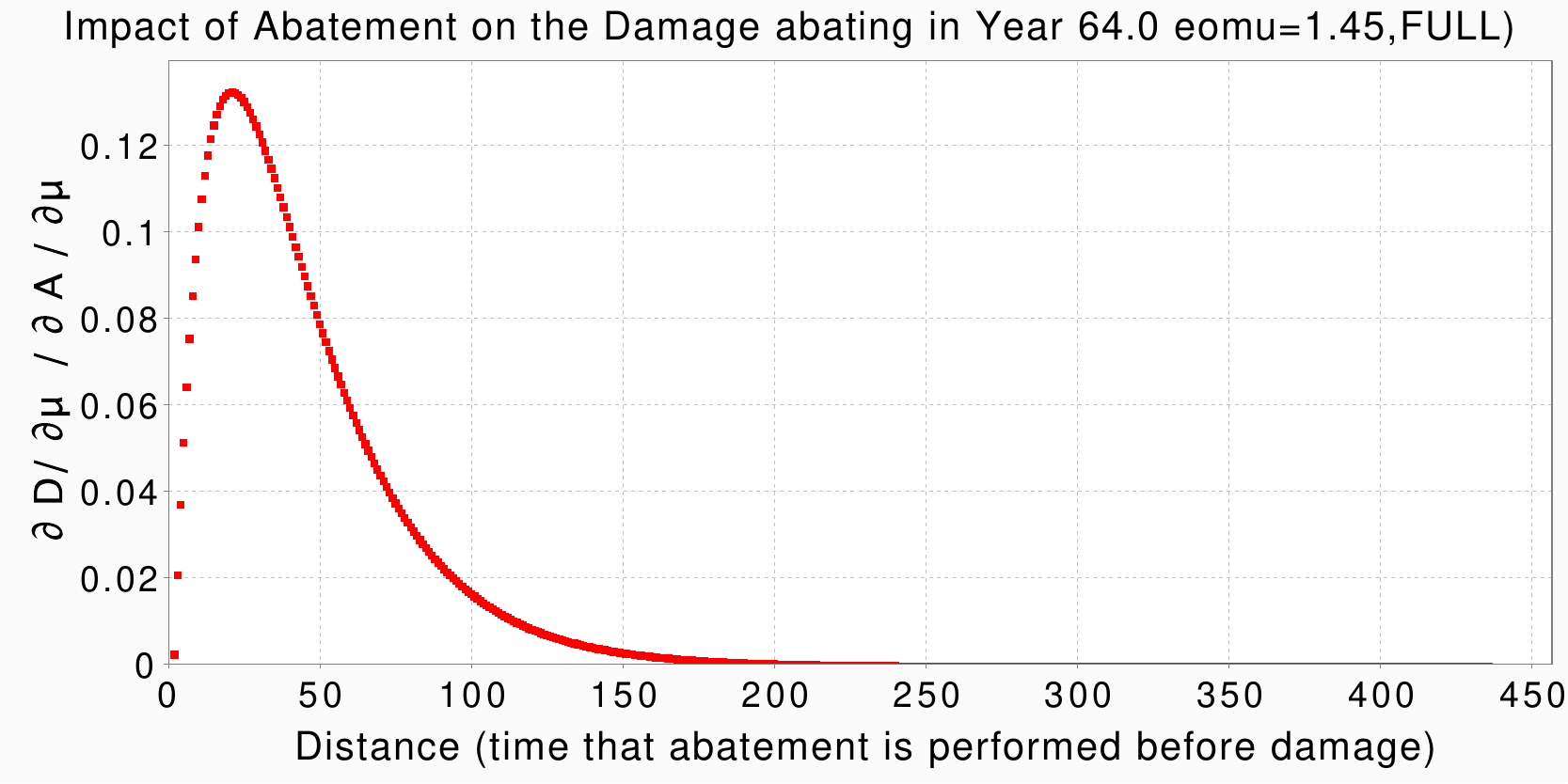}
    \caption{The value of $\frac{\partial C_{D}(s)}{\partial \mu(t)} / \frac{\partial C_{A}(t)}{\partial \mu(t)} \cdot \frac{\frac{\partial V}{\partial C(s)}}{\frac{\partial V}{\partial C(t)}}$ for $t = 64.0$: how much damage cost per abatement cost are seen in time $s \geq t$. Translated to the value function (including the effects of utility saturation and discounting.}
    \label{fig:sensitivityDamageToAbatement:full:64}
\end{figure}

For completeness, we state that the observed time-shift structure remains stable if the observation time $t$ is changed, but the effect is reduced. \cref{fig:sensitivityDamageToAbatement:full:64} depicts the damage cost per abatement cost, weighted with the cost-to-value sensitivity for a change in the abatement policy at $t=64$.

In summary: Abatement policy changes are more effective when done earlier, however, discounting is reducing the range, and hence the overall effectiveness to a time-horizon for 20 to 100 years. We see that discounting has a strong impact. The effect is twofold:
\begin{itemize}
    \item discounting and utility are localising-in-time the effect of a policy change

    \item discounting is reducing the sensitivity for future times, compare \cref{fig:sensitivityDamageToAbatement:full:64} to \cref{fig:sensitivityDamageToAbatement:full}.
\end{itemize}

The results of this section are important with respect to inter-temporal effort sharing, i.e., intergenerational equity. We see that changes in the abatement policy create an uneven distribution of damages over time. The damage peaks in approximately 20 year (\cref{fig:sensitivityDamageToAbatement:full}), but if discounting is removed and utility is considered only the peak resides in 70 years and the decay is slow (\cref{fig:sensitivityDamageToAbatement:UTILITY}). This shows that applying an interest rate significantly shifts the impact of abatement policy to future generations.

The previous numerical experiment showed the distribution of changes future damage cost for times $s \geq t$ for a change in the abatement policy at a single fixed time $t$. Optimising the full abatement policy $\mu$, the objective function is an the (discounted) sum of all those damage costs impacts plus the corresponding abatement cost.

\medskip
\clearpage

To get an understanding of the complete distribution of changes in the total cost (abatement cost and damage cost), we use our simplified abatement model, as this allows to show the sensitivity with respect to a single parameter, here $T^{\mu=1}$.

\begin{figure}
    \centering
    \includegraphics[width=0.9\textwidth]{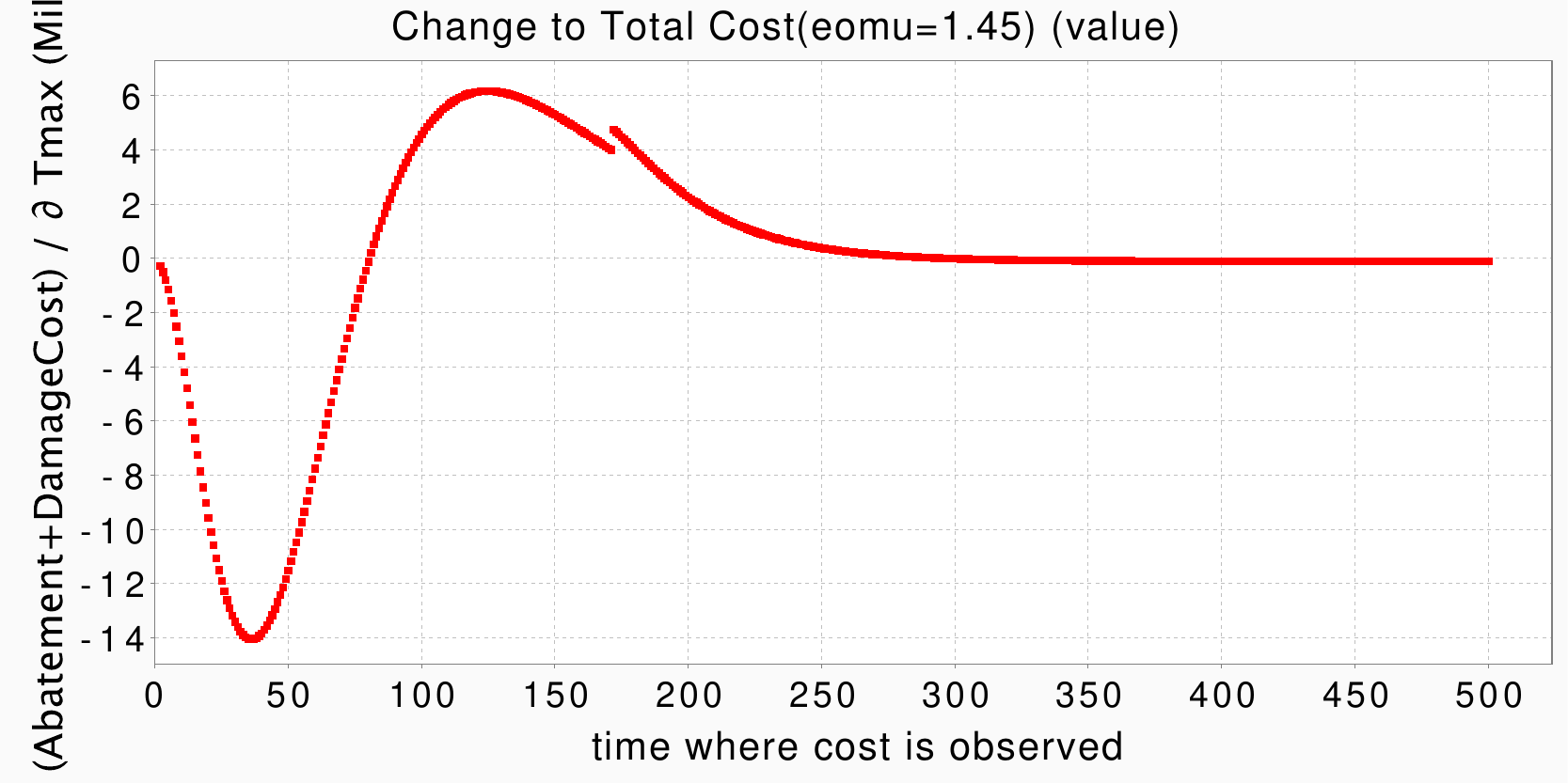}
    \caption{The value of $\frac{\partial C(t)}{\partial T^{\mu=1}} \cdot \frac{\partial V}{\partial C(t)}$ for different time $t$: what is the net change in (welfare-impact-weighted) cost (abatement cost + damage cost) seen in $t$, if the 100\% abatement is reached one year later.}
    \label{fig:SensitivityCostToAbatementTime}
\end{figure}

\cref{fig:SensitivityCostToAbatementTime} depicts now $\frac{\partial C(t)}{\partial T^{\mu=1}} \cdot \frac{\partial V}{\partial C(t)}$, where $C(t)$ is the sum of abatement cost and damage cost.
\begin{figure}
    \centering
    \includegraphics[width=0.9\textwidth]{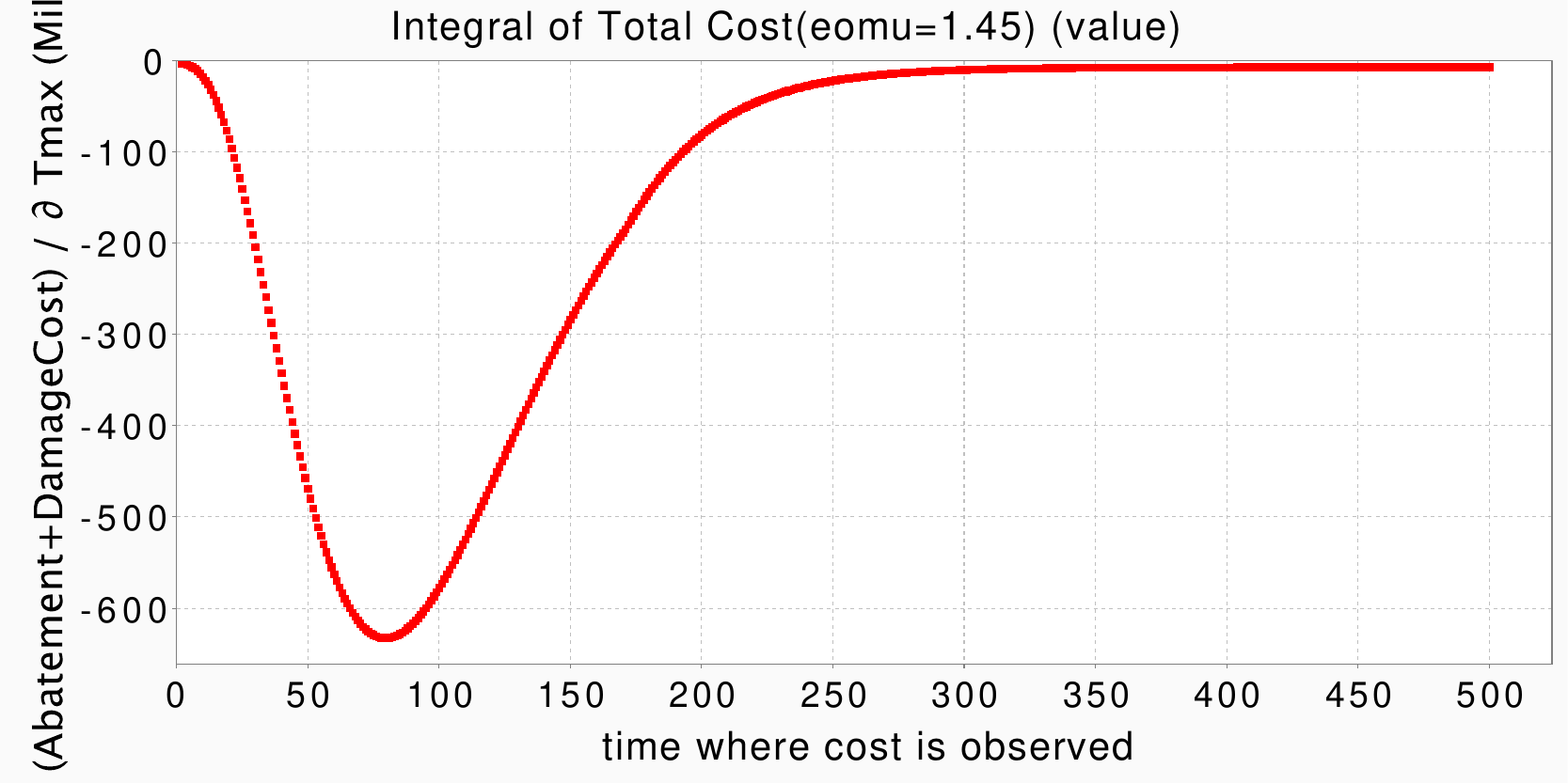}
    \caption{The integral of the function in \cref{fig:SensitivityCostToAbatementTime}.}
    \label{fig:SensitivityCostToAbatementTime:Integral}
\end{figure}
The first sharp decline is due to the decrease of abatement cost (if the policy is delaying), the positive part is due to the increase in damage cost. \cref{fig:SensitivityCostToAbatementTime:Integral} shows the integral of the function from \cref{fig:SensitivityCostToAbatementTime}. It goes to zero, because the model is in its equilibrium (calibrated). However, we also see that costs changes are zero only in a time-average sense, netting near-time benefits with far-time cost and vice-versa.

The function $\frac{\partial C(t)}{\partial T^{\mu=1}} \cdot \frac{\partial V}{\partial C(t)}$ depicted in \cref{fig:SensitivityCostToAbatementTime} is important for the understanding of the intergenerational inequality created by the calibration. The decline of cost due to abatement is located in a small temporal region (20 to 70 years), while the rise in cost is distributed across a large time frame (70 to 250 years). Hence, the model balances strong relieves for a few generations with burdens for many generations.

The uneven distribution of cost changes over time raises a question if we can equalise this distribution. As abatement cost and damage cost impact are offsetting (paying for abatement reduced damage), this motivates deferring abatement cost by allowing funding of abatement cost.

%
%

\clearpage
\subsection{Funding of Abatement Cost}
\label{sec:dice:results:fundingofabatementcost}

The starting point of our analysis is the function $\frac{\partial C(t)}{\partial T^{\mu=1}} \cdot \frac{\partial V}{\partial C(t)}$ depicted in the previous \cref{fig:SensitivityCostToAbatementTime}. It shows us essentially the cash-flow structure induced by change in abatement. It resembles a swap or a forward contract, where we pay in time $t_{1}$ (additional abatement cost) and in return receive in time $t_{2}$ (lower damage cost). In the model we considered for \cref{fig:SensitivityCostToAbatementTime} the maturity is approximately 100 years.

We now consider time-shifting the abatement cost by introducing a loan. We allow that abatement cost are funded over a predefined period using our model modification from Section~\ref{sec:diceExtension:fundingOfAbatement}.
\begin{figure}
    \centering
    \includegraphics[width=0.9\textwidth]{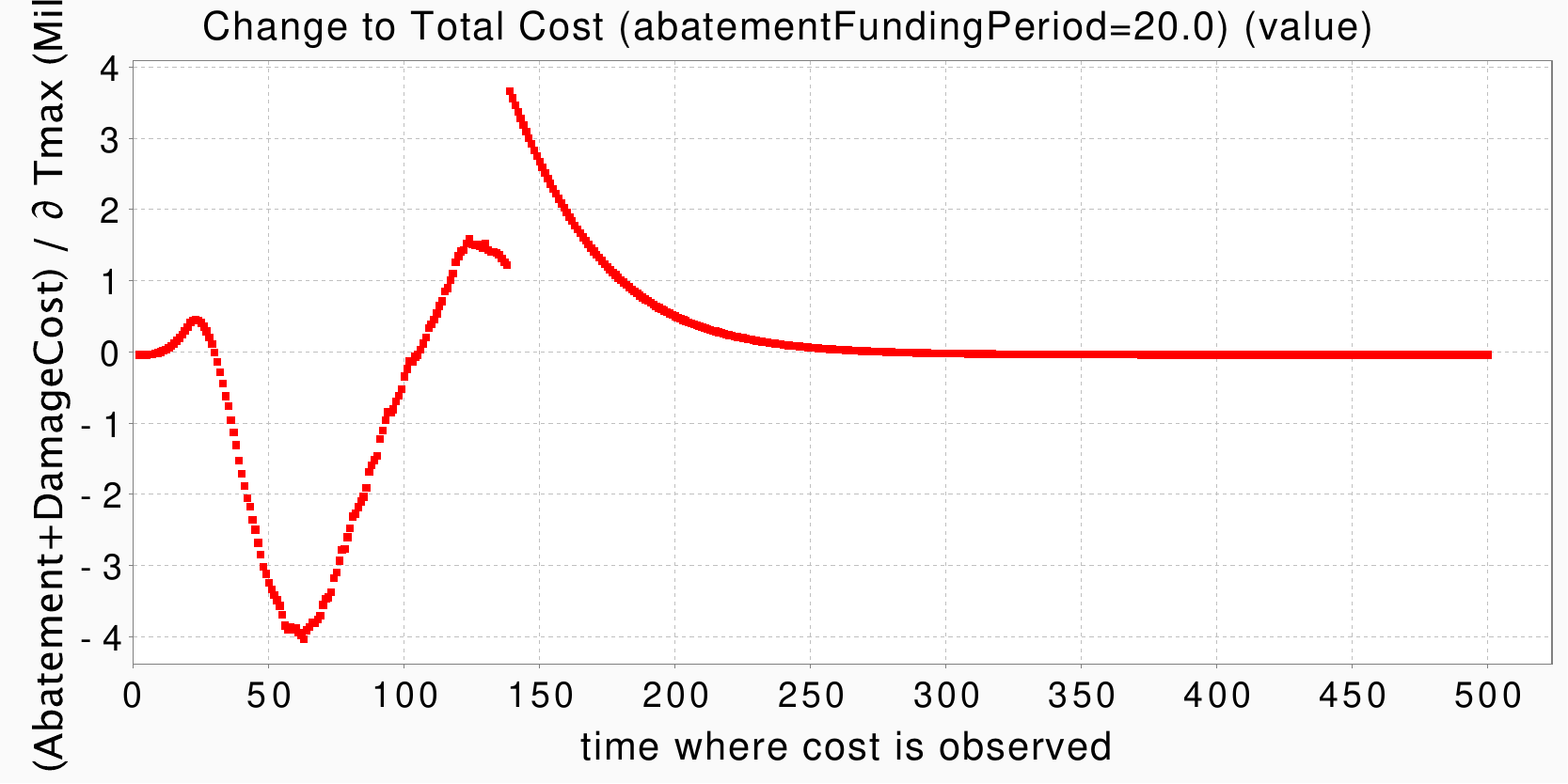}
    \caption{The value of $\frac{\partial C(t)}{\partial T^{\mu=1}} \cdot \frac{\partial V}{\partial C(t)}$ for different time $t$: what is the net change in (welfare-impact-weighted) cost (abatement cost + damage cost) seen in $t$, if the 100\% abatement is reached one year later.}
    \label{fig:SensitivityCostToAbatementTime:Funding}
\end{figure}
\begin{figure}
    \centering
    \includegraphics[width=0.9\textwidth]{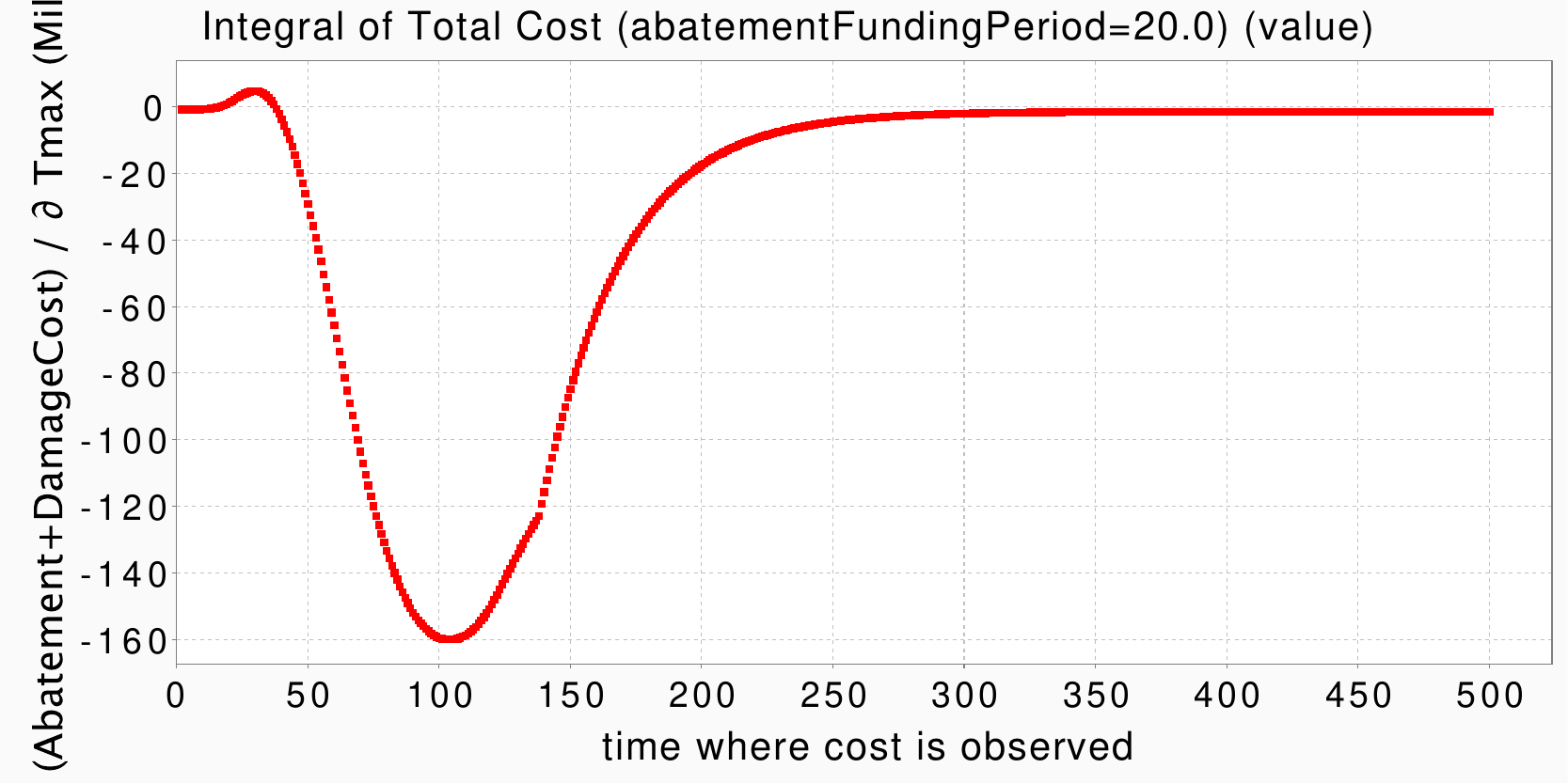}
    \caption{The integral of the function in \cref{fig:SensitivityCostToAbatementTime:Funding}.}
    \label{fig:SensitivityCostToAbatementTime:Funding:Integral}
\end{figure}

The \cref{fig:SensitivityCostToAbatementTime:Funding,fig:SensitivityCostToAbatementTime:Funding:Integral} plot the function and the integral for a funding period of 20 years. This shows that a funding of the abatement cost will improve the distribution of the cost structure (the amplitude is reduced). A benefit in reduced damage cost corresponds to the re-payment of abatement cost.
Compare the \cref{fig:SensitivityCostToAbatementTime:Funding,fig:SensitivityCostToAbatementTime:Funding:Integral} to \cref{fig:SensitivityCostToAbatementTime,fig:SensitivityCostToAbatementTime:Integral}.

The funding of abatement cost includes the application of the market interest rate (the discount rate). So with respect to the net-present-value of the cost, there is no change. The change in the model results from the change in the utility.

\begin{figure}
    \centering
    \includegraphics[width=0.9\textwidth]{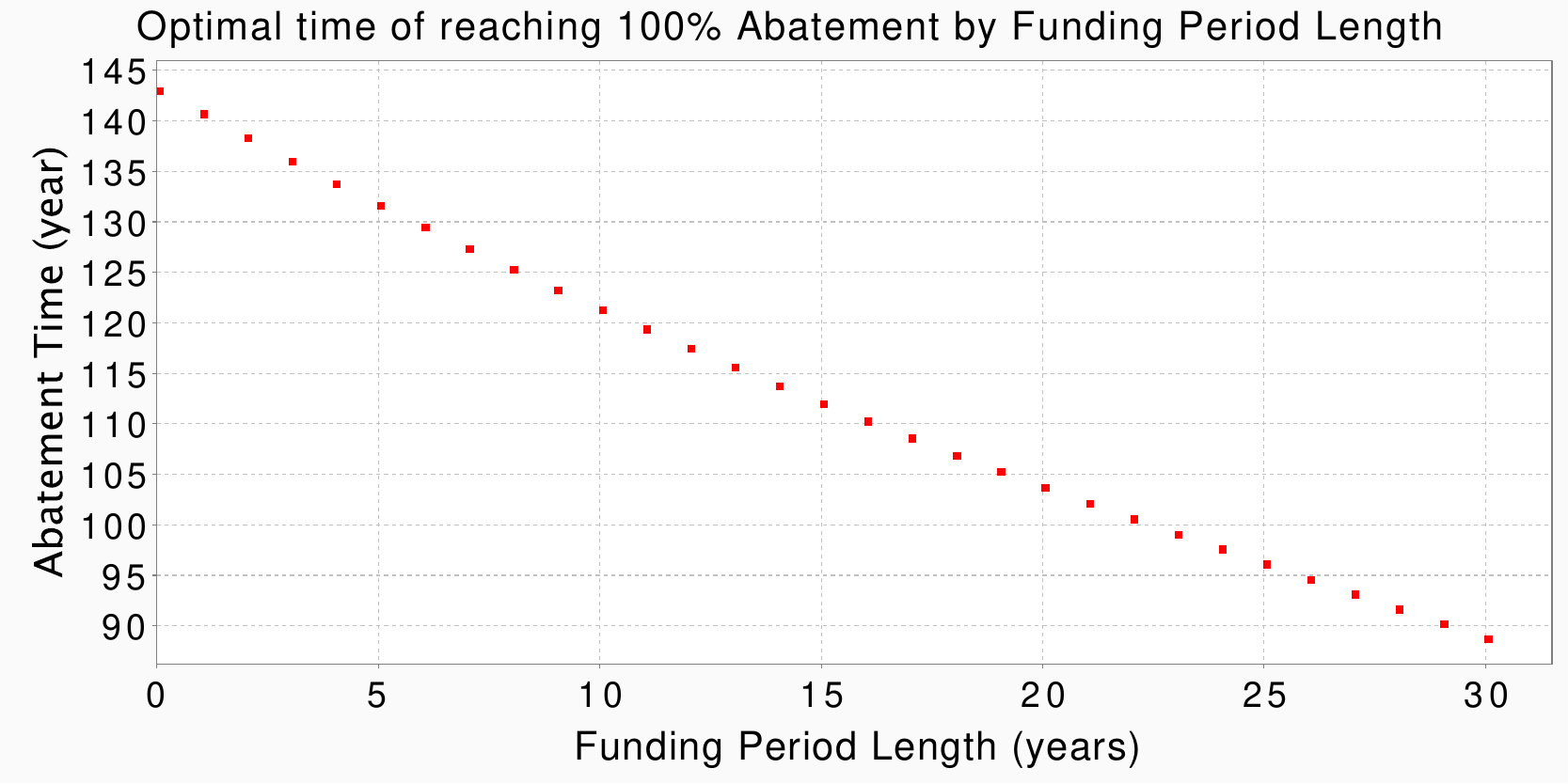}
    \caption{The calibrated (optimal) parameter $T^\mu$ for models using different funding periods.}    \label{fig:ClimateModelExperimentFunding:TimeOfMaxAbatementByAbatementFundingPeriod}
\end{figure}

In this funded model, we now observe that there is a much stronger preference for early abatement. \cref{fig:ClimateModelExperimentFunding:TimeOfMaxAbatementByAbatementFundingPeriod} shows the dependency of the calibrated $T^{\mu=1}$ of our simple (non-stochastic) abatement model on the funding length of the funding period.
A 20 year funding period for the abatement cost would induce a preference of performing abatement roughly 40 years earlier.

\begin{figure}[hp]
    \centering
    \includegraphics[width=0.9\textwidth]{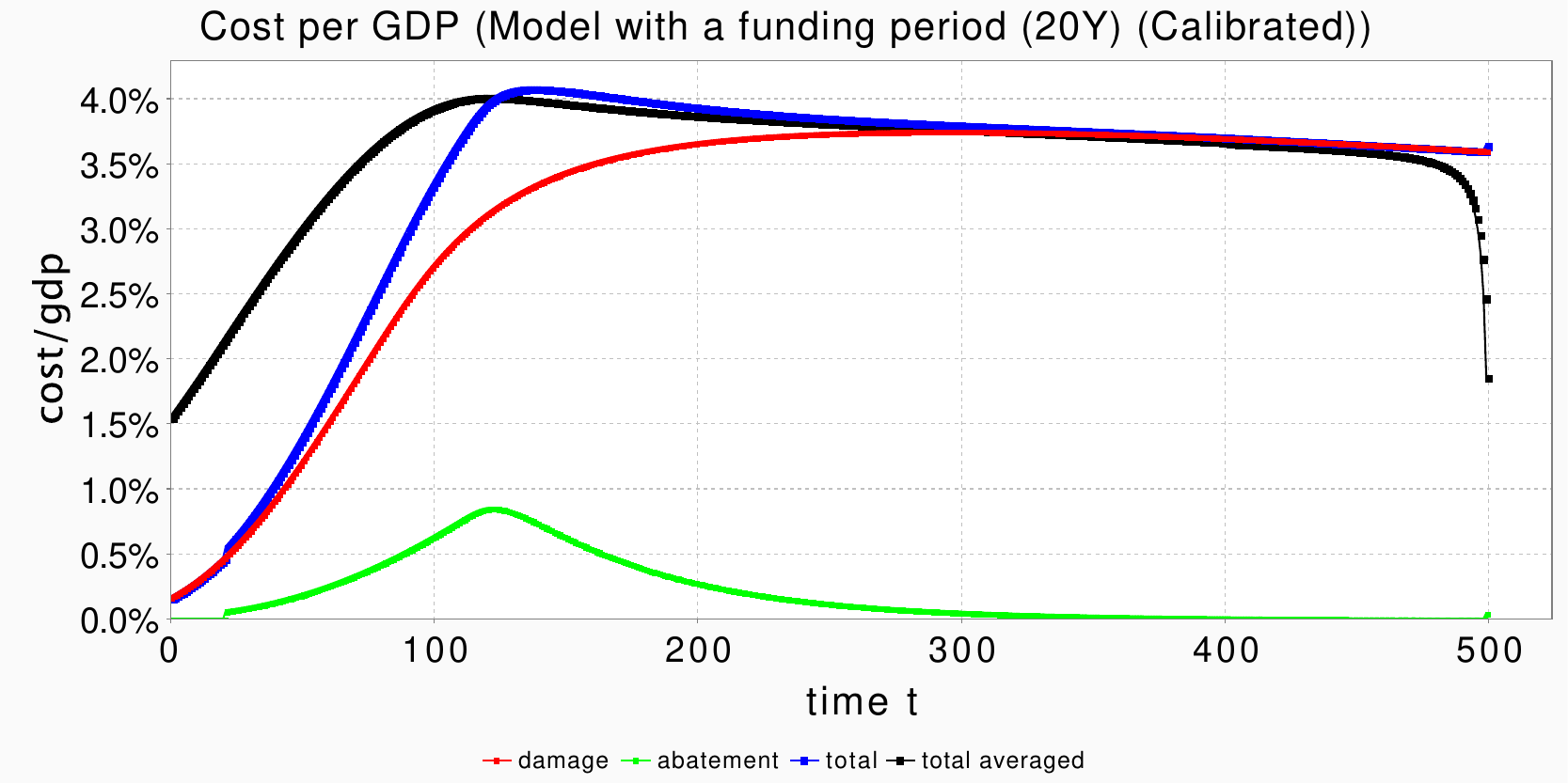}
    \caption{The cost per GDP in the calibrated model using a model with a 20 year funding period for the abatement. Damage cost (red), abatement cost (green) and the sum of the two (blue). Black shows a forward running average over a time window of 100 years.}
    \label{fig:costOverTimePerGDP-funding20}
\end{figure}

Introducing a funding period improves the intergenerational equity in the sense that the cost-per-GDP has a more even distribution.
\cref{fig:costOverTimePerGDP-funding20} shows the cost per GDP for a calibrated model where abatement cost were funded over a period of 20 years. Compare this to \cref{fig:costOverTimePerGDP-reduced}. The effect is not a strong as in \cref{fig:costOverTimePerGDP-fixed}, but it is significant and, it is the calibrated equilibrium state of this model.

%
%

\clearpage
\subsection{Non-Linear Financing Cost / Non-Linear Discounting}
\label{sec:dice:results:nonlineardiscounting}

Introducing a non-linear discounting, i.e., a discount factor that depends on the size of the value, also regulates the intergenerational equity of cost as large cost are more penalised.

We use the model modifications described in Section~\ref{sec:nonlinear_discounting}.

\subsubsection{Non-Linear Discounting Cost as a Function of Damage Cost}

\begin{figure}
    \centering
    \includegraphics[width=0.9\textwidth]{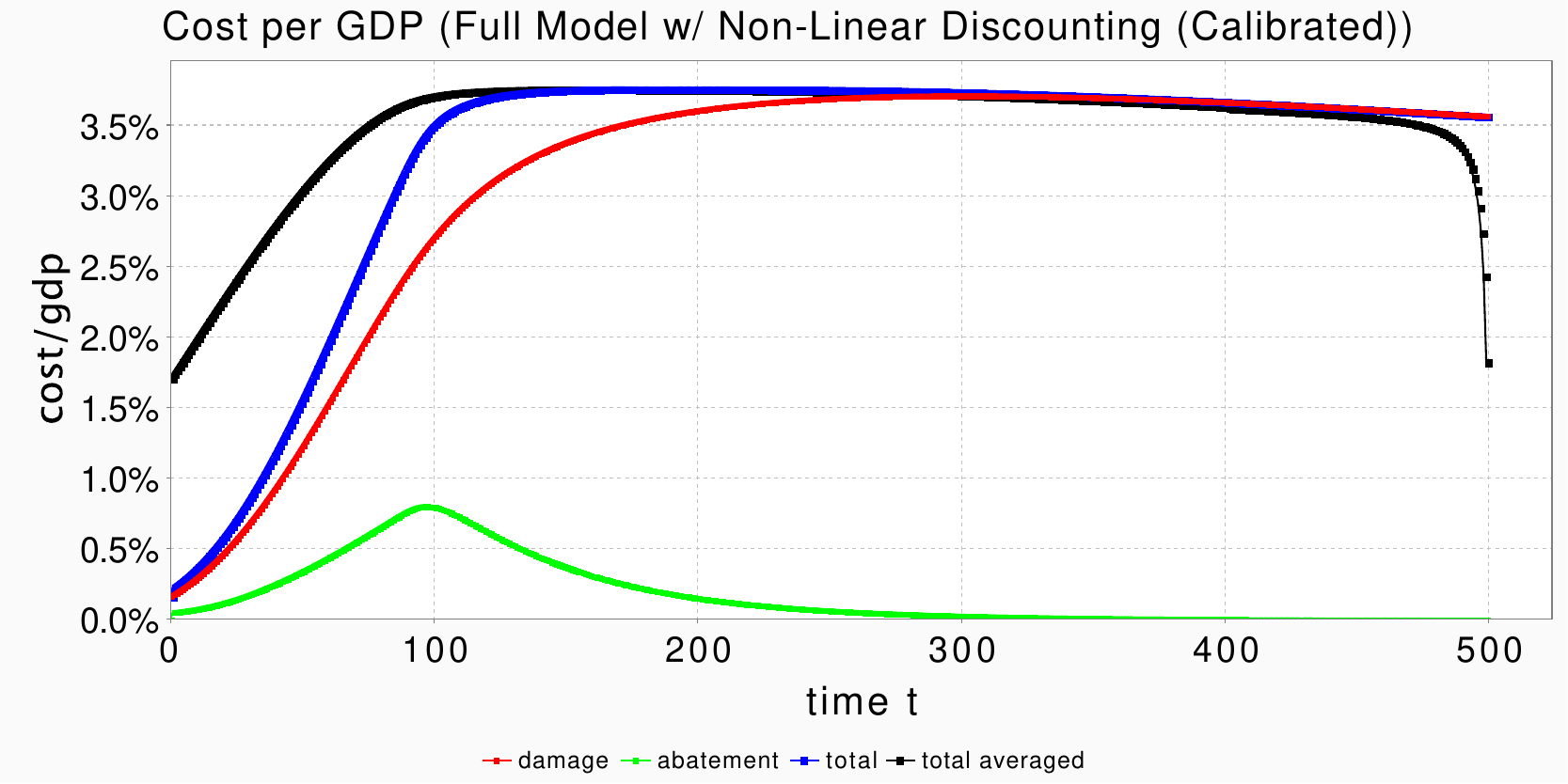}
    \caption{The temporal distribution of cost measured per GDP when using a model with a (strong) non-linear discounting applied to damage cost.}    \label{fig:CostOverTimePerGDP-non-linear-discounting}
\end{figure}

\cref{fig:CostOverTimePerGDP-non-linear-discounting} shows the cost-per-GDP in a calibrated model with non-linear discounting, where the default compensation factor is a function of the numéraire-relative damage cost. Compare this figure to \cref{fig:costOverTimePerGDP-reduced}. With a non-linear discounting, the policy optimisation will further levels the cost structure leading to more intergenerational equity. The generational moving average of the cost per GDP starts at 1.7\%. The total cost per GDP remains below 3. 7\%, while the unconstrained model shows up to 4.5\%.

The approach is also very transparent as one may explicitly attribute additional cost to large damages.

\subsubsection{Non-Linear Discounting Cost as a Function of Cost per GDP}

\begin{figure}
    \centering
    \includegraphics[width=0.9\textwidth]{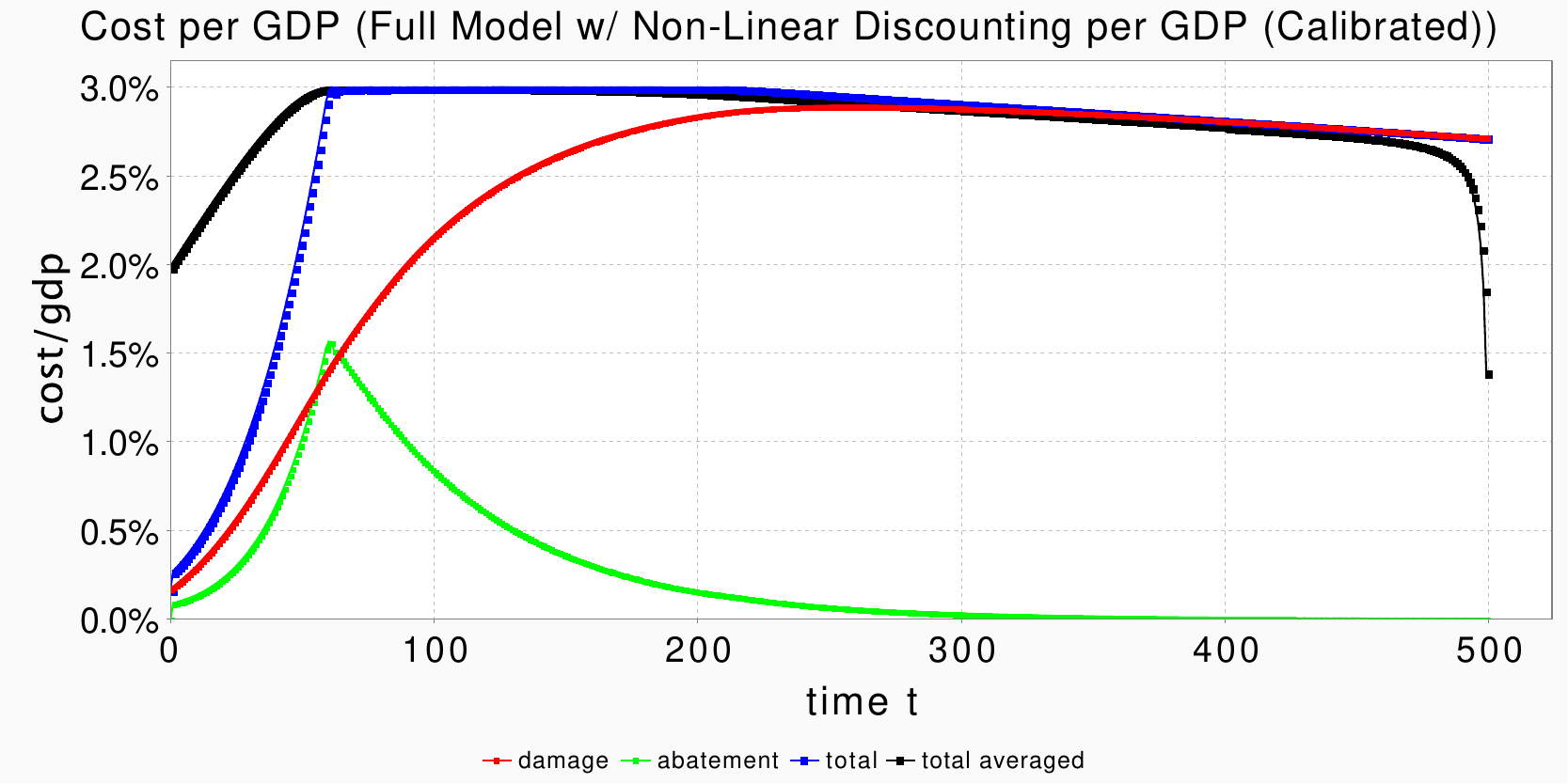}
    \caption{The temporal distribution of cost measured per GDP when using a model with a (strong) non-linear discounting applied to total cost per GDP.}  \label{fig:CostOverTimePerGDP-non-linear-discounting-per-gdp}
\end{figure}

Alternatively, we may define the default compensation factor as a function of the total cost per GDP. This allows to explicitly penalise pathways with large costs per GDP. For example, setting a boundary at 3.0 \% results in \cref{fig:CostOverTimePerGDP-non-linear-discounting-per-gdp}.
This experiment shows that a non-linear discounting (that is, a penalising of large cost) can be used to limit the intergenerational inequality, that is, improve the intergenerational equity. The method is particularly appealing because it directly addresses equalising the cost over time. In addition, it may be economically justified by the observation that large projects or large damages usually come with a kind of amplification of the cost \cite{hallegatte_why_2007,shane_construction_2009}.

\medskip

Our model modifications, the funding period and the non-linear discounting, can be applied simultaneously: the funding of abatement aligns abatement and damage in time, avoiding to separate benefits and costs in time. The non-linear discounting (of the funding) favours an even distribution of small cost over large singular costs in the future.

%
%

\clearpage
\subsection{Stochastic Interest Rates: Interest Rate Level and Interest Rate Risk}

We introduce stochasticity to the model that allows time-dependent stochastic interest rates.

First, this will affect the discount rate only. As noted in Section~\ref{sec:introducingStochInterestRates}, this does not alter the model, since applying the expectation or even a risk measure factors out to a deterministic discount factor (with time-dependent interest rate).

Within this model, we investigate the interplay of the choice of the objective function (expectation or risk measure), the interest rate level and the interest rate volatility.

Note that in this section, the objective function is the classical sum of the discounted utilities, which is a random variable, followed by the application of either an expectation or an expected shortfall (\cref{eq:expectedShortfall}) operator.

\subsubsection{Impact of Interest Level}

\begin{figure}[hp]
    \centering
    \includegraphics[width=0.9\textwidth]{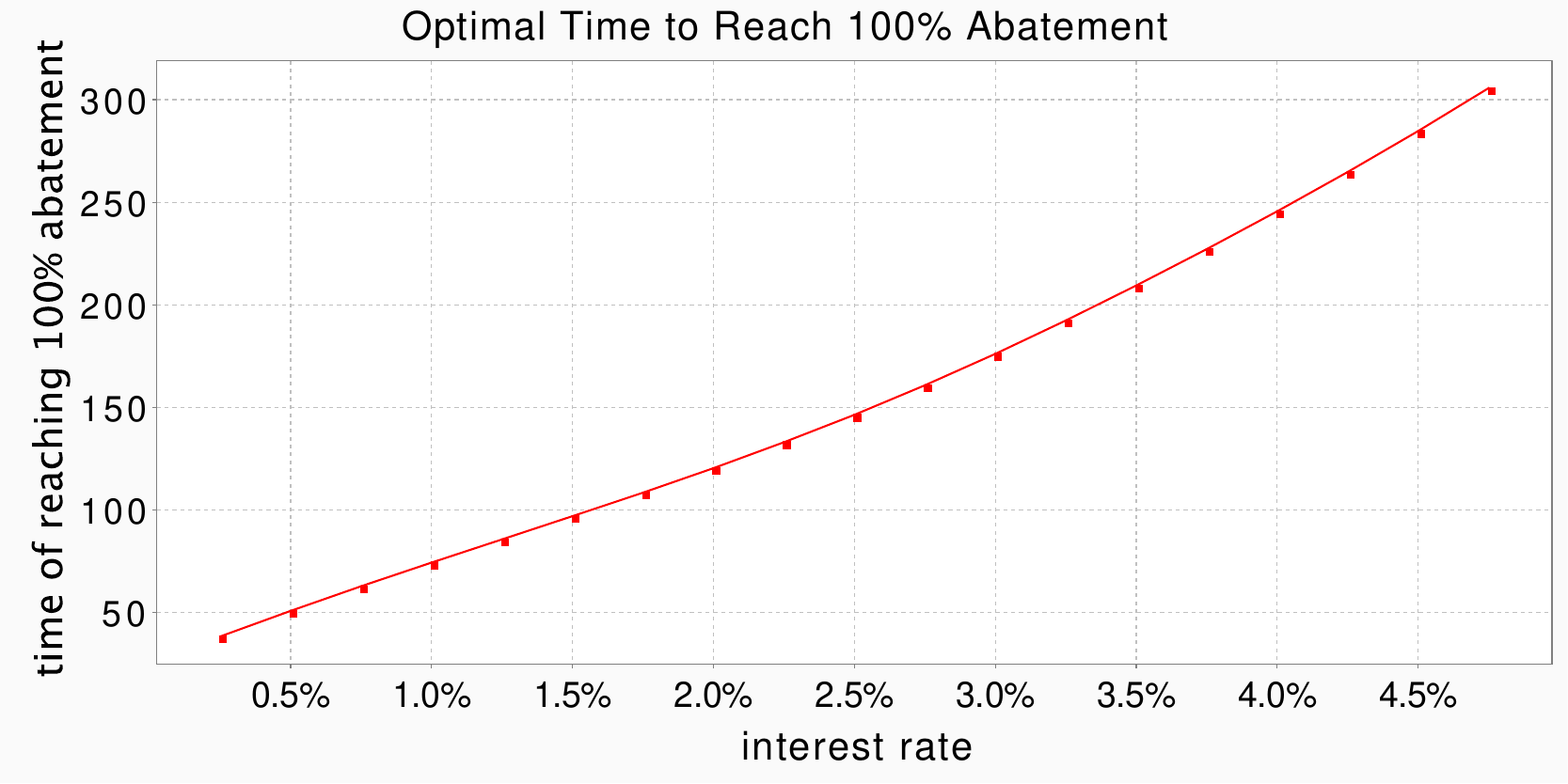}
    \caption{The dependency of $T^{\mu=1}$ on a the interest rate level (for calibrated constant savings rate $s_{0}$.}
    \label{fig:maxAbateTimeByRateLevel}
\end{figure}

To analyse the impact of different model parameters, we calculate the \textit{time of reaching maximum abatement}
\begin{equation*}
    T^{\mu=1} \ = \ \left( 1.0 - \mu_{0} \right) / a_{0} \text{.}
\end{equation*}
\cref{fig:maxAbateTimeByRateLevel} depicts the dependency of $T^{\mu=1}$ on a constant discount rate $r$ and states the strong dependency on that parameter. The parameter $s_{0}$ is calibrated for every $a_{0}$.

We reproduce the known effect that \emph{the lower the interest rate level, the earlier one should achieve 100\% abatement.}\footnote{The figures of this subsection can be reproduced by class \texttt{ClimateModelExperimentOptimalAbatementByInterestRate}.}

\clearpage

\subsubsection{Impact of Risk}

Simulating a model with a time-dependent stochastic discount rate $(t,\omega) \mapsto r(t,\omega)$, we analyse the optimal abatement model (and savings rate model), that produces the smallest welfare risk. Our objective function is the expected shortfall at level $\alpha$ \cref{eq:expectedShortfall} of the welfare.\footnote{The figures of this subsection can be reproduced by class \texttt{ClimateModelExperimentOptimalAbatementByInterestRateRisk}.}

As the objective function we may consider the left-tail expected shortfall, which is the expectation of $X$ conditional $X \leq \mathrm{VaR}_{\alpha}(X)$ (the small values), or, the right-tail expected shortfall, which is the expectation of $X$ conditional $X \geq \mathrm{VaR}_{\alpha}(X)$ (the large values).

\begin{figure}[hp]
    \centering
    \includegraphics[width=0.9\textwidth]{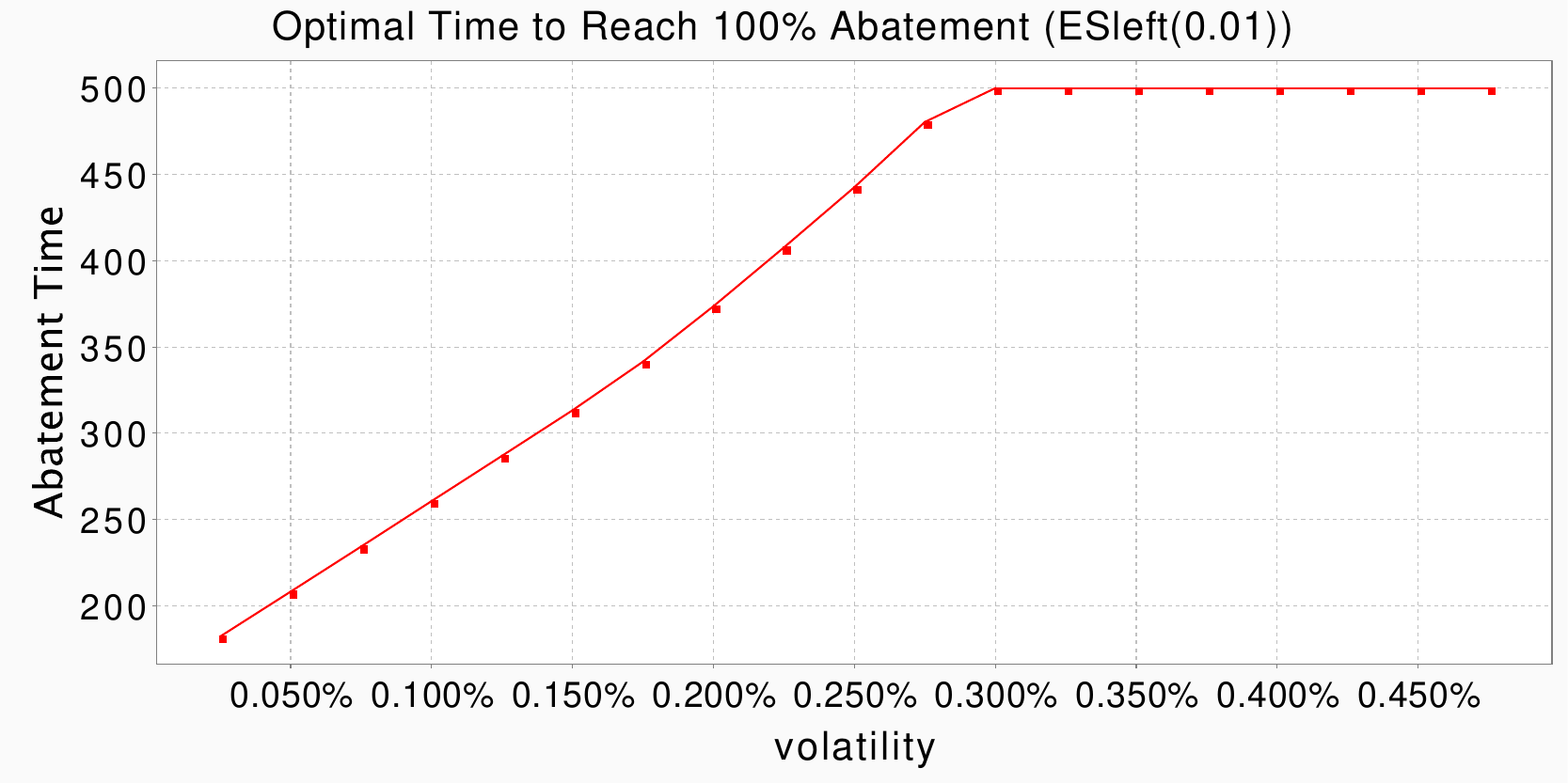}
    \caption{The dependency of the optimal $T^{\mu=1}$ on a the interest rate volatility (for calibrated constant savings rate $s_{0}$ maximising the left-tail expected short fall. \emph{The higher interest rate risk, the later one should achieve 100\% abatement.}}
    \label{fig:maxAbateTimeByRateVolatility-left}
\end{figure}

\cref{fig:maxAbateTimeByRateVolatility-left} depicts the dependency of $T^{\mu=1}$ on the interest rate volatility parameter, if the objective function is the left-tail expected shortfall.

\begin{figure}[hp]
    \centering
    \includegraphics[width=0.9\textwidth]{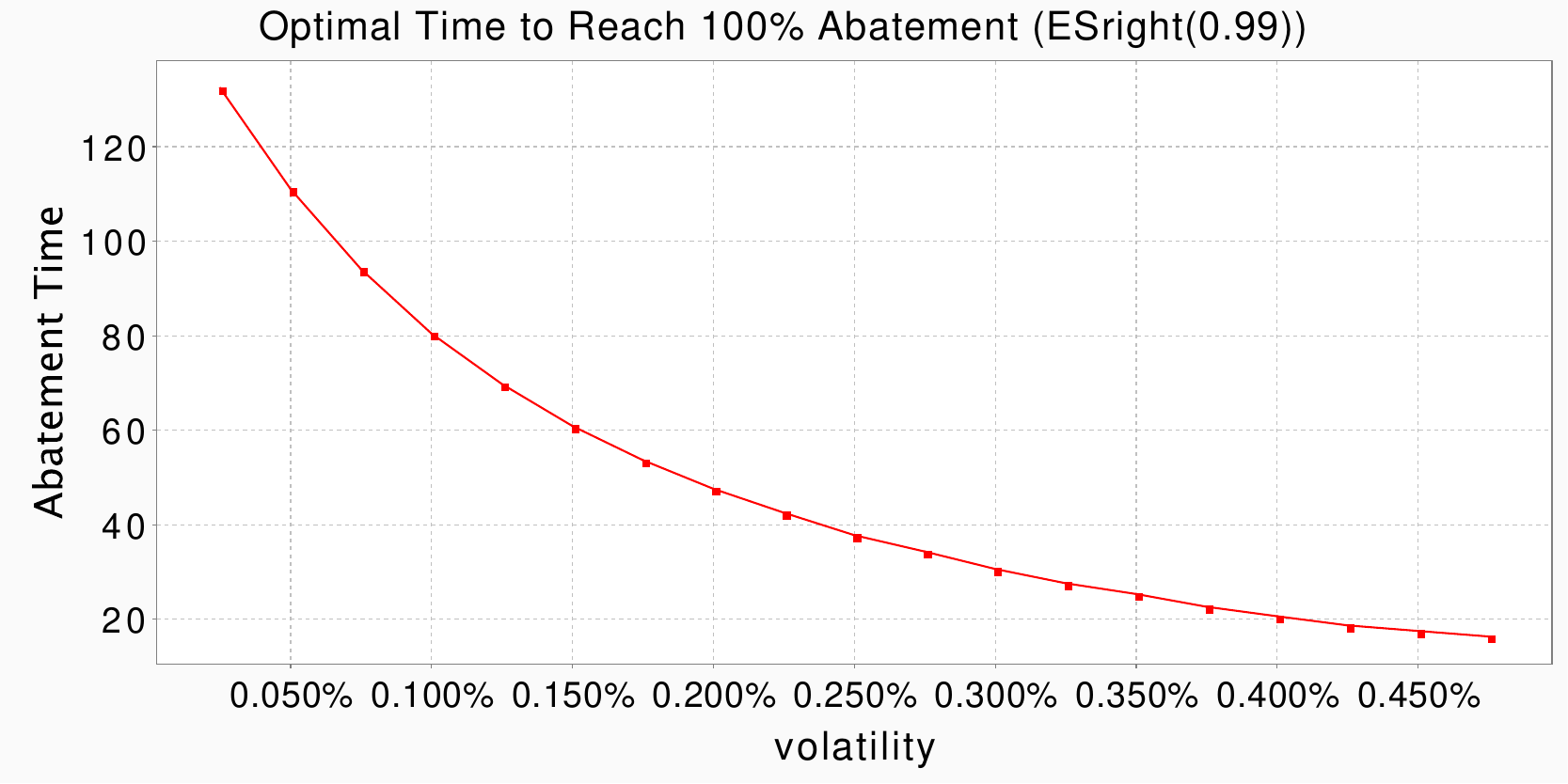}
    \caption{The dependency of the optimal $T^{\mu=1}$ on a the interest rate volatility (for calibrated constant savings rate $s_{0}$ maximising the right-tail expected short fall. \emph{The higher interest rate risk, the earlier one should achieve 100\% abatement.}
}
    \label{fig:maxAbateTimeByRateVolatility-right}
\end{figure}

\cref{fig:maxAbateTimeByRateVolatility-right} depicts the dependency of $T^{\mu=1}$ on the interest rate volatility parameter, if the objective function is the right-tail expected shortfall.

With the left-tail expected short fall we have that higher volatility will lead to significantly later abatement, whereas the right-tail expected short fall we that higher volatility will lead to significantly earlier abatement.

This effect is easy to understand but is important for our further investigations: The expected shortfall on the welfare will condition on scenarios with high (right tail) or low (left tail) welfare. These scenarios are perfectly correlated with low or high interest rates.

Hence, changing the objective function from expectation to expected shortfall is just a modification of the effective interest rate curve. The effective discount factor is the expected shortfall of the zero-coupon bond.

\begin{figure}[ph]
    \centering
    \includegraphics[width=0.9\textwidth]{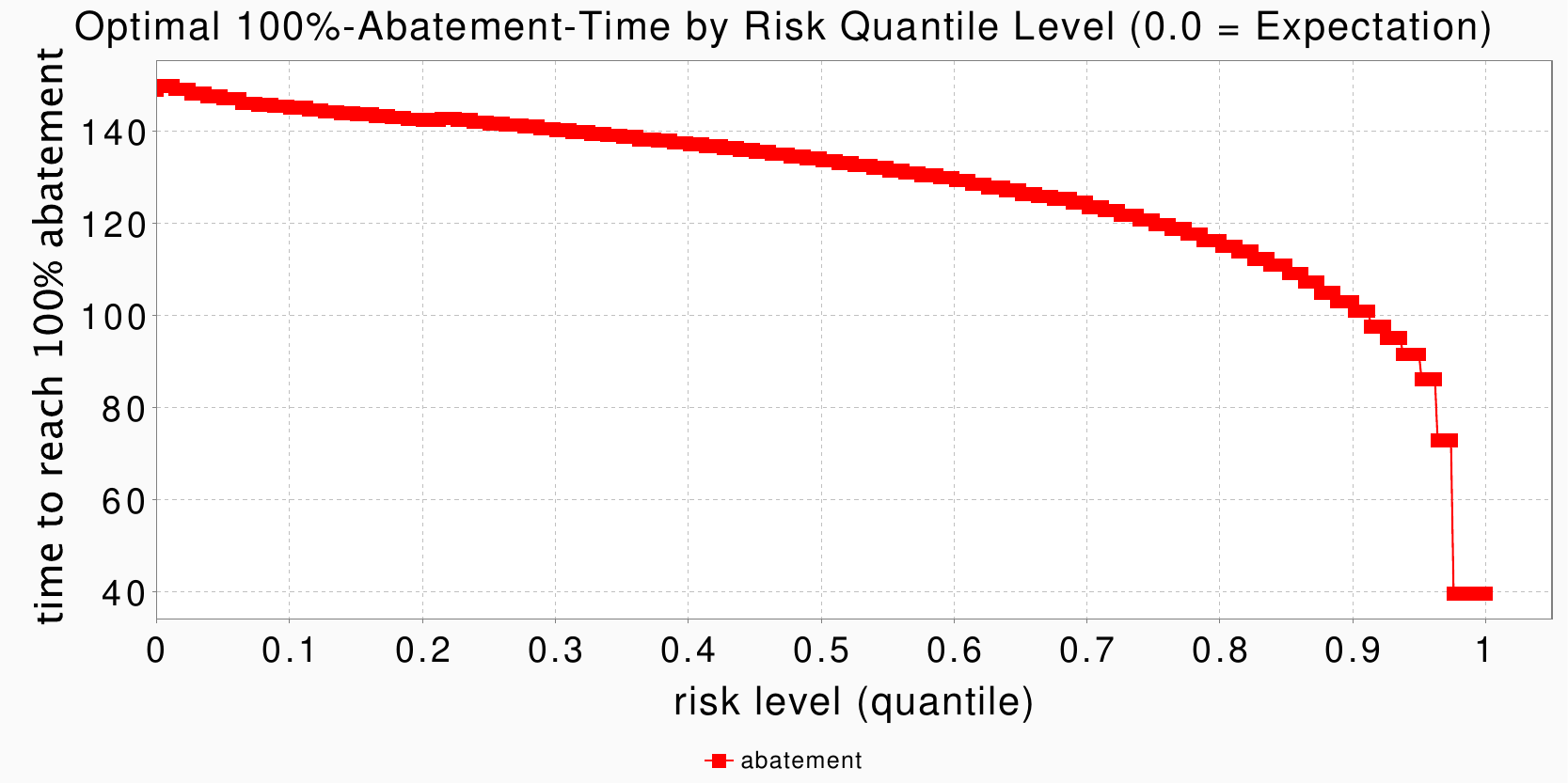}
    \caption{The dependency of $T^{\mu=1}$ on the quantile level $q$ of the risk measure (expected shortfall) for stochastic interest rates.}
    \label{fig:maxAbateTimeByRiskLevel}
\end{figure}

\cref{fig:maxAbateTimeByRiskLevel} shows the optimal $T^{\mu=1}$ as a function of the quantile level $q$ of the objective function.\footnote{The figures of this subsection can be reproduced by class \texttt{ClimateModelExperimentAbatementAndSavingsRateByQuantile}.}
\emph{The more risk averse, the earlier one should achieve 100\% abatement.}

\smallskip

This effect has some relevance: the scenarios that are associated with higher welfare are those associated with higher damages (since damages are proportional). In these scenarios, a higher risk would imply an earlier abatement.

\bigskip

Using stochastic interest rates for the discount rate only, keeping all other quantities deterministic, has a flaw: we do know that the abatement strategy depends on the interest rate level, however, the model assumes stochastic interest rates but considers a deterministic abatement strategy.

We will fix this an introduce a stochastic abatement strategy.

\clearpage
\subsection{Stochastic Abatement Model}

The previous experiments showed that the abatement strategy depends on the interest rate level. If interest rates are stochastic, this implies that the optimal abatement policy will be stochastic.

\subsubsection{Stochastic Parametric Abatement Model}

The observation, that the optimal time to reach $100\%$ abatement depends on the interest rate level (and, for risk, on the interest rate volatility) implies that in general the abatement strategy should depend on the current observed interest rate level and risk.

Thus, the abatement strategy would depend on the interest rates.

Restricted to our simplified model, the speed of increasing the abatement, \ie $a_{0}$, should depend on the interest rate., 

By this, the abatement function will become stochastic, and by that, all quantities (also the physical quantities temperature and carbon concentration) become stochastic.

One could also consider to adjust abatement to observed volatility - while this would yield more optimal pathways, it it debatable if such the implementation of such a complex strategy is realistic under political and societal constraints.

We analyse the effect by considering a simple parametric stochastic model for abatement and savings rate.\footnote{The figures of this subsection can be reproduced by class \texttt{ClimateModelExperimentOptimalParametricStochasticAbatement}.}
We consider a linear model
\begin{equation}
    \label{eq:stochAbatementParametricLinear}
    \mu(t,\omega) = \min\left( \mu_{0} + \left( a_{0} + a_{1} \cdot r(t,\omega) \right) \cdot t , 1.0 \right) \text{.}
\end{equation}
and a cubic model
\begin{equation}
    \label{eq:stochAbatementParametricCubic}
    \mu(t,\omega) = \min\left( \mu_{0} + \left( a_{0} + a_{1} \cdot r(t,\omega) + a_{2} \cdot r(t,\omega)^{2} \right) \cdot t , 1.0 \right) \text{.}
\end{equation}

The simplicity of the models allows us to interpret the parameters. We have that $a_{1}$ in \eqref{eq:stochAbatementParametricLinear} describes the change of the abatement speed by interest rate change,
\begin{equation*}
    a_{1} \ = \ \frac{\partial}{\partial r} \ \frac{\partial \mu(t)}{\partial t} \text{.}
\end{equation*}

Obviously, the stochastic abatement models may achieve a larger-or-equal welfare or lower-or-equal welfare-risk than the corresponding deterministic model, simple due to the additional degree of freedom $a_{1}$ in the abatement optimisation.


\subsubsection{Impact on the Abatement Strategy: Expectation}

\begin{figure}[hp]
    \centering
    \includegraphics[width=0.9\textwidth]{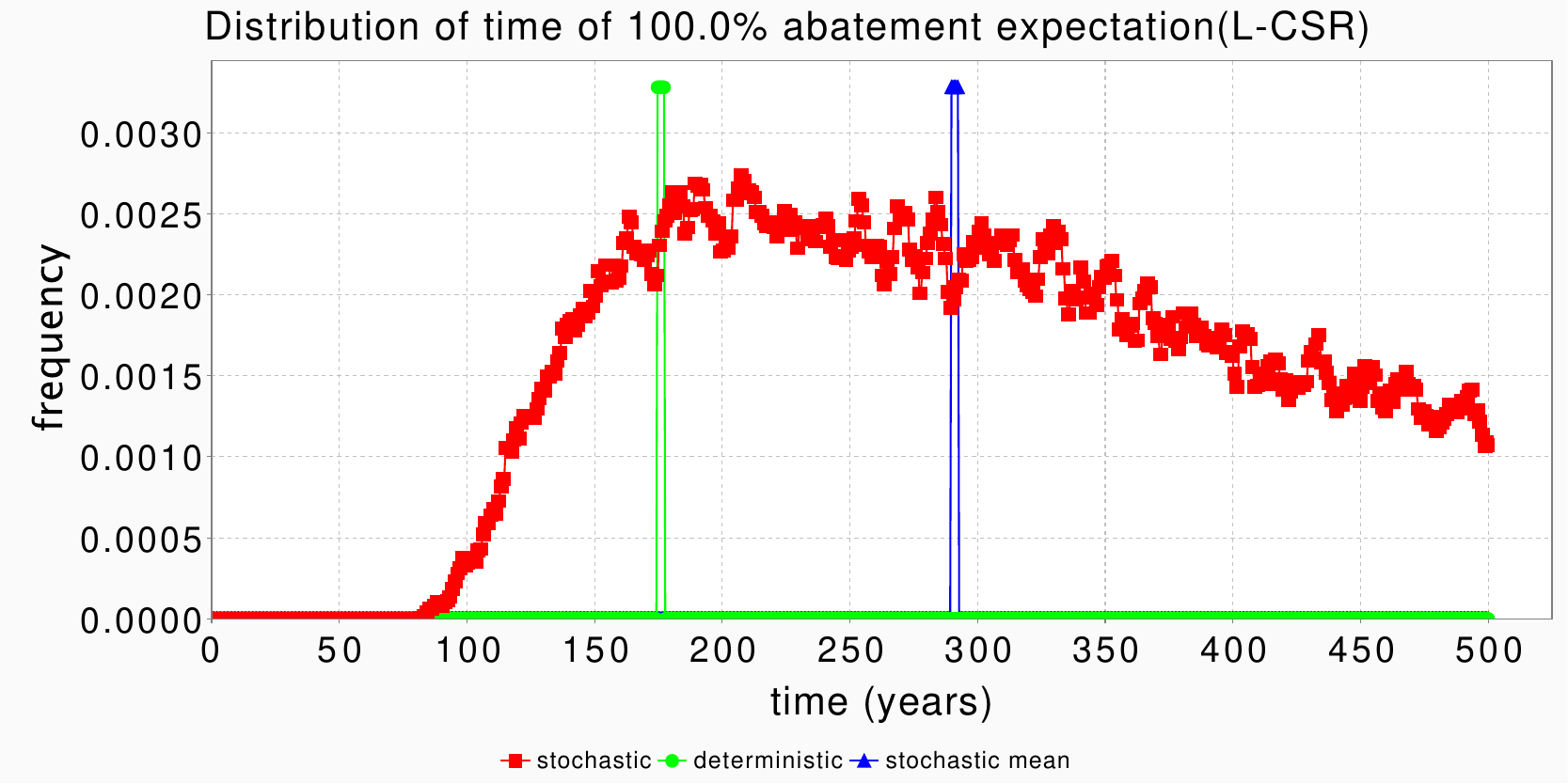}
    \caption{The distribution of $T^{\mu=1}$ for stochastic interest rates and the abatement model \cref{eq:stochAbatementParametricLinear}.}
    \label{fig:maxAbateTimeDistributionLinearModel}
\end{figure}

Again, we depict the impact of the new model parameters, by considering $T^{\mu=1}$, which now becomes a random variable. \cref{fig:maxAbateTimeDistributionLinearModel} shows the distribution of the random variable $T^{\mu=1}$ under the linear model \eqref{eq:stochAbatementParametricLinear} when the objective function is the expected welfare.

The distribution of $T^{\mu=1}$ shows how the model adapts the abatement policy. The calibrated (optimal) parameter $a_{2}$ is negative, i.e., the abatement speed is reduced once interest rates increase.

The distribution has a fat tail, such that the average abatement speed is smaller (the average max.~abatement time is larger) than that of the deterministic model.

We also see, that on some sample paths a much faster abatement speed is chosen.

\subsubsection{Discussion of the Result}

The result expose a risk of the strategy to optimise the expectation of the discounted future welfare: The strategy will significantly reduce the abatement speed if interest rates rise.

This will then result in a sub-optimal paths if thereafter interest rate will fall again. The path is optimal in expectation.

This raises the question if the optimality in expectation is the correct criteria: is it justified to raise the risk of future losses if they are compensated by the possibility of future gains?

\subsubsection{Impact on the Abatement Strategy: Expected Shortfall}

\begin{figure}[hp]
    \centering
    \includegraphics[width=0.9\textwidth]{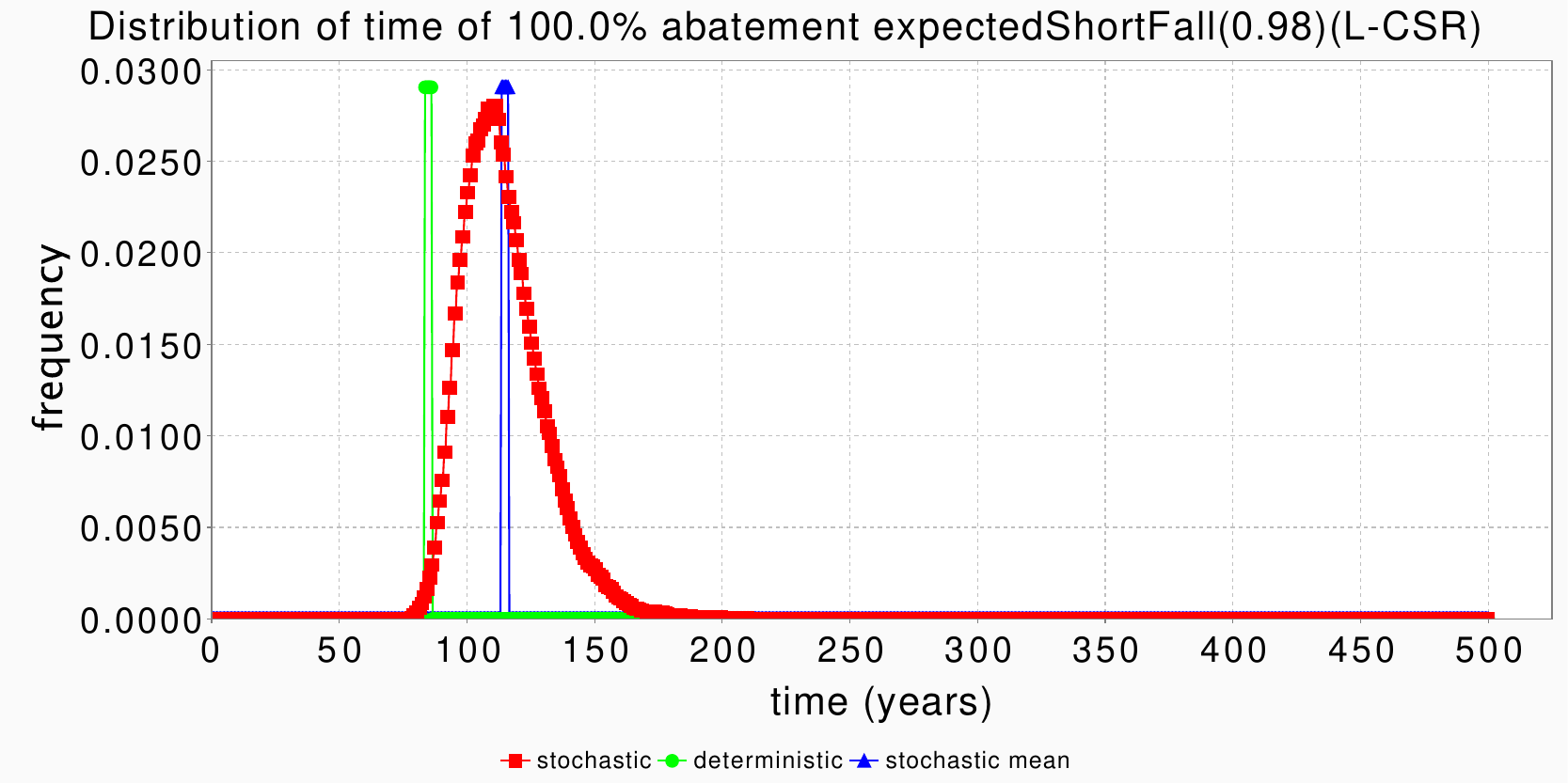}
    \caption{The distribution of $T^{\mu=1}$ for stochastic interest rates and the abatement model \cref{eq:stochAbatementParametricLinear} with the ES(0.98) objective function.}
    \label{fig:maxAbateTimeDistributionLinearModelES}
\end{figure}

\cref{fig:maxAbateTimeDistributionLinearModelES} shows the distribution when the objective function is the welfare risk.

\medskip

\begin{figure}[ph]
    \centering
    \includegraphics[width=0.9\textwidth]{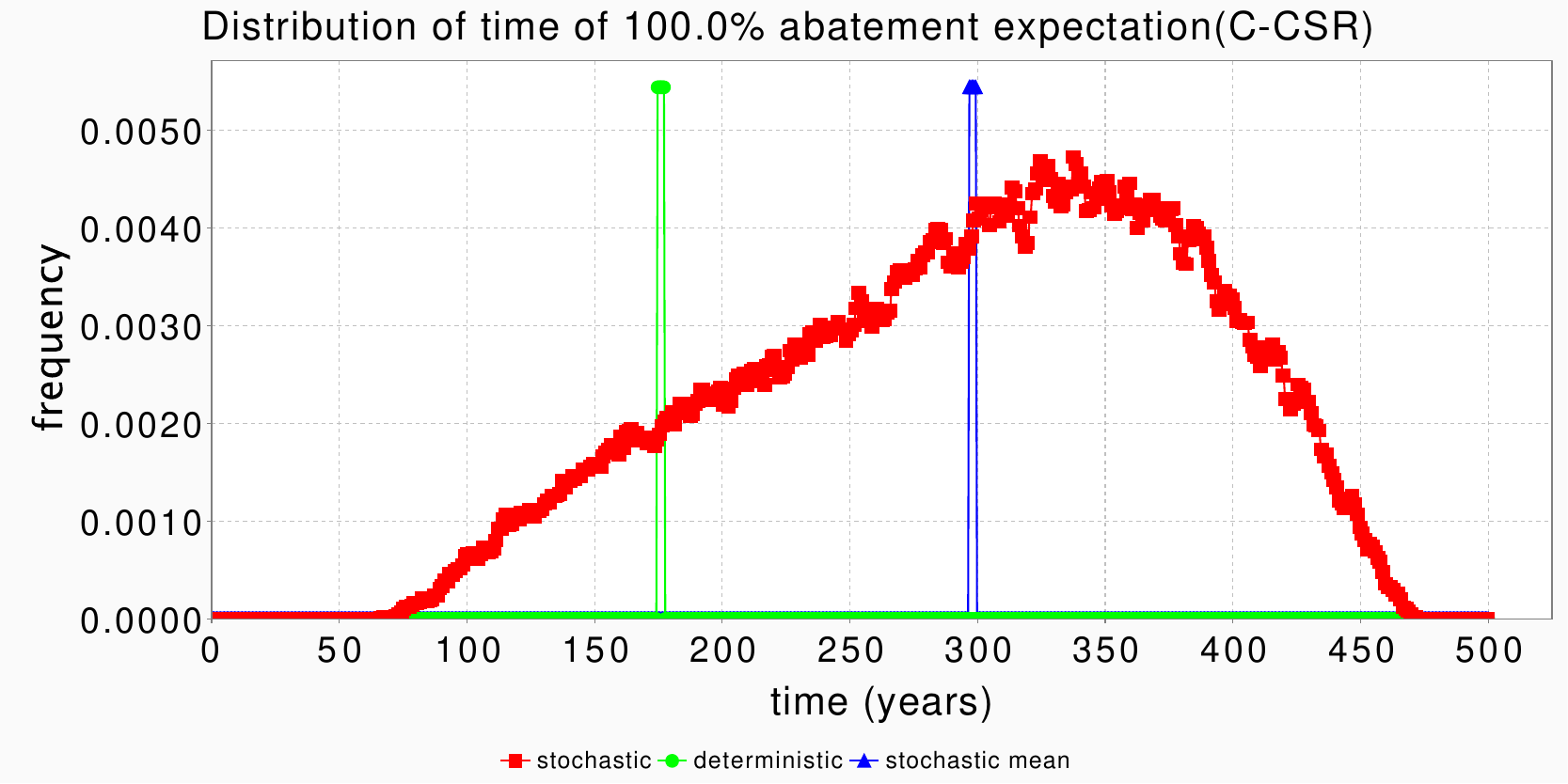}
    \caption{The distribution of $T^{\mu=1}$ for stochastic interest rates and the abatement model \cref{eq:stochAbatementParametricLinear}.}
    \label{fig:maxAbateTimeDistributionCubicModel}
\end{figure}

\begin{figure}[ph]
    \centering
    \includegraphics[width=0.9\textwidth]{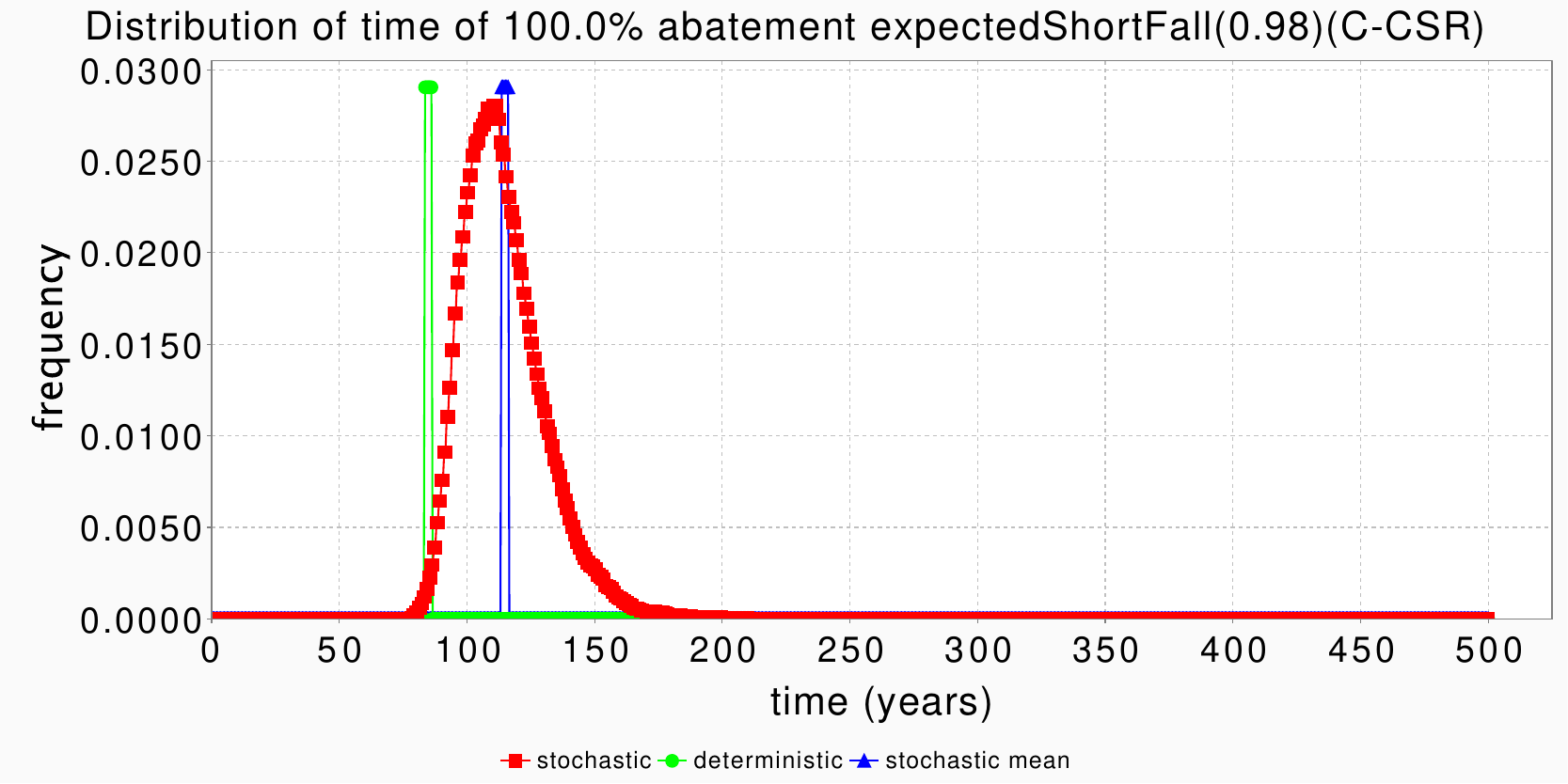}
    \caption{The distribution of $T^{\mu=1}$ for stochastic interest rates and the abatement model \cref{eq:stochAbatementParametricCubic} with the ES(0.98) objective function.}
    \label{fig:maxAbateTimeDistributionCubicModelES}
\end{figure}

The fat tail of the distribution is due to our model assumption: the interest rate model combined with the linear abatement model.
If we calibrate a quadratic abatement model, the distribution changes, increasing the abatement speed in the slow abatement scenarios, see \cref{fig:maxAbateTimeDistributionCubicModel} and \cref{fig:maxAbateTimeDistributionCubicModelES}


\clearpage
\subsubsection{Impact on the Cost (Damage and Abatement)}

While for the deterministic model we analysed the temporal distribution of the cost $C(t)$, for the stochastic model an additional metric may be analysed first: the spacial distribution of the cost, i.e., the distribution of the (now) random variable of the aggregated discounted cost.

We calibrate the three models \cref{eq:abatementModelDeterministic,eq:stochAbatementParametricLinear,eq:stochAbatementParametricCubic} using the standard objective function with an expectation operator (Figure x,y,z). We then calibrate the models using the standard objective function with an expected shortfall operator applied to the objective function.

\subsubsection{Temporal Distribution of Cost}

We investigate how the switch to a stochastic interest rate and a stochastic abatement policy changes the temporal distribution of cost. We look at the discounted cost and the cost per GDP.

\medskip

\begin{figure}
    \centering
    \includegraphics[width=0.9\linewidth]{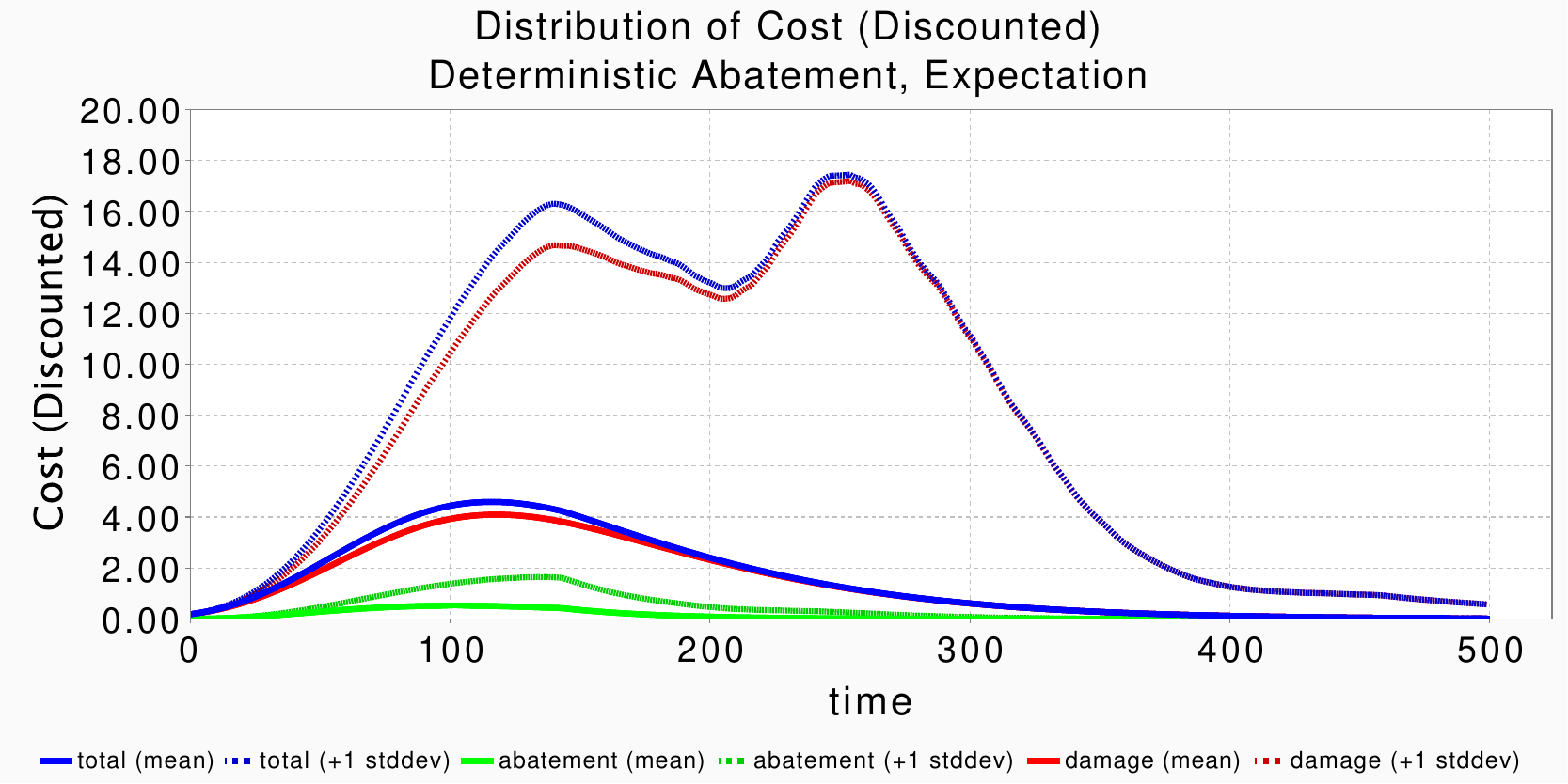}
    \caption{Discounted cost distribution for stochastic interest rates and the
deterministic abatement model with the expectation as objective function.}
    \label{fig:CostOverTime-deterministic-abatement-exp-DISCOUNTED}
\end{figure}

\begin{figure}
    \centering
    \includegraphics[width=0.9\linewidth]{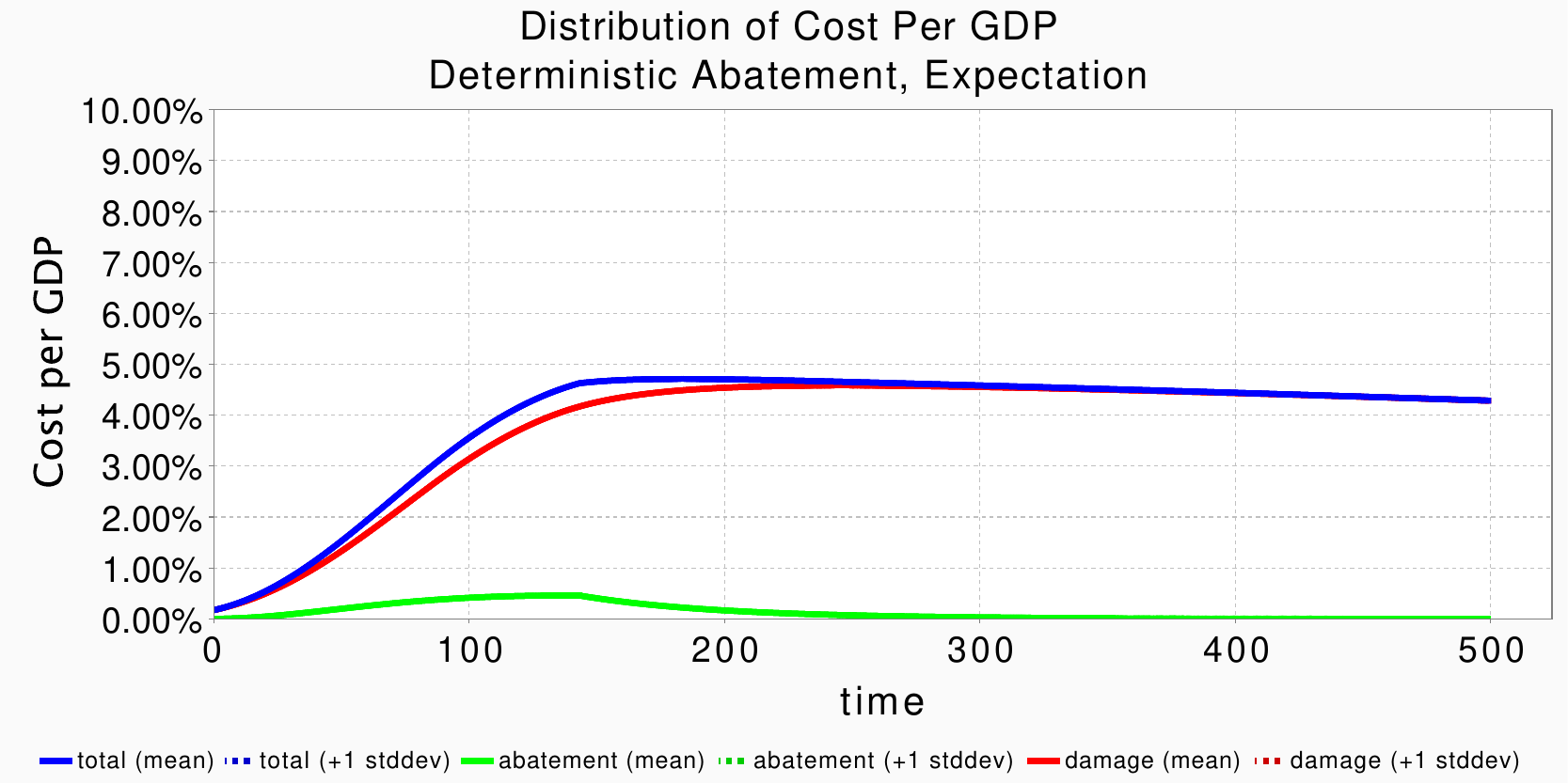}
    \caption{Undiscounted cost per GDP distribution for stochastic interest rates and the
deterministic abatement model with the expectation as objective function.}
    \label{fig:CostOverTime-deterministic-abatement-exp-PER_GDP}
\end{figure}

\cref{fig:CostOverTime-deterministic-abatement-exp-DISCOUNTED,fig:CostOverTime-deterministic-abatement-exp-PER_GDP} show the cost, discounted and per GDP, for the deterministic abatement model. 

For the deterministic abatement model with stochastic interest rate, the discounted cost are a random variable, but cost per GDP are still deterministic. The randomness is only due to the discount factor.


\medskip

\begin{figure}
    \centering
    \includegraphics[width=0.9\linewidth]{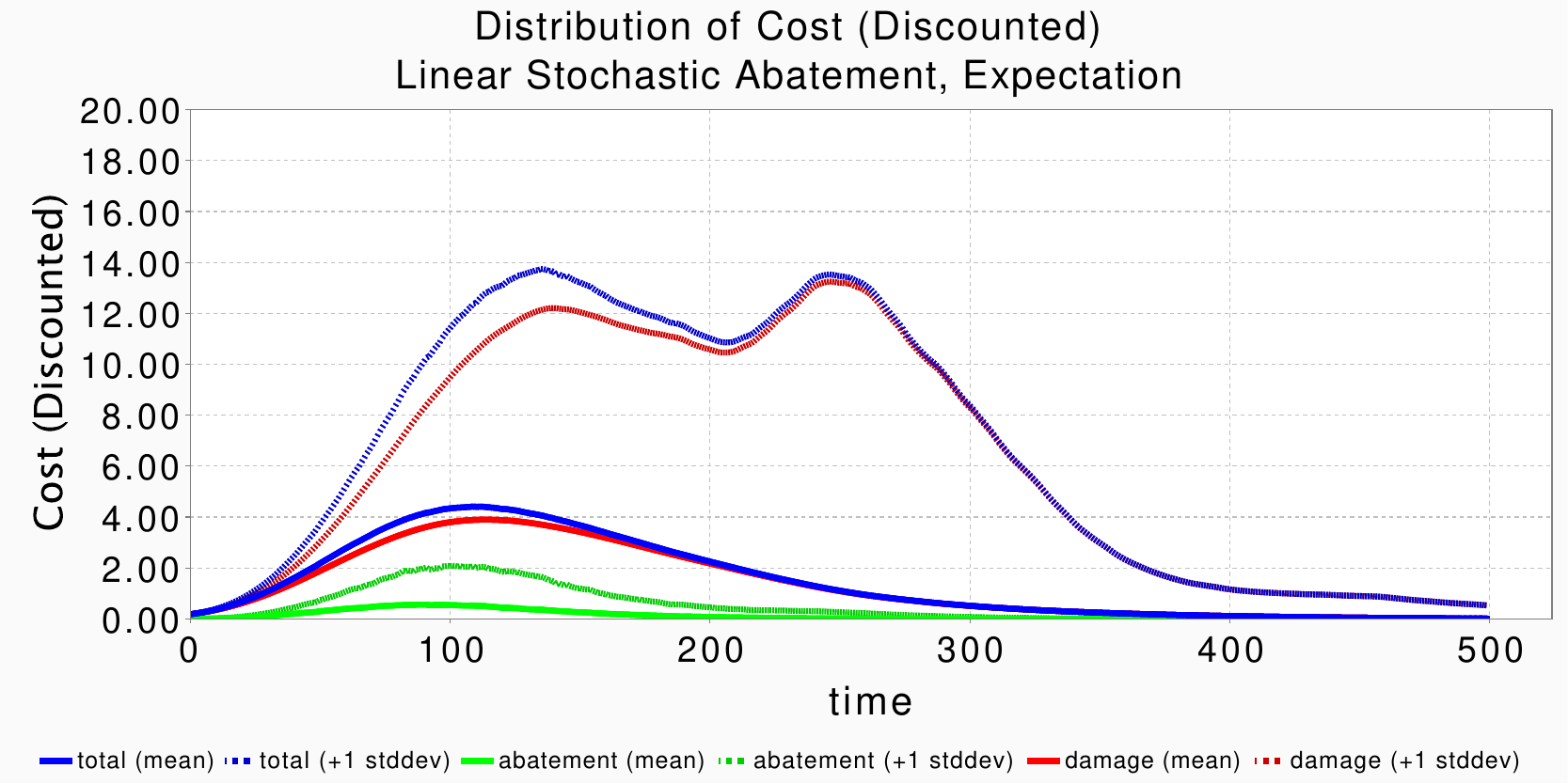}
    \caption{Discounted cost distribution for stochastic interest rates and the abatement model \cref{eq:stochAbatementParametricCubic} with the expectation as objective function.}
    \label{fig:CostOverTime-lin-stoch-abatement-exp-DISCOUNTED}
\end{figure}

\begin{figure}
    \centering
    \includegraphics[width=0.9\linewidth]{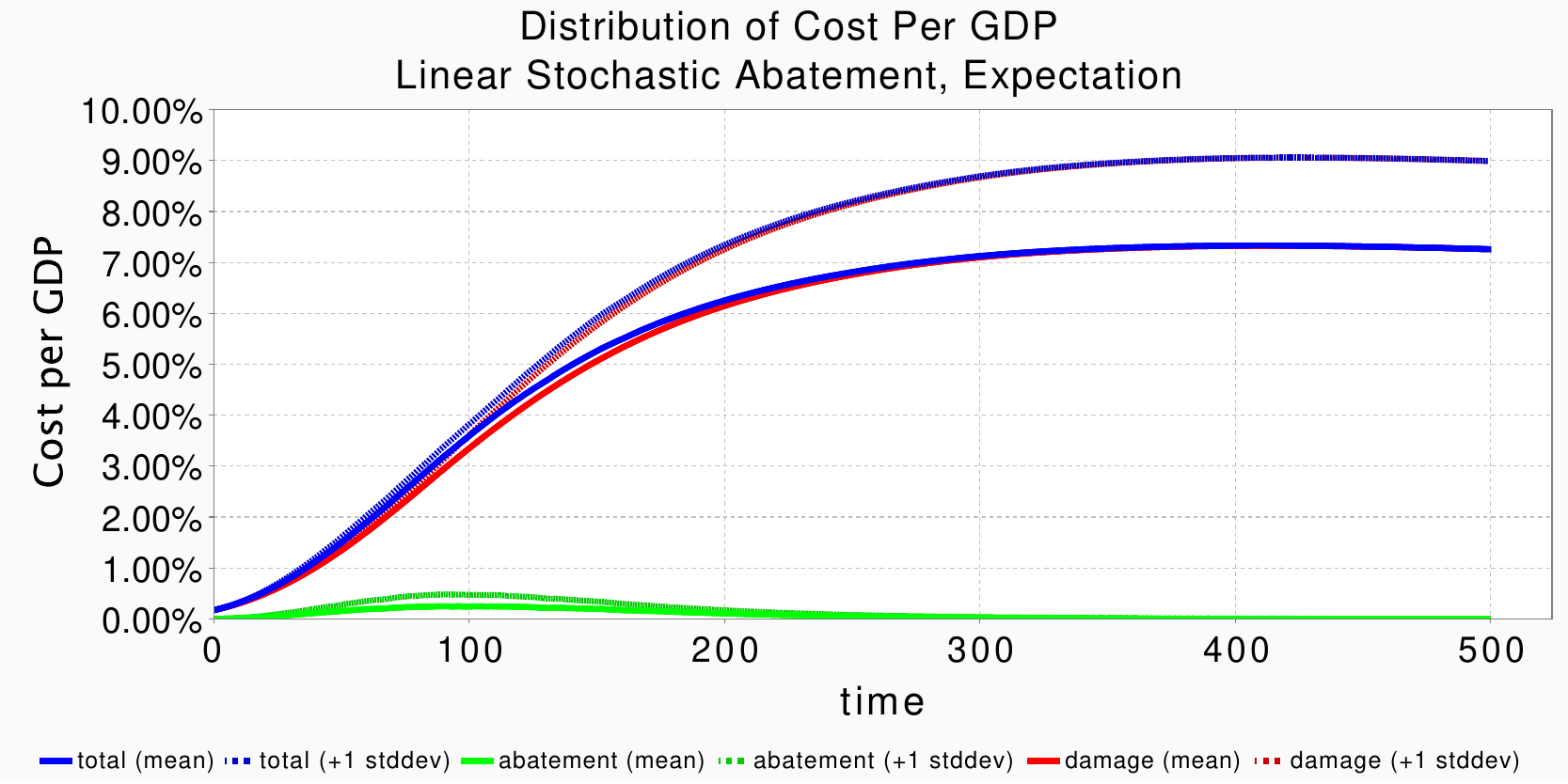}
    \caption{Undiscounted cost per GDP distribution using stochastic interest rates and the abatement model \cref{eq:stochAbatementParametricCubic} with the expectation as objective function.}
    \label{fig:CostOverTime-lin-stoch-abatement-exp-PER_GDP}
\end{figure}

\cref{fig:CostOverTime-lin-stoch-abatement-exp-DISCOUNTED,fig:CostOverTime-lin-stoch-abatement-exp-PER_GDP} show the cost, discounted and per GDP, for the (linear) stochastic abatement model. 

A stochastic policy, i.e., adapting the policy to the interest rate level) reduces the discounted cost, both in the mean and in the risk.
The cost per GDP will become stochastic under this model too. Surprisingly, the cost per GDP is increased, when moving to the stochastic abatement strategy.


This effect is clearly visible beyond $t=100$, but it is also visible prior $t=100$.

\subsubsection{Conclusion}

Moving to stochastic interest rates it is reasonable to consider an abatement policy that adapts to the interest rate level. This could be interpreted as a re-calibration of the policy to the interest rate level.
Our simple two parametric stochastic abatement model indeed improves the social welfare (in expectation).
However, adapting the abatement policy to the interest rate level apparently worsens the intergenerational equity of the cost (in terms of cost per GDP). In addition it introduces a significant risk of higher cost for future generations.

\clearpage
\subsection{Funding in a Model of Stochastic Interest Rates}

\begin{figure}
    \centering
    \includegraphics[width=0.9\linewidth]{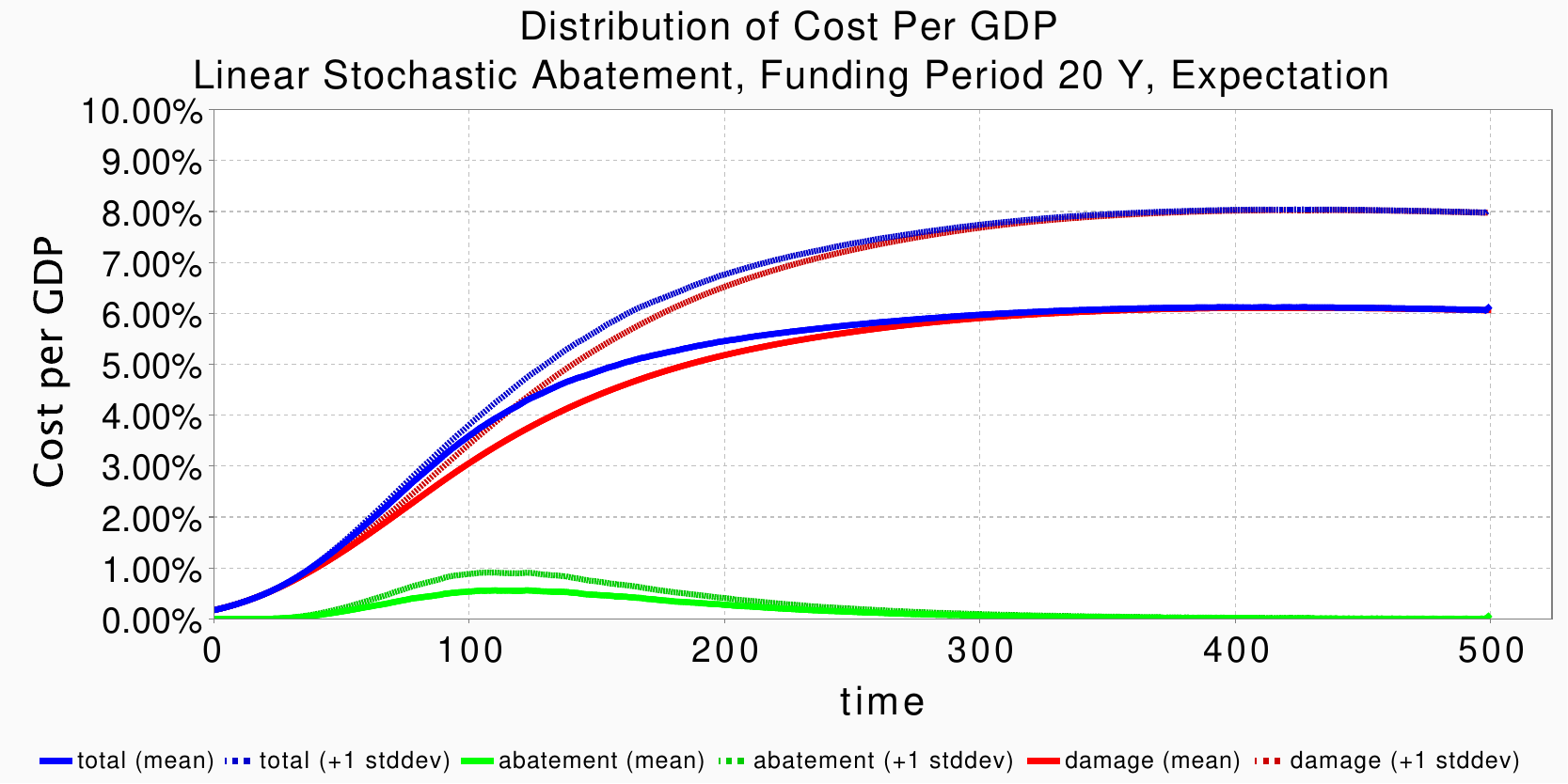}
    \caption{Discounted cost distribution using stochastic interest rates and the abatement model \cref{eq:stochAbatementParametricCubic} considering funding of abatement cost with the expectation as objective function.}
    \label{fig:CostOverTime-lin-stoch-abatement-funding-PER_GDP}
\end{figure}

We now consider the introduction of a funding period for the abatement cost to the stochastic model. We take a look at the random variables $C_{\mathrm{A}}(t)$, $C_{\mathrm{D}}(t)$ and $C(t)$. We measure the mean and the variance. We observe that a funding period will reduce both, the mean and the variance of the cost distribution. While we already observed in the deterministic model that the introduction of the funding period reduces the maximum cost, we now see that is also reduces the risk of cost. \cref{fig:CostOverTime-lin-stoch-abatement-funding-PER_GDP} shows the cost per GDP for the calibrated model with funding period. This result includes funding of damages adding to \cref{fig:CostOverTime-lin-stoch-abatement-exp-PER_GDP}.


\clearpage
\subsection{Non-Linear Discounting in a Model of Stochastic Interest Rates}

\begin{figure}
    \centering
    \includegraphics[width=0.9\linewidth]{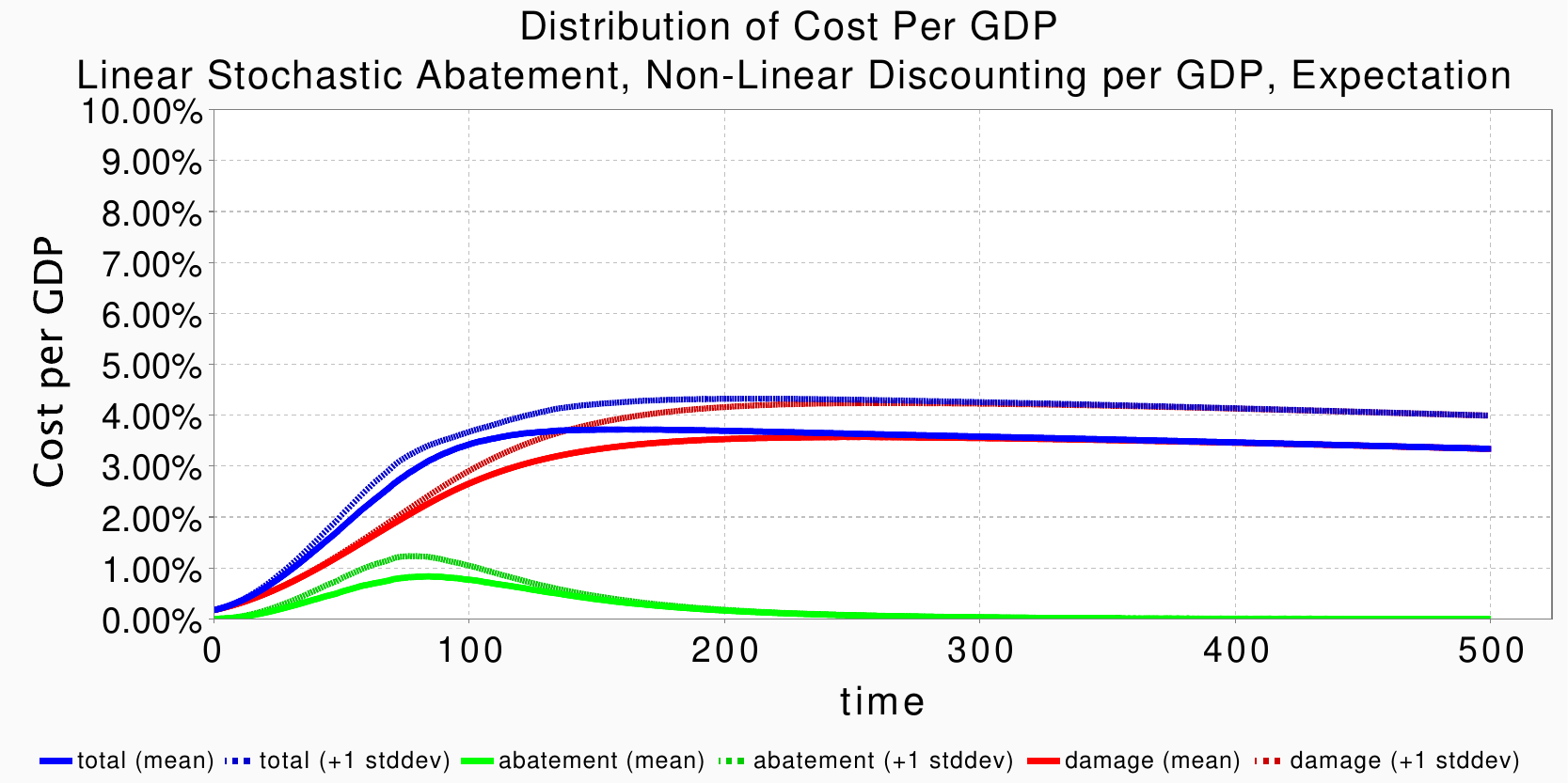}
    \caption{Undiscounted cost per GDP distribution using stochastic interest rates and the abatement model \cref{eq:stochAbatementParametricCubic} considering funding of abatement cost and non-linear discounting of damage cost with the expectation as objective function.}
    \label{fig:costOverTime-lin-stoch-abatement-nonlineardiscounting-PER_GDP}
\end{figure}

We add non-linear funding of damage cost, i.e., non-linear discounting to the model with stochastic interest rates and a stochastic abatement strategy.  \cref{fig:costOverTime-lin-stoch-abatement-nonlineardiscounting-PER_GDP} shows the cost-per-GDP in the calibrated model. This result has to be compared to  \cref{fig:CostOverTime-lin-stoch-abatement-exp-PER_GDP}.

As before, we observe that the addition of a non-linear discounting allows to level the cost, here to roughly 3.5\% of the GDP. In addition we see that the non-linear discounting of the cost leads to a reduction of the risk (depicted as +1 stddev). The first standard deviation stays around 4.0\% of the GDP. The risk is +0.5\% of the GDP, while the unconstrained model exhibits a risk around +2\% of the GDP, see \cref{fig:CostOverTime-lin-stoch-abatement-exp-PER_GDP}.

\clearpage
\section{Conclusion}
\label{sec:discussion_conclusion}

We investigate the classical DICE integrated assessment model and compare several extensions related to intergenerational equity. Our investigations and extensions are general in nature and could be applied to other integrated assessment models. We provide a complete open source implementation of the model illustrating all our extensions, utilising (stochastic) algorithmic differentiation to efficiently calculate sensitivities of model quantities against each other.
We use algorithmic differentiation to perform a detailed sensitivity analysis, analysing the temporal changes of the cost structure to policy changes and the dependency of damage cost change on abatement cost changes. The findings motivate some of our model modifications.

Our first observation is that the DICE model calibration does not consider any intergenerational equity as it is balancing the marginal cost changes induced by a policy change. Absolute cost and its distribution over time are irrelevant in the objective function. This effect is driven by the aggregation of discounted welfare into intergenerational welfare. This discounting scheme implies that gains of one generation are exchanged with burdens of another generation. Considering the cost relative to the current GDP (\cref{fig:costOverTimePerGDP-reduced}) shows that later generations have to bear higher damages relative to their economic wealth. We find that adjusting the objective function using a $p$-norm can mitigate this (\cref{fig:costOverTimePerGDP-pnorm4}).
While the $p$-norm has no obvious economic motivation, we use a more applied scenario considering funding of abatement cost, i.e., allowing abatement to be finance over usual time frames. This also improves the intergenerational equity (\cref{fig:costOverTimePerGDP-funding20}) by decreasing the burden for many future generations, while some present generations face higher costs. 
Further, endowing the model with non-linear discounting, which is a discount factor depending on the size of the amount borrowed due to increased default risk, fosters equal distribution of cost in the calibration (\cref{fig:CostOverTimePerGDP-non-linear-discounting}). It may be a reasonable extension to improve intergenerational equity, since funding costs of damages would add to the burden of future generations. Additionally, penalising costs above a certain level of GDP could be used to equalise the burden relative to GDP for future generations (\cref{fig:CostOverTimePerGDP-non-linear-discounting-per-gdp}).
     
We also extend the DICE model with stochastic interest rates. Introducing stochastic interest rates without any further modification does not significantly alter the model. In fact, one can comprise any risk measure applied to a risky discounting in a deterministic discount factor. However, introducing stochastic interest rates motivates allowing for a stochastic abatement policy. This introduces stochasticity to all state variables and hence another aspect in the intergenerational equity: the distribution of risk.
We repeat our analysis in the model with stochastic interest rates. We observe a general reduction of risk (e.g. risk as cost per GDP).

Using a simple IAM includes several caveats. First, the representation of damages could be improved \cite{Burke2015,Glanemann2020,Hansel2020}. Second, the lack of regional compared to, e.g., econometric damage estimates \cite{kotz_day--day_2021,kotz_effect_2022} ignores regional inequalities and potential distributional effects of projected costs. Since our study aims to introduce methods such as limiting total costs of climate change by a proportion of GDP to the debate on mitigation pathways, these limitations could be addressed by including our concepts in more complex assessments of mitigation pathways \cite{rennert_comprehensive_2022,van_der_wijst_new_2023}.

Modifying the DICE model in a modular way, we address the issue that the classical model does not consider equity between generations. The proposed extensions accounting for funding cost of abatement and increased financing cost of large damages can improve this aspect of the intertemporal optimisation, even more amplified if we also consider stochastic interest rate risk.
\smallskip

\newpage
\section*{List of Symbols}
\addcontentsline{toc}{section}{List of Symbols}
\footnotesize
\begin{tabular}{|c|p{0.8\textwidth}|}
    \hline
    $t$ & Simulation time. $[t] = \mathrm{years}$. \\
    \hline
    $t_{i}$ & Simulation time from the time discretization $\{t_{i}\}. $ \\
    \hline
    $\Delta t_{i}$ & Simulation time step $\Delta t_{i} = t_{i+1} - t_{i}$. \\
    \hline
    $\omega$ & The sample path in a stochastic model (a scenario). \\
    \hline
    \hline
    $T$ & Temperature vector above pre-industrial level. \\
    \hline
    $T_\mathrm{AT}$ & Atmospheric temperature above pre-industrial level. $[T_\mathrm{AT}] = \mathrm{K}$. \\
    \hline
    $F$ & Temperature forcing. $[F] = \mathrm{K}/year$. \\
    \hline
    $M$ & Carbon concentration vector. \\
    \hline
    $M_\mathrm{AT}$ & Carbon concentration in atmosphere. $[M_\mathrm{AT}] = \mathrm{GtC}$. \\
    \hline
    $E$ & Emission. $[E] = \mathrm{GtCO}_{2}$. \\
    \hline
    $c_{\mathrm{C/CO2}}$ & Conversion factor from $\mathrm{GtCO}_{2}$ to GtC, \texttt{conversionGtCperGtCO2}. \\
    \hline
    \hline
    $\Omega(t)$ & Damage cost as percentage of the GDP.  \\
    \hline
    $C_{\mathrm{D}}(t)$ & Damage cost in time $t$.  \\
    \hline
    $\Lambda(t)$ & Abatement cost as percentage of the emissions.  \\
    \hline
    $\Delta T_{A}$ & Time span of the funding period of abatement cost.\\
    \hline
    $C_{\mathrm{A}}(t)$ & Abatement cost (maturing) in time $t$.\\
    \hline
    $C(t)$ & Total cost (maturing) in time $t$. $C(t) = C_{\mathrm{A}}(t) + C_{\mathrm{D}}(t)$.\\
    \hline
    $C_C(t)$ & Consumption in time $t$.\\
    \hline
    \hline
    $\mu(t)$ & Abatement percentage in time $t$ (free parameter of the model), may be a random variable in a stochastic abatement model.\\
    \hline
    $s(t)$ & Savings rate in time $t$ (free parameter of the model).\\
    \hline
    \hline
    $V(t)$ & Time $t$ utility (welfare). In the original model often denoted by $U(t)$.\\
    \hline
    $r$ & Interest rate (short rate), may be a stochastic process. $[r] = \frac{1}{\mathrm{year}}$ \\
    \hline
    $N(t)$ & The numéraire $N(t) := \exp(\int_{0}^{t} r(s) \mathrm{d}s)$.\\
    \hline
    $V^{*}$ & Discounted welfare. $V^{*} = \sum_{i=0}^{n-1} U(t) \frac{N(0)}{N(t)}$. May be a random variable.\\
    \hline
    $W$ & The objective function. The expectation of $V^{*}$ or a risk measure applied to $V^{*}$.\\
    \hline
    \hline
    $T^{\mu=1}$ & The time $t$ at which the (simplified) abatement model reaches 100\% abatement.\\
    \hline
    $\mu_{0}$ & The initial value of the abatement model.\\
    \hline
    $a_{0}$ & The slope of the (simplified) abatement model.\\
    \hline
    $a_{1}$ & The additional slope per interest rate level of the (simplified) stochastic abatement model.\\
    \hline
    \hline
    $SCC$ & The social cost of carbon.\\
    \hline
    $SCC^{N}$ & The numéraire-relative (discounted) social cost of carbon.\\
    \hline
\end{tabular}

\newpage
\phantomsection
\addcontentsline{toc}{section}{References}
\bibliographystyle{naturemag-custom}
\bibliography{references.bib}

\clearpage

\begin{appendix}

\section{Long Time Horizons - Closing the Model}
\label{sec:closingthemodel}

Since we perform numerical optimisations of the abatement and savings rate function, it is important to verify that the optimisation algorithm cannot cheat upon the policy, by moving relevant costs beyond the models time-horizon.

For positive interest rates, the exponential decay of the discount factor could be sufficient to ensure this, if output and cost do not exhibit an exponential growth and the time-horizon is sufficiently high.

However, one can observe that the time horizon affects the calibration looking at the savings rates: shortly before the time-horizon the savings rate drops as it becomes optimal to stop investing and consume everything, see \cref{fig:calibration}.

Hence, the question arises up to which time the calibration is not affected by the arbitrarily set time horizon.

To show robustness of the model with respect to the time horizon, we check that all quantities exhibit some plausible asymptotic for $T \rightarrow \infty$. We compare the calibration of a 100 year model, 500 year model and 2000 year model, to check when the deviant behaviour occurs.\footnote{Figures of this subsection can be reproduced with class \texttt{ClimateModelExperimentTimeConvergence}.}

\begin{figure}
    \centering
    \includegraphics[width=0.9\linewidth]{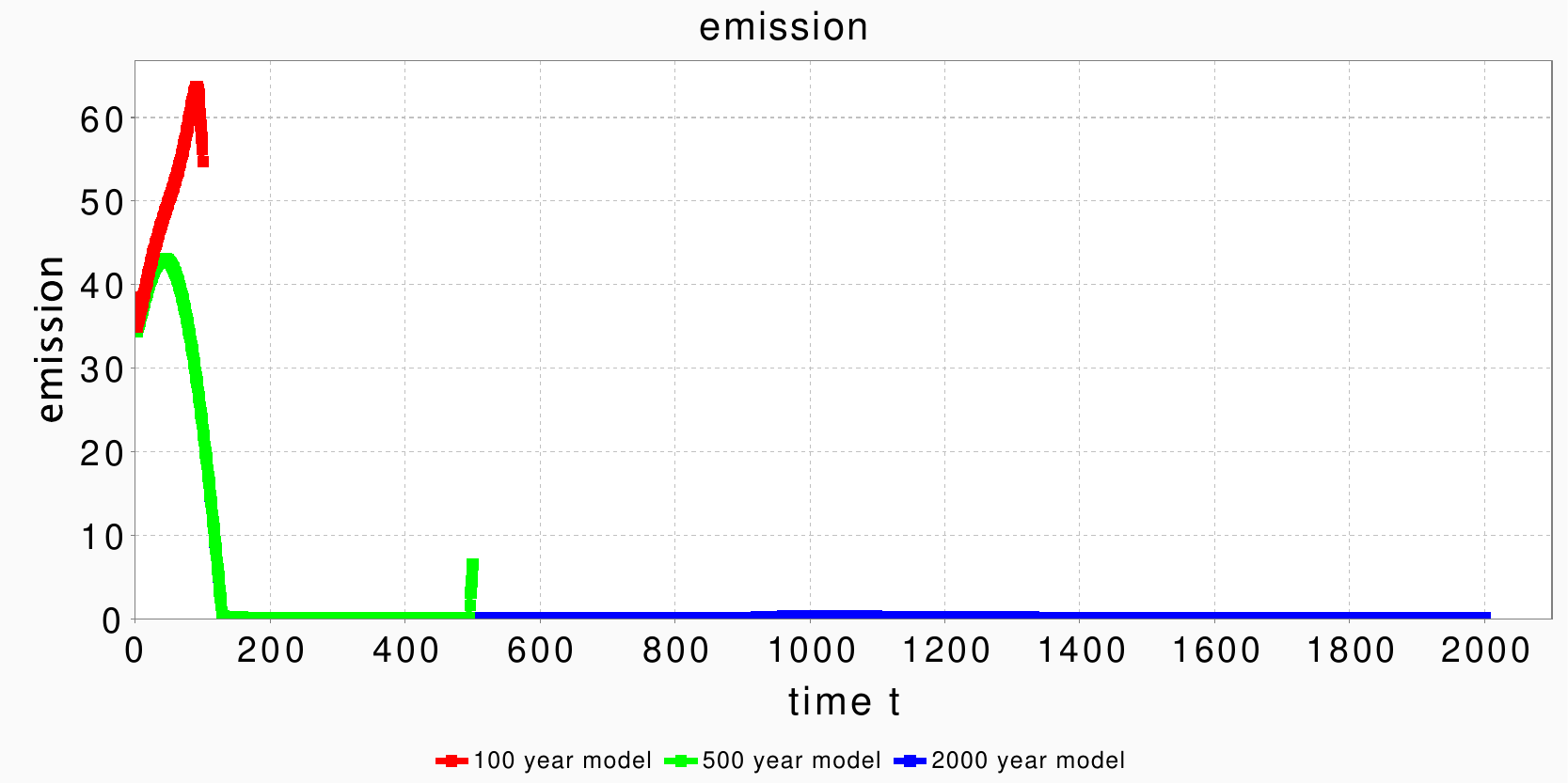}
    \caption{Optimal emission pathway in a calibrated model with a time horizon of 100, 500 or 2000 years.}
    \label{fig:timeConvergence-emission}
\end{figure}

\begin{figure}
    \centering
    \includegraphics[width=0.9\linewidth]{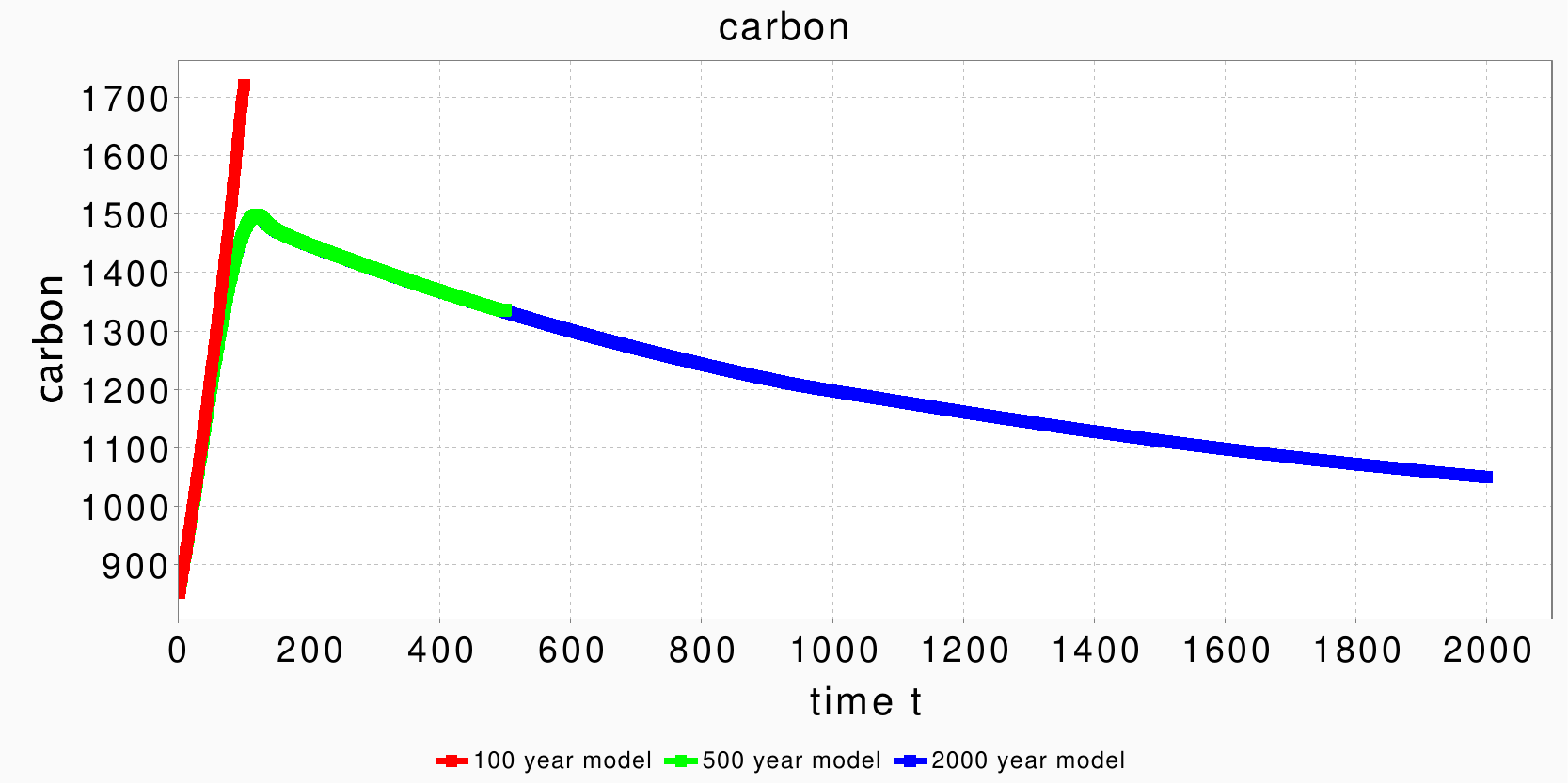}
    \caption{Optimal carbon in atmosphere pathway in a calibrated model with a time horizon of 100, 500 or 2000 years.}
    \label{fig:timeConvergence-carbon}
\end{figure}

\begin{figure}
    \centering
    \includegraphics[width=0.9\linewidth]{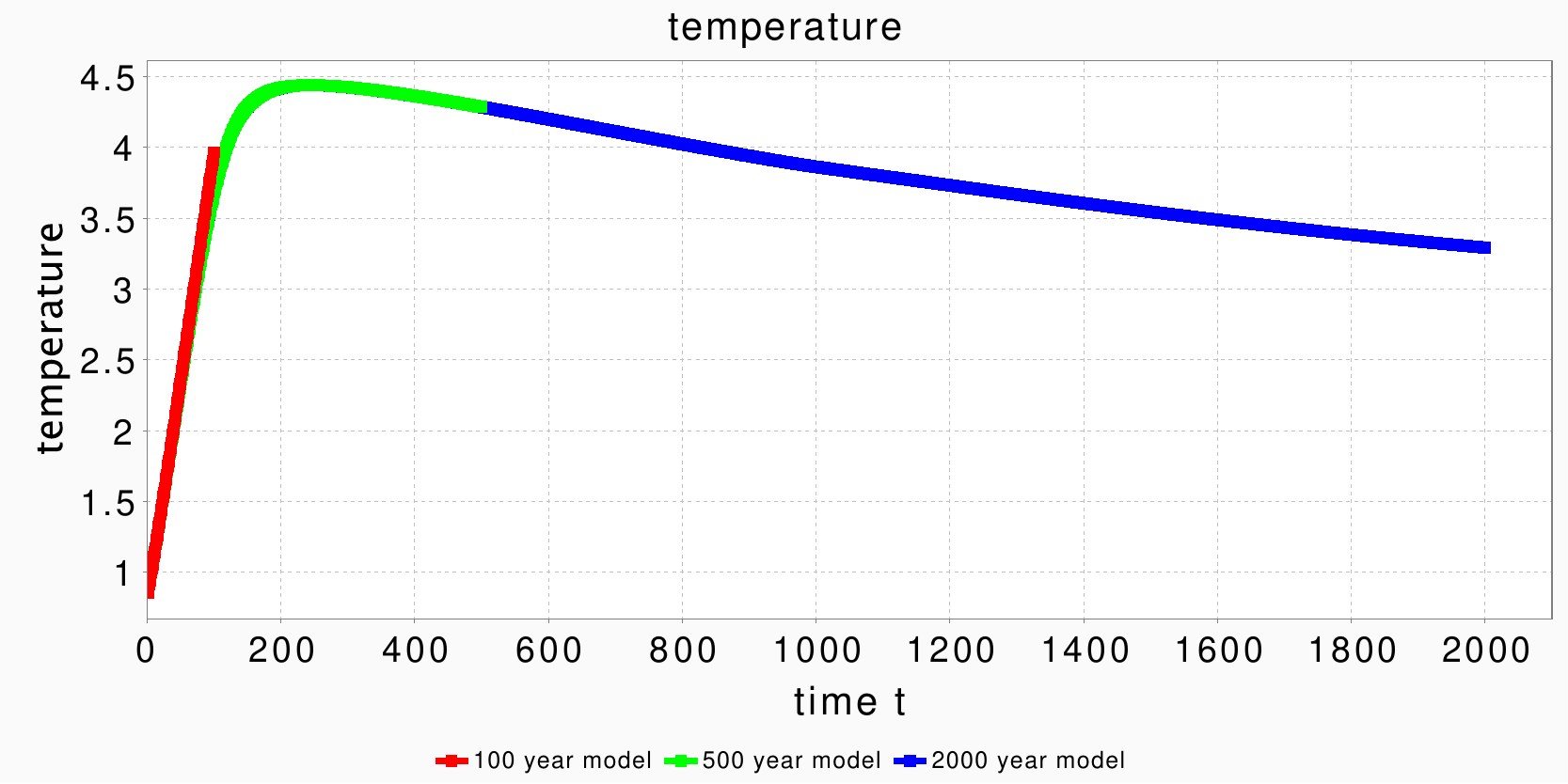}
    \caption{Optimal temperature in atmosphere pathway in a calibrated model with a time horizon of 100, 500 or 2000 years.}
    \label{fig:timeConvergence-temperature}
\end{figure}

\begin{figure}
    \centering
    \includegraphics[width=0.9\linewidth]{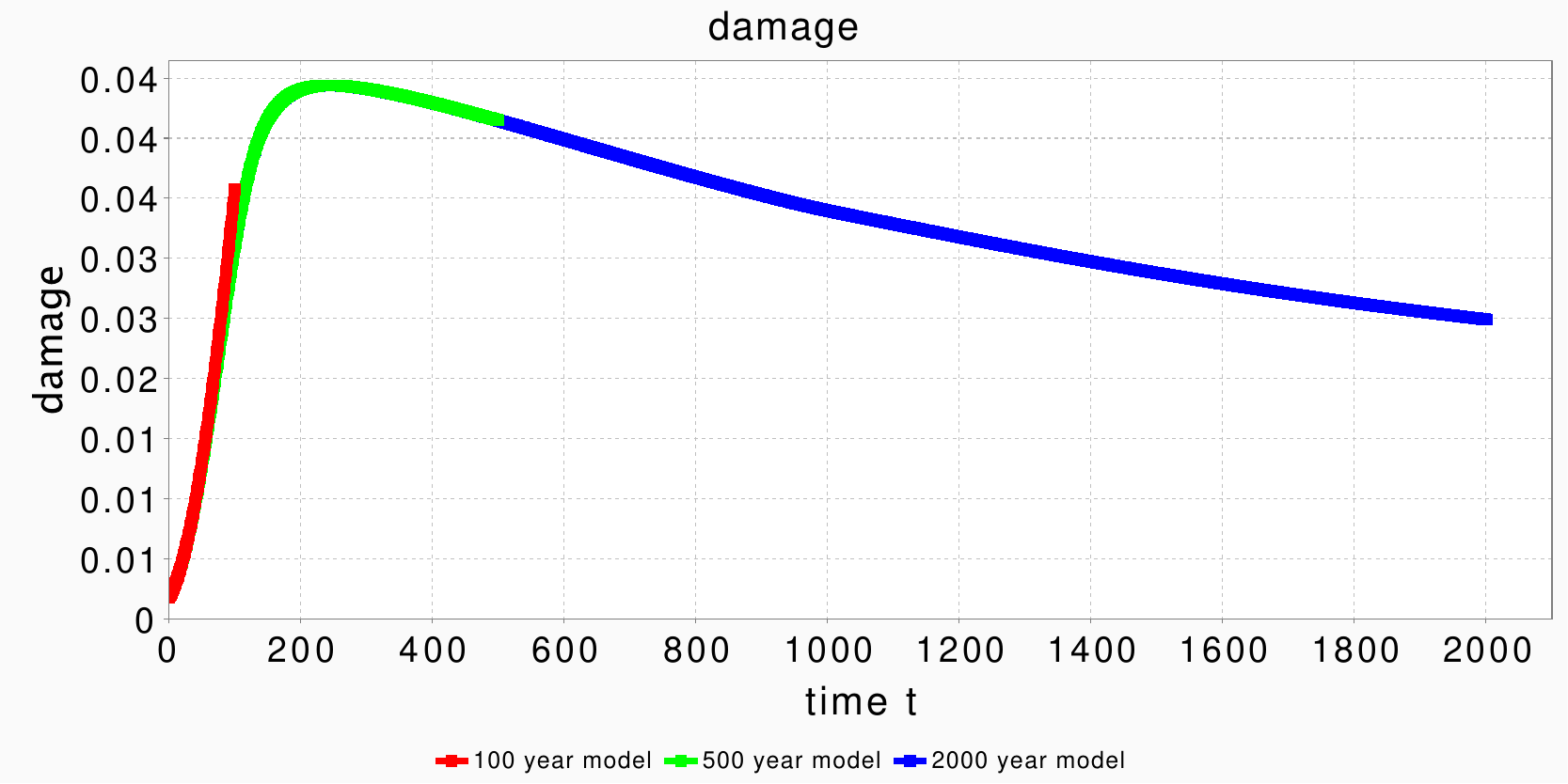}
    \caption{Damage in a calibrated model with a time horizon of 100, 500 or 2000 years.}
    \label{fig:timeConvergence-damage}
\end{figure}

\begin{figure}
    \centering
    \includegraphics[width=0.9\linewidth]{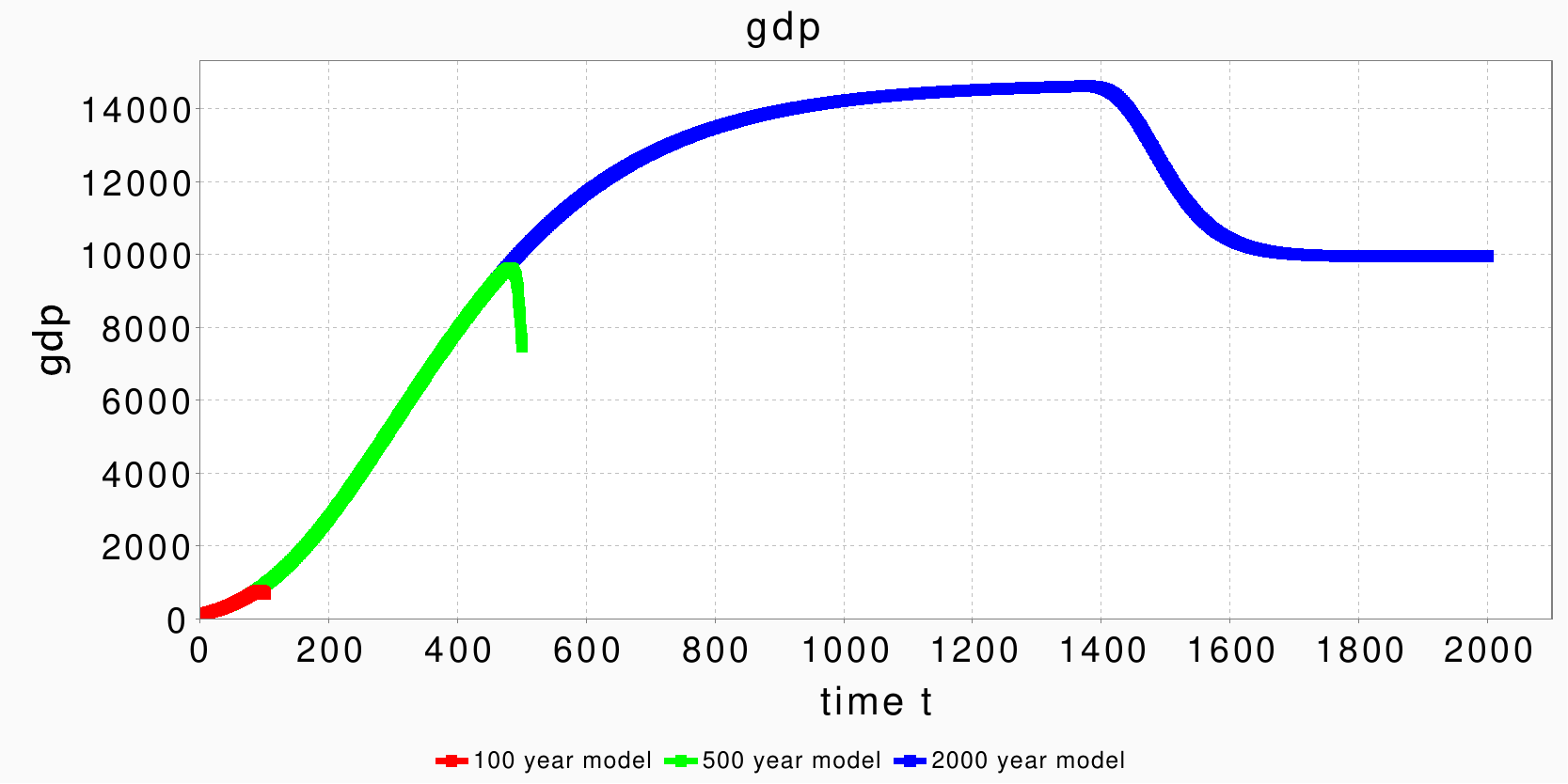}
    \caption{GDP in a calibrated model with a time horizon of 100, 500 or 2000 years.}
    \label{fig:timeConvergence-gdp}
\end{figure}

\cref{fig:timeConvergence-emission,fig:timeConvergence-carbon,fig:timeConvergence-temperature,fig:timeConvergence-damage,fig:timeConvergence-gdp} show different model entities (emission, carbon in atmosphere, temperature in atmosphere, damage and gdp) for a 100 year, 500 year and 2000 year model.
Apparently, the states start deviating among the models shortly before the individual time-horizon, but otherwise the models agree.

\clearpage
\section{Implementation}

Our model framework allows to calculate derivatives via automatic algorithmic differentiation (AAD). Since our model is stochastic, we utilize a stochastic algorithmic differentiation\footnote{As long a only pathwise operators are concerned, this is no different from applying AAD to the Monte-Carlo simulation. The algorithm differs if conditional expectations are involved.}. Since the framework allows to inject the implementation of the random variable arithmetic, this is a one-line modification to the code.

A reference implementation of our modified DICE model can be found in \cite{Fries2019}. The calibration of the model is performed using an ADAM optimiser (provided by Max Singhoff). All numerical experiments are available in a code repository \href{https://gitlab.com/finmath/finmath-climate-nonlinear}{gitlab.com/finmath/finmath-climate-nonlinear}, allowing to reproduce all figures from Section~\ref{sec:numerical_experiments}.

\end{appendix}

\end{document}